# General Relativistic 1 + 3 Orthonormal Frame Approach Revisited

Henk van Elst[1*] & Claes Uggla[2†]

1 *Astronomy Unit, Queen Mary & Westfield College, University of London, Mile End Road London E1 4NS, United Kingdom*

2 *Department of Physics, Stockholm University, Box 6730 S–113 85 Stockholm, Sweden*

and

*Department of Physics, Luleå University of Technology S–951 87 Luleå, Sweden*

February 25, 1996

**Abstract**

The equations of the 1 + 3 orthonormal frame approach are explicitly presented and discussed. Natural choices of local coordinates are mentioned. A dimensionless formulation is subsequently given. It is demonstrated how one can obtain a number of interesting problems by specializing the general equations. In particular, equation systems for "silent" dust cosmological models also containing magnetic Maxwell fields, locally rotationally symmetric spacetime geometries and spatially homogeneous cosmological models are presented. We show that while the 3-Cotton–York tensor is zero for Szekeres dust models, it is nonzero for a generic representative within the "silent" class.

*e-mail: H.van.Elst@maths.qmw.ac.uk

†e-mail: uggla@vanosf.physto.se

# 1 Introduction

In spacetime geometry scenarios of general relativistic astrophysics and cosmology it often happens that there exists a preferred timelike vector field. For example, one might have a source corresponding to a stress-energy-momentum tensor with a timelike eigendirection, perfect fluids being perhaps the most familiar example. Other cases arise when symmetries give rise to a preferred timelike direction. Spatially self-similar and spatially homogeneous models are examples of this situation. If there exists a preferred timelike vector field on the spacetime manifold then the associated unit timelike vector field $\mathbf{u}$ (such that $u_\mu u^\mu = -1$) determines tensors $U^\mu{}_\nu$ and $h^\mu{}_\nu$, respectively projecting parallel and orthogonal to $\mathbf{u}$, according to

$$\begin{aligned} U^\mu{}_\nu &:= -u^\mu u_\nu \, , \\ h^\mu{}_\nu &:= \delta^\mu{}_\nu + u^\mu u_\nu \, . \end{aligned} \tag{1.1}$$

It follows that

$$\begin{aligned} U^\mu{}_\rho U^\rho{}_\nu &= U^\mu{}_\nu \, , \quad U^\mu{}_\nu u^\nu = u^\mu \, , \quad U^\mu{}_\mu = 1 \, , \\ h^\mu{}_\rho h^\rho{}_\nu &= h^\mu{}_\nu \, , \quad h^\mu{}_\nu u^\nu = 0 \, , \quad h^\mu{}_\mu = 3 \, . \end{aligned} \tag{1.2}$$

Due to the structure imposed on the spacetime manifold by the existence of $\mathbf{u}$, a $1+3$ tensor decomposition of all geometrical objects of physical interest can be made with the help of the projection tensors $U^\mu{}_\nu$ and $h^\mu{}_\nu$. For example, it is standard to decompose the covariant derivative $\nabla_\mu u_\nu$ into its irreducible parts according to [1]

$$\nabla_\mu u_\nu = -u_\mu \dot{u}_\nu + \sigma_{\mu\nu} + \tfrac{1}{3}\Theta\, h_{\mu\nu} - \omega_{\mu\nu} \, , \tag{1.3}$$

where $\sigma_{\mu\nu}$ is symmetric and tracefree, $\omega_{\mu\nu}$ is antisymmetric, and $0 = \dot{u}_\mu u^\mu = \sigma_{\mu\nu} u^\nu = \omega_{\mu\nu} u^\nu$. The kinematical fields associated with the timelike congruence $\mathbf{u}$ are defined by

$$\begin{aligned} \dot{u}^\mu &:= u^\nu \nabla_\nu u^\mu \, , \\ \Theta &:= \nabla_\mu u^\mu \, , \\ \sigma_{\mu\nu} &:= \dot{u}_{(\mu} u_{\nu)} + \nabla_{(\mu} u_{\nu)} - \tfrac{1}{3}\Theta\, h_{\mu\nu} \, , \\ \omega_{\mu\nu} &:= -u_{[\mu} \dot{u}_{\nu]} - \nabla_{[\mu} u_{\nu]} \, , \end{aligned} \tag{1.4}$$

where $\dot{u}^\mu$ denotes the acceleration vector; $\Theta$ the (volume) rate of expansion scalar; $\sigma_{\mu\nu}$ is the rate of shear tensor, with magnitude

$$\sigma^2 := \tfrac{1}{2}\sigma_{\mu\nu}\sigma^{\mu\nu} \, , \tag{1.5}$$

and $\omega_{\mu\nu}$ is the vorticity tensor. It is convenient to define a vorticity vector

$$\omega^\mu := \tfrac{1}{2}\eta^{\mu\nu\rho\sigma}\omega_{\nu\rho} u_\sigma \quad \Longleftrightarrow \quad \omega_{\mu\nu} = \eta_{\mu\nu\rho\sigma}\omega^\rho u^\sigma \, , \tag{1.6}$$

where $\eta^{\mu\nu\rho\sigma}$ is the totally antisymmetric permutation tensor. The sign convention $\eta^{0123} = 1/\sqrt{-g}$, $\eta_{0123} = -\sqrt{-g}$ (where $g$ denotes the determinant of the spacetime metric tensor, $g_{\mu\nu}$, with Lorentzian signature), will be employed throughout, following the conventions established in Ref. [2][1]. The magnitude $\omega$ of the vorticity is defined by

$$\omega^2 := \omega_\mu \omega^\mu = \tfrac{1}{2}\omega_{\mu\nu}\omega^{\mu\nu} \, . \tag{1.7}$$

The vector field $\mathbf{u}$ is hypersurface forming if $\omega = 0$.

It is useful to define a representative length $\ell$ along the worldlines of $\mathbf{u}$, describing the volume expansion (contraction) behavior of the congruence completely, by the equation

$$\frac{u^\mu \nabla_\mu \ell}{\ell} := \tfrac{1}{3}\Theta \, . \tag{1.8}$$

[1] At this point there exists a lot of confusion in the literature. We think that in general it would be sensible to follow the conventions of Misner, Thorne and Wheeler [3], such that $\eta_{0123} = \sqrt{-g}$, $\eta^{0123} = -1/\sqrt{-g}$; hence $\eta_{\mu\nu\rho\sigma}$ is the natural canonical 4-form representing the future pointing, righthanded unit volume element in four dimensions. Then $-\eta_{\mu\nu\rho\sigma}u^\sigma = u^\sigma \eta_{\sigma\mu\nu\rho}$ translates into the unit 3-volume element $\epsilon_{ijk}$, and the decomposition of the gradient of the 4-velocity $\mathbf{u}$ reads $\nabla_\mu u_\nu = -u_\mu \dot{u}_\nu + \sigma_{\mu\nu} + \frac{1}{3}\Theta\, h_{\mu\nu} + \omega_{\mu\nu}$. Overall a *minus sign* would occur in all subsequent definitions, which involve $\eta_{\mu\nu\rho\sigma}u^\sigma$ or its orthonormal frame form $\eta_{abcd}u^d$. Furthermore, applying these conventions, in all equations terms *linear* in the vorticity would come with the opposite sign.

It is also useful to define the Hubble parameter $H$ and the dimensionless (cosmological) deceleration parameter $q$:

$$H := \frac{u^\mu \nabla_\mu \ell}{\ell} = \tfrac{1}{3}\,\Theta \ , \quad q := -\,\frac{\ell\, u^\mu \nabla_\mu (u^\nu \nabla_\nu \ell)}{(u^\rho \nabla_\rho \ell)^2} = 3\, u^\mu \nabla_\mu \left[ \frac{1}{\Theta} \right] - 1 \ . \tag{1.9}$$

Work on the covariant $1+3$ splitting of fluid spacetime geometries [1] was initiated first by Eisenhart and Synge and continued by Gödel, Raychaudhuri, and the group comprized of Schücking, Ehlers, Sachs and Trümper (see e.g. Ref. [4]). Extensions are due to, besides others, Hawking [5]. Nice reviews of this approach with particular focus on applications in relativistic cosmology were subsequently given by Ellis [6, 7].

The outline of the paper is the following: In section 2 we present and discuss the basic equations in the $1+3$ orthonormal frame approach. General features of this approach and various applications have been discussed in the past by, for example, Pirani [8], Ellis [9] and MacCallum [10, 11]. Other useful references addressing general properties of the orthonormal frame approach are the books by Wald [12] and de Felice and Clarke [13] and the paper by Edgar [14]. The equations are given explicitly in terms of irreducibly decomposed quantities, including the Bianchi identities, which had not been given fully expanded in the aforementioned references. As the curvature is associated with, in principle, physical observations, it is of advantage to consider curvature related quantities as variables, as we discuss in this section. The present formulation is completely analogous to the Newman–Penrose approach [15]; we only replace a null congruence by a timelike congruence. It is also shown how one can naturally introduce local coordinates adapted to the existence of a preferred timelike vector field in the context of the orthonormal frame approach, following the work by Jantzen et al [16]. In section 3 we give a dimensionless formulation of the $1+3$ orthonormal frame equations based on "expansion-normalized" variables. This is a direct generalization of the formulations of Wainwright and coworkers in the context of Bianchi cosmology and models with two commuting spacelike Killing vector fields (see e.g., Ref. [2]). In section 4 we give a number of examples showing the usefulness of the formalism developed in sections 2 and 3. First we show how one can include magnetic Maxwell fields in the so-called "silent" cosmological models [17] - [19]. We also discuss further possible generalizations of this class and the connection between the "silent" models and the algebraically special spacetimes in the context of the Newman–Penrose formalism. Thereafter it is shown how one can easily obtain a useful form for the equations of locally rotationally symmetric models (LRS) [9, 20] and spatially homogeneous models [21, 22], simply by deleting terms from the equations given in sections 2 and 3. We finally conclude with some remarks in section 5.

We use the following index conventions for tensors: *covariant* spacetime indices are denoted by letters from the second half of the greek alphabet ($\mu$, $\nu$, $\rho$, $\ldots = 0-3$), with spatial coordinate indices symbolized by letters from the second half of the latin alphabet ($i$, $j$, $k$, $\ldots = 1-3$); *orthonormal frame* spacetime indices are denoted by letters from the first half of the latin alphabet ($a$, $b$, $c$, $\ldots = 0-3$), with spatial frame indices chosen from the first half of the greek alphabet ($\alpha$, $\beta$, $\gamma$, $\ldots = 1-3$).

# 2 The $1+3$ Orthonormal Frame Approach

In the orthonormal frame approach one chooses at each point of the spacetime manifold a set of four linearly independent 1-forms $\{\boldsymbol{\omega}^a\}$ such that the line element can locally be expressed as

$$ds^2 = \eta_{ab}\, \boldsymbol{\omega}^a\, \boldsymbol{\omega}^b \ , \tag{2.1}$$

where $\eta_{ab} = \mathrm{diag}\,[\,-1,1,1,1\,]$, $(\sqrt{-\eta} = 1)$, is a constant Minkowskian frame metric. The vectors $\{\mathbf{e}_a\}$ dual to the 1-forms $\{\boldsymbol{\omega}^a\}$ satisfy the relation

$$\langle\, \boldsymbol{\omega}^a,\, \mathbf{e}_b \,\rangle = \delta^a{}_b \ . \tag{2.2}$$

In the $1+3$ orthonormal frame approach to fluid spacetime geometries it is customary to choose the timelike frame vector $\mathbf{e}_0$ to be the 4-velocity of the matter fluid flow, although other choices are possible as well and will be mentioned later on. Along these lines we are now going to make a $1+3$ split of the commutator relations as well as of the curvature variables and their field equations, part of which are constituted by the Bianchi identities.

## 2.1 The commutators

The commutation functions, $\gamma^a{}_{bc}$, are defined by

$$[\,\mathbf{e}_a, \mathbf{e}_b\,] = \gamma^c{}_{ab}\,\mathbf{e}_c \ , \tag{2.3}$$

where the frame vectors $\mathbf{e}_a$ are understood to act as differential operators, $\mathbf{e}_a(T)$, on any geometrical objects $T$. The covariant derivative is given by

$$\nabla_b Y^a = \mathbf{e}_b(Y^a) + \Gamma^a{}_{cb}\,Y^c \ , \quad \nabla_b V_a = \mathbf{e}_b(V_a) - \Gamma^c{}_{ab}\,V_c \ . \tag{2.4}$$

If one assumes that there is no torsion and that the connection is related to the metric through

$$\nabla_c \eta_{ab} = 0 = -\,\Gamma^d{}_{ac}\,\eta_{db} - \Gamma^d{}_{bc}\,\eta_{ad} \ , \tag{2.5}$$

one can express the Ricci rotation coefficients, $\Gamma^a{}_{bc}$, in terms of the commutation functions by the relation

$$\Gamma_{abc} = \tfrac{1}{2}\,[\ \eta_{ad}\,\gamma^d{}_{cb} + \eta_{bd}\,\gamma^d{}_{ac} - \eta_{cd}\,\gamma^d{}_{ba}\ ] \quad \Longleftrightarrow \quad \gamma^a{}_{bc} = -\,[\ \Gamma^a{}_{bc} - \Gamma^a{}_{cb}\ ] \ , \tag{2.6}$$

where $\Gamma_{abc} = \eta_{ad}\,\Gamma^d{}_{bc}$ and $\Gamma_{(ab)c} = 0$. The latter property arises as a consequence of Eq. (2.5).

Aligning the timelike direction of the orthonormal frame with the tangent of the preferred timelike congruence, $\mathbf{e}_0 = \mathbf{u}$ ( $u^a = \delta^a{}_0$, $u_a = -\,\delta^0{}_a$ ), the commutation functions with one or two indices equal to zero can be expressed in terms of the frame components of the kinematic quantities associated with the timelike congruence as defined in Eq. (1.4), and the quantity

$$\Omega^a := \tfrac{1}{2}\,\eta^{abcd}\,\mathbf{e}_b \cdot \dot{\mathbf{e}}_c\,u_d \ , \tag{2.7}$$

( where $\dot{\mathbf{e}}_a := u^b \nabla_b \mathbf{e}_a$ ), which can be interpreted as the local angular velocity of the (to be chosen) spatial frame $\{\,\mathbf{e}_\alpha\,\}$ with respect to a second spatial frame $\{\,\tilde{\mathbf{e}}_\alpha\,\}$, which is *Fermi-propagated* along $\mathbf{e}_0 = \mathbf{u}$. Secondly, the purely spatial components, $\gamma^\alpha{}_{\beta\gamma}$, are decomposed, following work by Schücking, Kundt and Behr ( see e.g. Ref. [21] ), into an object $a_\alpha$ and a symmetric object $n_{\alpha\beta}$ as follows:

$$\gamma^\alpha{}_{\beta\gamma} := 2\,a_{[\beta}\,\delta^\alpha{}_{\gamma]} + \epsilon_{\beta\gamma\delta}\,n^{\delta\alpha} \ . \tag{2.8}$$

$\epsilon_{\alpha\beta\gamma}$ is the totally antisymmetric 3-D permutation tensor with $\epsilon_{123} = 1 = \epsilon^{123}$. The commutators are described by the expressions for $\gamma^a{}_{bc}$. Their $1+3$ decomposition leads to

$$[\,\mathbf{e}_0, \mathbf{e}_\alpha\,] \ = \ \dot{u}_\alpha\,\mathbf{e}_0 - [\ \tfrac{1}{3}\,\Theta\,\delta^\beta{}_\alpha + \sigma^\beta{}_\alpha + \epsilon^\beta{}_{\alpha\gamma}\,(\,\omega^\gamma - \Omega^\gamma\,)\ ]\ \mathbf{e}_\beta \ , \tag{2.9}$$

$$[\,\mathbf{e}_\alpha, \mathbf{e}_\beta\,] \ = \ -\,2\,\epsilon_{\alpha\beta\gamma}\,\omega^\gamma\ \mathbf{e}_0 + [\ 2\,a_{[\alpha}\,\delta^\gamma{}_{\beta]} + \epsilon_{\alpha\beta\delta}\,n^{\delta\gamma}\ ]\ \mathbf{e}_\gamma \ . \tag{2.10}$$

As an aside we here mention the conditions that a particular spatial frame vector $\mathbf{e}_\alpha$ be *hypersurface orthogonal* (HSO) [12]. They are given by

$$\begin{array}{rcll} 0 & = & \sigma_{\alpha\beta} + \epsilon_{\alpha\beta\gamma}\,(\,\omega^\gamma - \Omega^\gamma\,) & \alpha \neq \beta \neq \gamma \ , \\ 0 & = & n_{\alpha\alpha} & \text{no summation} \ . \end{array} \tag{2.11}$$

## 2.2 The curvature

The relation between the Riemann curvature tensor and the Ricci rotation coefficients is given by [2]

$$R^a{}_{bcd} := \mathbf{e}_c(\Gamma^a{}_{bd}) - \mathbf{e}_d(\Gamma^a{}_{bc}) + \Gamma^a{}_{ec}\,\Gamma^e{}_{bd} - \Gamma^a{}_{ed}\,\Gamma^e{}_{bc} - \Gamma^a{}_{be}\,\gamma^e{}_{cd} \ , \tag{2.12}$$

while the Ricci curvature tensor and Ricci curvature scalar are defined by

$$R_{bd} := R^a{}_{bad} \ , \qquad R := R^{ab}{}_{ab} \ . \tag{2.13}$$

The definition of the Riemann curvature tensor and the symmetry property $\Gamma_{(ab)c} = 0$ imply $R_{abcd} = -\, R_{abdc}$ and $R_{abcd} = -\, R_{bacd}$. The 16 Jacobi identities

$$[\,[\mathbf{e}_a, \mathbf{e}_b\,], \mathbf{e}_c\,] + [\,[\mathbf{e}_b, \mathbf{e}_c\,], \mathbf{e}_a\,] + [\,[\mathbf{e}_c, \mathbf{e}_a\,], \mathbf{e}_b\,] = 0 \quad \Longleftrightarrow \quad \mathbf{e}_{[a}(\gamma^d{}_{bc]}) + \gamma^e{}_{[ab}\,\gamma^d{}_{c]e} = 0 \ , \tag{2.14}$$

correspond to the 16 equations

$$R^a{}_{[bcd]} = 0 \ , \quad R_{abcd} = R_{cdab} \ , \tag{2.15}$$

where the anticyclic relation $R^a{}_{[bcd]} = 0$, together with the antisymmetric properties of the first and last index pairs of the curvature tensor, leads to $R_{abcd} = R_{cdab}$.

The Einstein field equations with a nonzero cosmological constant can be written as

$$R^a{}_b = T^a{}_b - \tfrac{1}{2}\,T\,\delta^a{}_b + \Lambda\,\delta^a{}_b \ . \tag{2.16}$$

Here (and in the remainder of this paper) we use geometrized units, $c = 1 = 8\pi\,G/c^2$ [3], such that the dimensions of all dynamical variables can be expressed as integer powers of the single remaining nontrivial dimension [ length ].

In the phenomenological fluid description of a general matter source the standard decomposition of the stress-energy-momentum tensor $T_{ab}$ with respect to a timelike vector field $\mathbf{u}$ is given by

$$T_{ab} = \mu\,u_a\,u_b + 2\,q_{(a}\,u_{b)} + p\,h_{ab} + \pi_{ab} \ . \tag{2.17}$$

Here the following matter fields arise: $\mu$ denotes the total energy density scalar, $p$ the isotropic pressure scalar, $q^a$ the energy current density vector, and $\pi_{ab}$ the anisotropic pressure tensor. We have that

$$q_a\,u^a = 0 \ , \quad \pi_{ab}\,u^b = 0 \ , \quad \pi^a{}_a = 0 \ , \quad \pi_{ab} = \pi_{ba} \ , \tag{2.18}$$

and the matter fields need to be linked by an appropriate thermodynamic equation of state in order to provide a coherent picture of the physics underlying a fluid spacetime geometry scenario.

The completely tracefree Weyl conformal curvature tensor is defined by [2]

$$C^{ab}{}_{cd} := R^{ab}{}_{cd} - 2\,\delta^{[a}{}_{[c}\,R^{b]}{}_{d]} + \tfrac{1}{3}\,R\,\delta^a{}_{[c}\,\delta^b{}_{d]} \ . \tag{2.19}$$

When there exists a preferred timelike vector field $\mathbf{u}$, it is also useful to decompose the Weyl tensor into its "electric part"

$$E_{ac} := C_{abcd}\,u^b\,u^d \ , \tag{2.20}$$

and its "magnetic part"

$$H_{ac} := {}^*C_{abcd}\,u^b\,u^d \ , \tag{2.21}$$

where the dual is defined by

$${}^*C_{abcd} := \tfrac{1}{2}\,\eta_{ab}{}^{ef}\,C_{efcd} \ . \tag{2.22}$$

The "electric" and "magnetic" tensors are symmetric and tracefree and satisfy $E_{ab}\,u^b = 0 = H_{ab}\,u^b$. It follows that $E_{ab} = 0 = H_{ab} \Leftrightarrow C_{abcd} = 0$. These definitions lead to [6]

$$C^{ab}{}_{cd} = \Big[\,4\,\delta^{[a}{}_e\,\delta^{b]}{}_f\ \delta^g{}_{[c}\,\delta^h{}_{d]} - \eta^{ab}{}_{ef}\,\eta^{gh}{}_{cd}\,\Big]\ u^e\,u_g\,E^f{}_h - 2\,\Big[\,\eta^{ab}{}_{ef}\ \delta^g{}_{[c}\,\delta^h{}_{d]} + \delta^{[a}{}_e\,\delta^{b]}{}_f\ \eta^{gh}{}_{cd}\,\Big]\ u^e\,u_g\,H^f{}_h \tag{2.23}$$

Then, from Eq. (2.19), using Eqs. (2.16), (2.17) and (2.23), we obtain for the Riemann curvature tensor the expression

$$\begin{aligned} R^{ab}{}_{cd} &= \Big[\,4\,\delta^{[a}{}_e\,\delta^{b]}{}_f\ \delta^g{}_{[c}\,\delta^h{}_{d]} - \eta^{ab}{}_{ef}\,\eta^{gh}{}_{cd}\,\Big]\ u^e\,u_g\,E^f{}_h - 2\,\Big[\,\eta^{ab}{}_{ef}\ \delta^g{}_{[c}\,\delta^h{}_{d]} + \delta^{[a}{}_e\,\delta^{b]}{}_f\ \eta^{gh}{}_{cd}\,\Big]\ u^e\,u_g\,H^f{}_h \\ &\quad + 2\,\delta^{[a}{}_{[c}\,\Big[\,(\mu + p)\,u^{b]}\,u_{d]} + q^{b]}\,u_{d]} + u^{b]}\,q_{d]} + \pi^{b]}{}_{d]}\,\Big] + \tfrac{2}{3}\,(\mu + \Lambda)\,\delta^a{}_{[c}\,\delta^b{}_{d]} \ . \end{aligned} \tag{2.24}$$

Inserting expression (2.12) for $R^{ab}{}_{cd}$ into the lefthand side of Eq. (2.24) implies that one can obtain the Einstein field equations (by contraction on the indices $a$ and $c$), the 16 Jacobi identities and also expressions for $E_{ab}$ and $H_{ab}$ in terms of the frame derivatives $\mathbf{e}_a$ and the basic variables

$$\{\ \Theta, \dot{u}_\alpha, \sigma_{\alpha\beta}, \omega_\alpha, \Omega_\alpha, a_\alpha, n_{\alpha\beta}, \mu, p, q_\alpha, \pi_{\alpha\beta}\ \} \ . \tag{2.25}$$

**The field equations**

$$\mathbf{e}_0(\Theta) \;=\; -\tfrac{1}{3}\,\Theta^2 + (\mathbf{e}_\alpha + \dot{u}_\alpha - 2\,a_\alpha)\,(\dot{u}^\alpha) - 2\,\sigma^2 + 2\,\omega^2 - \tfrac{1}{2}\,(\mu + 3p) + \Lambda \;, \tag{2.26}$$

$$\begin{aligned}\mathbf{e}_0(\sigma^{\alpha\beta}) \;=\; & -\,\Theta\,\sigma^{\alpha\beta} + (\delta^{\gamma(\alpha}\,\mathbf{e}_\gamma + \dot{u}^{(\alpha} + a^{(\alpha})\,(\dot{u}^{\beta)}) + 2\,\omega^{(\alpha}\,\Omega^{\beta)} + \pi^{\alpha\beta} - {}^*\!S^{\alpha\beta} \\ & -\,\tfrac{1}{3}\,\delta^{\alpha\beta}\,[\,(\mathbf{e}_\gamma + \dot{u}_\gamma + a_\gamma)\,(\dot{u}^\gamma) + 2\,\omega_\gamma\,\Omega^\gamma\,] + \epsilon^{\gamma\delta(\alpha}\left[\,2\,\Omega_\gamma\,\sigma^{\beta)}{}_\delta - n^{\beta)}{}_\gamma\,\dot{u}_\delta\,\right] \;,\end{aligned} \tag{2.27}$$

$$0 \;=\; \mu - \tfrac{1}{3}\,\Theta^2 + \sigma^2 - \omega^2 - 2\,\omega_\alpha\,\Omega^\alpha - \tfrac{1}{2}\,{}^*\!R + \Lambda \;, \tag{2.28}$$

$$\begin{aligned}0 \;=\; & q^\alpha - \tfrac{2}{3}\,\delta^{\alpha\beta}\,\mathbf{e}_\beta(\Theta) + (\mathbf{e}_\beta - 3\,a_\beta)\,(\sigma^{\alpha\beta}) + n^\alpha{}_\beta\,\omega^\beta \\ & -\,\epsilon^{\alpha\beta\gamma}\,[\,(\mathbf{e}_\beta + 2\,\dot{u}_\beta - a_\beta)\,(\omega_\gamma) + n_{\beta\delta}\,\sigma^\delta{}_\gamma\,] \;,\end{aligned} \tag{2.29}$$

where

$${}^*\!S_{\alpha\beta} \;:=\; \mathbf{e}_{(\alpha}(a_{\beta)}) + b_{\alpha\beta} - \tfrac{1}{3}\,\delta_{\alpha\beta}\,[\,\mathbf{e}_\gamma(a^\gamma) + b^\gamma{}_\gamma\,] - \epsilon^{\gamma\delta}{}_{(\alpha}\,(\mathbf{e}_{|\gamma|} - 2\,a_{|\gamma|})\,(n_{\beta)\delta}) \;, \tag{2.30}$$

$${}^*\!R \;:=\; 2\,(2\,\mathbf{e}_\alpha - 3\,a_\alpha)\,(a^\alpha) - \tfrac{1}{2}\,b^\alpha{}_\alpha \;, \tag{2.31}$$

$$b_{\alpha\beta} \;:=\; 2\,n_{\alpha\gamma}\,n^\gamma{}_\beta - n^\gamma{}_\gamma\,n_{\alpha\beta} \;. \tag{2.32}$$

Equation (2.26), which is the (00)-part of Eq. (2.16), is commonly called the Raychaudhuri equation. The symmetric tracefree $(\alpha\beta)$-part of Eq. (2.16) corresponds to the shear evolution Eq. (2.27), while Eq. (2.28) constitutes a generalized Friedmann equation. Finally, Eq. (2.29) is the $(0\alpha)$-part of Eq. (2.16).

The objects ${}^*\!S_{\alpha\beta}$ and ${}^*\!R$ are the tracefree part and the trace of ${}^*\!R_{\alpha\beta}$, which is just the intrinsic 3-Ricci curvature of spacelike 3-surfaces orthogonal to the matter flow tangent $\mathbf{u}$, if $\omega^\alpha = 0$ [10]. However, when $\omega^\alpha \neq 0$, then ${}^*\!R_{\alpha\beta}$ is *not* even a tensor; it is just a symbol defined by its relation to the objects $a_\alpha$ and $n_{\alpha\beta}$. There are a number of ways one can define a "3-curvature" on the local rest 3-spaces orthogonal to $\mathbf{u}$ for the case when $\omega^\alpha \neq 0$, but we will not discuss these here. Instead we refer to Ref. [23].

**The Jacobi identities**

$$\begin{aligned}\mathbf{e}_0(a^\alpha) \;=\; & -\tfrac{1}{3}\,(\delta^{\alpha\beta}\,\mathbf{e}_\beta + \dot{u}^\alpha + a^\alpha)\,(\Theta) + \tfrac{1}{2}\,(\mathbf{e}_\beta + \dot{u}_\beta - 2\,a_\beta)\,(\sigma^{\alpha\beta}) \\ & -\,\tfrac{1}{2}\,\epsilon^{\alpha\beta\gamma}\,(\mathbf{e}_\beta + \dot{u}_\beta - 2\,a_\beta)\,(\omega_\gamma - \Omega_\gamma) \;,\end{aligned} \tag{2.33}$$

$$\begin{aligned}\mathbf{e}_0(n^{\alpha\beta}) \;=\; & -\tfrac{1}{3}\,\Theta\,n^{\alpha\beta} - (\delta^{\gamma(\alpha}\,\mathbf{e}_\gamma + \dot{u}^{(\alpha})\,(\omega^{\beta)} - \Omega^{\beta)}) + 2\,\sigma^{(\alpha}{}_\gamma\,n^{\beta)\gamma} + \delta^{\alpha\beta}\,(\mathbf{e}_\gamma + \dot{u}_\gamma)\,(\omega^\gamma - \Omega^\gamma) \\ & -\,\epsilon^{\gamma\delta(\alpha}\left[\,(\mathbf{e}_\gamma + \dot{u}_\gamma)\,(\sigma^{\beta)}{}_\delta) - 2\,n^{\beta)}{}_\gamma\,(\omega_\delta - \Omega_\delta)\,\right] \;,\end{aligned} \tag{2.34}$$

$$\mathbf{e}_0(\omega^\alpha) \;=\; -\tfrac{2}{3}\,\Theta\,\omega^\alpha + \sigma^\alpha{}_\beta\,\omega^\beta + \tfrac{1}{2}\,n^\alpha{}_\beta\,\dot{u}^\beta - \epsilon^{\alpha\beta\gamma}\,[\,\tfrac{1}{2}\,(\mathbf{e}_\beta - a_\beta)\,(\dot{u}_\gamma) + \omega_\beta\,\Omega_\gamma\,] \;, \tag{2.35}$$

$$0 \;=\; (\mathbf{e}_\beta - 2\,a_\beta)\,(n^{\alpha\beta}) - \tfrac{2}{3}\,\Theta\,\omega^\alpha - 2\,\sigma^\alpha{}_\beta\,\omega^\beta + \epsilon^{\alpha\beta\gamma}\,[\,\mathbf{e}_\beta(a_\gamma) + 2\,\omega_\beta\,\Omega_\gamma\,] \;, \tag{2.36}$$

$$0 \;=\; (\mathbf{e}_\alpha - \dot{u}_\alpha - 2\,a_\alpha)\,(\omega^\alpha) \;. \tag{2.37}$$

**The "electric" and "magnetic parts" of the Weyl tensor**

Converting the covariant equation (4.16) in Ref. [6] into a relation with respect to an orthonormal frame gives

$$\begin{aligned}E_{\alpha\beta} \;=\; & -\mathbf{e}_0(\sigma_{\alpha\beta}) - \tfrac{2}{3}\,\Theta\,\sigma_{\alpha\beta} + (\mathbf{e}_{(\alpha} + \dot{u}_{(\alpha} + a_{(\alpha})\,(\dot{u}_{\beta)}) - \sigma_{\alpha\gamma}\,\sigma^\gamma{}_\beta - \omega_\alpha\,\omega_\beta + \tfrac{1}{2}\,\pi_{\alpha\beta} \\ & -\,\tfrac{1}{3}\,\delta_{\alpha\beta}\,[\,(\mathbf{e}_\gamma + \dot{u}_\gamma + a_\gamma)\,(\dot{u}^\gamma) - 2\,\sigma^2 - \omega^2\,] + \epsilon^{\gamma\delta}{}_{(\alpha}\,[\,2\,\Omega_{|\gamma|}\,\sigma_{\beta)\delta} - n_{\beta)\gamma}\,\dot{u}_\delta\,] \;.\end{aligned} \tag{2.38}$$

Then, after combining this relation with the field equation (2.27), one easily derives

$$E_{\alpha\beta} + \tfrac{1}{2}\,\pi_{\alpha\beta} \;=\; \tfrac{1}{3}\,\Theta\,\sigma_{\alpha\beta} - \sigma_{\alpha\gamma}\,\sigma^{\gamma}{}_{\beta} - \omega_{\alpha}\,\omega_{\beta} - 2\,\omega_{(\alpha}\,\Omega_{\beta)} + \tfrac{1}{3}\,\delta_{\alpha\beta}\,[\,2\,\sigma^{2} + \omega^{2} + 2\,\omega_{\gamma}\,\Omega^{\gamma}\,] + {}^{*}S_{\alpha\beta}\;. \tag{2.39}$$

From this expression it can be seen that the "electric part" of the Weyl tensor is closely related to ${}^{*}S_{\alpha\beta}$.

To obtain an explicitly tracefree equation for the "magnetic part" of the Weyl tensor one can combine the covariant equation (4.19) in Ref. [6] with the Jacobi identity $h^{\mu}{}_{\nu}\nabla_{\mu}\omega^{\nu} = \dot{u}_{\mu}\,\omega^{\mu}$. This leads to the covariant expression

$$\begin{aligned} H_{\mu\nu} \;=\;& 2\,\dot{u}_{(\mu}\,\omega_{\nu)} + h^{\rho}{}_{(\mu}\,h^{\sigma}{}_{\nu)}\nabla_{\rho}\omega_{\sigma} - \tfrac{1}{3}\,[\,2\,\dot{u}_{\rho}\,\omega^{\rho} + h^{\rho}{}_{\sigma}\nabla_{\rho}\omega^{\sigma}\,]\,h_{\mu\nu} \\ &+\; h^{\rho}{}_{(\mu}\,h^{\sigma}{}_{\nu)}\,\eta_{\rho\tau\kappa\lambda}(\nabla^{\tau}\sigma^{\kappa}{}_{\sigma})\,u^{\lambda}\;, \end{aligned} \tag{2.40}$$

which, when formulated with respect to an orthonormal frame, gives

$$\begin{aligned} H_{\alpha\beta} \;=\;& (\mathbf{e}_{(\alpha} + 2\,\dot{u}_{(\alpha} + a_{(\alpha})\,(\omega_{\beta)}) + \tfrac{1}{2}\,n^{\gamma}{}_{\gamma}\,\sigma_{\alpha\beta} - 3\,n^{\gamma}{}_{(\alpha}\,\sigma_{\beta)\gamma} - \tfrac{1}{3}\,\delta_{\alpha\beta}\,[\,(\mathbf{e}_{\gamma} + 2\,\dot{u}_{\gamma} + a_{\gamma})\,(\omega^{\gamma}) - 3\,n_{\gamma\delta}\,\sigma^{\gamma\delta}\,] \\ &+\; \epsilon^{\gamma\delta}{}_{(\alpha}\,[\,(\mathbf{e}_{|\gamma|} - a_{|\gamma|})\,(\sigma_{\beta)\delta}) - n_{\beta)\gamma}\,\omega_{\delta}\,]\;. \end{aligned} \tag{2.41}$$

Concluding this subsection, we briefly review the deviation equation for timelike geodesics. This equation gives a direct measure of the focusing or diverging of two initially parallel fiducial timelike geodesics due to gravitational effects, as one follows along one of them towards say increasing values of its affine parameter. It was discussed in some length by, for example, Szekeres [24] (see also the earlier discussion given by Synge and Schild [25]). The separation of the geodesics is encoded in a spacelike vector field $\boldsymbol{\xi}$ (using frame components: $\xi_a\,\xi^a > 0$), orthogonal to the tangent $\mathbf{u}$ ($\xi_a\,u^a = 0$) of the reference geodesic (with $\dot{u}^a = 0$). In principle the latter represents a potential observer. Then, on using the decomposition of the Riemann curvature tensor given in Eq. (2.24), the acceleration of the spacelike vector field $\boldsymbol{\xi}$ is expressed by

$$\begin{aligned} u^{b}\nabla_{b}\,[\,u^{c}\nabla_{c}\xi^{a}\,] \;&=\; -\,R^{a}{}_{bcd}\,u^{b}\,\xi^{c}\,u^{d} \\ &=\; -\,\left[E^{a}{}_{b}(\mathbf{u}) - \tfrac{1}{2}\,\pi^{a}{}_{b}(\mathbf{u})\right]\,\xi^{b} - \tfrac{1}{6}\,[\,\mu(\mathbf{u}) + 3\,p(\mathbf{u})\,]\,\xi^{a} + \tfrac{1}{3}\,\Lambda\,\xi^{a}\;. \end{aligned} \tag{2.42}$$

Supposing that $\boldsymbol{\xi}$ represents three eigendirections of a reference volume forming body, this relation provides a nice illustration of the different possible distortions the body may experience as it moves geodesically in the presence of nonzero spacetime curvature; specifically as measured in the local rest 3-spaces orthogonal to $\mathbf{u}$. In these rest spaces three different effects contributing linearly to the acceleration of $\boldsymbol{\xi}$ can be distinguished: (i) the volume-preserving, shearing effects characteristic of tidal forces as exerted by the "electric part" of the Weyl tensor, $E_{ab}$, and the anisotropic pressure of the matter fluid, $\pi_{ab}$ (Note that $\pi_{ab}$ acts in diametrically opposed directions to $E_{ab}$ and thus tries to balance the shearing action of the latter.), (ii) the volume-shrinking effects of the total energy density, $\mu$, and scalar pressure, $p$, (given that $(\mu + 3\,p) > 0$), and (iii) the volume-inflating, antigravitating effects induced by a nonzero, positive cosmological constant, $\Lambda$ (which is an invariant spacetime constant and thus does not depend on the choice of $\mathbf{u}$). The remaining information on the components of the Riemann curvature tensor as expressed by Eq. (2.24), namely the values of $H_{ab}$ and $q^a$, can be found from measurements with respect to two further orthonormal frames $\{\,\tilde{\mathbf{e}}_a\,\}$ and $\{\,\bar{\mathbf{e}}_a\,\}$ that move in different directions relative to $\{\,\mathbf{e}_a\,\}$, which is aligned with the tangent $\mathbf{u}$ [24, 3].

## 2.3 The Bianchi identities

The Bianchi identities are given by [26]

$$\nabla_{[a}R_{bc]de} = 0 \;\;\Leftrightarrow\;\; \nabla_{d}C_{ab}{}^{cd} \;=\; -\nabla_{[a}R^{c}{}_{b]} - \tfrac{1}{6}\,\delta^{c}{}_{[a}\,\nabla_{b]}R \;=\; -\nabla_{[a}T^{c}{}_{b]} - \tfrac{1}{3}\,\delta^{c}{}_{[a}\,\nabla_{b]}T\;. \tag{2.43}$$

We are now going to express these relations in terms of $E_{\alpha\beta}$, $H_{\alpha\beta}$ and the irreducible parts of $T_{ab}$.

**Bianchi identities for the Weyl tensor**

Using the decomposition of the Weyl curvature tensor in terms of its "electric" and "magnetic parts" in the above expression (2.43) and subtracting the twice contracted Bianchi identities (2.48) and (2.49) (see below) where appropriate, leads to

$$\begin{aligned}
\mathbf{e}_0(E^{\alpha\beta} + \tfrac{1}{2}\pi^{\alpha\beta}) \;=\; & -\tfrac{1}{2}(\mu+p)\,\sigma^{\alpha\beta} - \Theta\,(E^{\alpha\beta} + \tfrac{1}{6}\pi^{\alpha\beta}) - \tfrac{1}{2}\,(\delta^{\gamma(\alpha}\,\mathbf{e}_\gamma + 2\,\dot{u}^{(\alpha} + a^{(\alpha})\,(q^{\beta)}) \\
& + 3\,\sigma^{(\alpha}{}_\gamma\,(E^{\beta)\gamma} - \tfrac{1}{6}\pi^{\beta)\gamma}) + \tfrac{1}{2}\,n^\gamma{}_\gamma\,H^{\alpha\beta} - 3\,n^{(\alpha}{}_\gamma\,H^{\beta)\gamma} \\
& + \tfrac{1}{3}\,\delta^{\alpha\beta}\,\left[\, \tfrac{1}{2}\,(\mathbf{e}_\gamma + 2\,\dot{u}_\gamma + a_\gamma)\,(q^\gamma) - 3\,\sigma_{\gamma\delta}\,(E^{\gamma\delta} - \tfrac{1}{6}\pi^{\gamma\delta}) + 3\,n_{\gamma\delta}\,H^{\gamma\delta}\,\right] \\
& + \epsilon^{\gamma\delta(\alpha}\,\Big[\, (\mathbf{e}_\gamma + 2\,\dot{u}_\gamma - a_\gamma)\,(H^{\beta)}{}_\delta) \\
& \qquad - (\omega_\gamma - 2\,\Omega_\gamma)\,(E^{\beta)}{}_\delta + \tfrac{1}{2}\pi^{\beta)}{}_\delta) + \tfrac{1}{2}\,n^{\beta)}{}_\gamma\,q_\delta \,\Big] \;,
\end{aligned} \tag{2.44}$$

$$\begin{aligned}
\mathbf{e}_0(H^{\alpha\beta}) \;=\; & -\Theta\,H^{\alpha\beta} + 3\,\sigma^{(\alpha}{}_\gamma\,H^{\beta)\gamma} - \tfrac{3}{2}\,\omega^{(\alpha}\,q^{\beta)} - \tfrac{1}{2}\,n^\gamma{}_\gamma\,(E^{\alpha\beta} - \tfrac{1}{2}\pi^{\alpha\beta}) \\
& + 3\,n^{(\alpha}{}_\gamma\,(E^{\beta)\gamma} - \tfrac{1}{2}\pi^{\beta)\gamma}) \\
& - \delta^{\alpha\beta}\,\left[\, \sigma_{\gamma\delta}\,H^{\gamma\delta} - \tfrac{1}{2}\,\omega_\gamma\,q^\gamma + n_{\gamma\delta}\,(E^{\gamma\delta} - \tfrac{1}{2}\pi^{\gamma\delta})\,\right] \\
& - \epsilon^{\gamma\delta(\alpha}\,\Big[\, (\mathbf{e}_\gamma - a_\gamma)\,(E^{\beta)}{}_\delta - \tfrac{1}{2}\pi^{\beta)}{}_\delta) + 2\,\dot{u}_\gamma\,E^{\beta)}{}_\delta \\
& \qquad - \tfrac{1}{2}\,\sigma^{\beta)}{}_\gamma\,q_\delta + (\omega_\gamma - 2\,\Omega_\gamma)\,H^{\beta)}{}_\delta \,\Big] \;,
\end{aligned} \tag{2.45}$$

$$\begin{aligned}
0 \;=\; & (\mathbf{e}_\beta - 3\,a_\beta)\,(E^{\alpha\beta} + \tfrac{1}{2}\pi^{\alpha\beta}) - \tfrac{1}{3}\,\delta^{\alpha\beta}\,\mathbf{e}_\beta(\mu) + \tfrac{1}{3}\,\Theta\,q^\alpha - \tfrac{1}{2}\,\sigma^\alpha{}_\beta\,q^\beta + 3\,\omega_\beta\,H^{\alpha\beta} \\
& - \epsilon^{\alpha\beta\gamma}\,\left[\, \sigma_{\beta\delta}\,H^\delta{}_\gamma + \tfrac{3}{2}\,\omega_\beta\,q_\gamma + n_{\beta\delta}\,(E^\delta{}_\gamma + \tfrac{1}{2}\pi^\delta{}_\gamma)\,\right] \;,
\end{aligned} \tag{2.46}$$

$$\begin{aligned}
0 \;=\; & (\mathbf{e}_\beta - 3\,a_\beta)\,(H^{\alpha\beta}) - (\mu+p)\,\omega^\alpha - 3\,\omega_\beta\,(E^{\alpha\beta} - \tfrac{1}{6}\pi^{\alpha\beta}) - \tfrac{1}{2}\,n^\alpha{}_\beta\,q^\beta \\
& + \epsilon^{\alpha\beta\gamma}\,\left[\, \tfrac{1}{2}\,(\mathbf{e}_\beta - a_\beta)\,(q_\gamma) + \sigma_{\beta\delta}\,(E^\delta{}_\gamma + \tfrac{1}{2}\pi^\delta{}_\gamma) - n_{\beta\delta}\,H^\delta{}_\gamma\,\right] \;.
\end{aligned} \tag{2.47}$$

**Bianchi identities for the source terms**

These arise after double contraction of Eq. (2.43), projecting the resultant relation respectively parallel and orthogonal to $\mathbf{u}$. We get

$$\mathbf{e}_0(\mu) \;=\; -\,(\mu+p)\,\Theta - (\mathbf{e}_\alpha + 2\,\dot{u}_\alpha - 2\,a_\alpha)\,(q^\alpha) - \sigma_{\alpha\beta}\,\pi^{\alpha\beta} \;, \tag{2.48}$$

$$\begin{aligned}
\mathbf{e}_0(q^\alpha) \;=\; & -\tfrac{4}{3}\,\Theta\,q^\alpha - \delta^{\alpha\beta}\,\mathbf{e}_\beta(p) - (\mu+p)\,\dot{u}^\alpha - (\mathbf{e}_\beta + \dot{u}_\beta - 3\,a_\beta)\,(\pi^{\alpha\beta}) - \sigma^\alpha{}_\beta\,q^\beta \\
& + \epsilon^{\alpha\beta\gamma}\,\left[\, (\omega_\beta + \Omega_\beta)\,q_\gamma + n_{\beta\delta}\,\pi^\delta{}_\gamma\,\right] \;.
\end{aligned} \tag{2.49}$$

If the fluid flow congruence is irrotational, $\omega^\mu = 0$, it is convenient to define a tensor encoding the conformal curvature properties of the 3-surfaces orthogonal to $\mathbf{u}$. Throughout the literature it is mainly the conformally invariant symmetric tracefree 3-Cotton–York tensor (density) [27, 28] which is employed for these purposes. This tensor is divergence-free. We present it defined as the spatial rotation of the symmetric tracefree 3-Ricci tensor, a form due to Jantzen et al [23], which is equivalent to its usual definition. The symmetric tracefree 3-Ricci tensor can be obtained by use of the Gauß embedding equation [29]. In covariant form one gets:

$${}^*S_{\mu\nu} = (E_{\mu\nu} + \tfrac{1}{2}\pi_{\mu\nu}) - \tfrac{1}{3}\,\Theta\,\sigma_{\mu\nu} + \sigma_{\mu\rho}\,\sigma^\rho{}_\nu - \tfrac{2}{3}\,\sigma^2\,h_{\mu\nu} \;. \tag{2.50}$$

Then it follows for the covariant components of the 3-Cotton–York tensor ($h$ denotes the determinant of $h_{\mu\nu}$):

$${}^*C_{\mu\nu} \;:=\; h^{1/3}\,h^\rho{}_{(\mu}\,h^\sigma{}_{\nu)}\,\eta_{\rho\tau\kappa\lambda}(\nabla^\tau\,{}^*S^\kappa{}_\sigma)\,u^\lambda$$

$$
\begin{aligned}
= & \; - h^{1/3} \left[ h^\rho{}_\mu \, h^\sigma{}_\nu \, u^\tau \nabla_\tau H_{\rho\sigma} + \tfrac{4}{3} \Theta \, H_{\mu\nu} - 3 \, \sigma^\rho{}_{(\mu} \, H_{\nu)\rho} + \sigma^\rho{}_\sigma \, H^\sigma{}_\rho \, h_{\mu\nu} \right] \\
& + \; h^{1/3} \, h^\rho{}_{(\mu} \, h^\sigma{}_{\nu)} \eta_{\rho\tau\kappa\lambda} \left[ \nabla^\tau (\sigma^\kappa{}_\xi \, \sigma^\xi{}_\sigma + \pi^\kappa{}_\sigma) - \tfrac{1}{3} (\nabla^\tau \Theta) \, \sigma^\kappa{}_\sigma \right. \\
& \qquad \left. - \; 2 \, \dot{u}^\tau \, E^\kappa{}_\sigma + \tfrac{1}{2} \, \sigma^\tau{}_\sigma \, q^\kappa \right] u^\lambda \; , \qquad (2.51)
\end{aligned}
$$

where we have made a substitution in terms of the covariant form of Eq. (2.45) [6]. Again, specializing this expression to an orthonormal frame we arive at

$$
{}^*C_{\alpha\beta} = \epsilon^{\gamma\delta}{}_{(\alpha} \, (\mathbf{e}_{|\gamma|} - a_{|\gamma|}) \, ({}^*S_{\beta)\delta}) - 3 \, n^\gamma{}_{(\alpha} \, {}^*S_{\beta)\gamma} + \tfrac{1}{2} \, n^\gamma{}_\gamma \, {}^*S_{\alpha\beta} + \delta_{\alpha\beta} \, n_{\gamma\delta} \, {}^*S^{\gamma\delta} \qquad (2.52)
$$

$$
\begin{aligned}
= & \; - \mathbf{e}_0(H_{\alpha\beta}) - \tfrac{4}{3} \Theta \, H_{\alpha\beta} + 3 \, \sigma^\gamma{}_{(\alpha} \, H_{\beta)\gamma} - 3 \, n^\gamma{}_{(\alpha} \, \sigma^\delta{}_{\beta)} \, \sigma_{\gamma\delta} + \tfrac{1}{2} \, n^\gamma{}_\gamma \, \sigma_{\alpha\delta} \, \sigma^\delta{}_\beta + 2 \, n_{\alpha\beta} \, \sigma^2 \\
& - \; 3 \, n^\gamma{}_{(\alpha} \, \pi_{\beta)\gamma} + \tfrac{1}{2} \, n^\gamma{}_\gamma \, \pi_{\alpha\beta} \\
& - \; \delta_{\alpha\beta} \left[ \sigma_{\gamma\delta} \, H^{\gamma\delta} - n^\gamma{}_\delta \, \sigma^\delta{}_\epsilon \, \sigma^\epsilon{}_\gamma + n^\gamma{}_\gamma \, \sigma^2 - n_{\gamma\delta} \, \pi^{\gamma\delta} \right] \\
& + \; \epsilon^{\gamma\delta}{}_{(\alpha} \left[ (\mathbf{e}_{|\gamma|} - a_{|\gamma|}) \, (\sigma^\epsilon{}_{\beta)} \, \sigma_{\delta\epsilon} + \pi_{\beta)\delta}) - \tfrac{1}{3} \, \mathbf{e}_{|\gamma|}(\Theta) \, \sigma_{\beta)\delta} - 2 \, \dot{u}_{|\gamma|} \, E_{\beta)\delta} \right. \\
& \qquad \left. + \; \tfrac{1}{2} \, \sigma_{\beta)\gamma} \, q_\delta + 2 \, \Omega_{|\gamma|} \, H_{\beta)\delta} \right] \; , \qquad (2.53)
\end{aligned}
$$

where Eq. (2.30) for ${}^*S_{\alpha\beta}$ is understood to be inserted into Eq. (2.52).

## 2.4 The source

There exists an abundance of possible choices for the matter source of the gravitational field. Here we will, however, consider only perfect fluids and Maxwell vacuum fields.

### 2.4.1 Perfect fluids

A perfect fluid is described by the stress-energy-momentum tensor

$$
T_{ab} = \tilde{\mu} \, \tilde{u}_a \, \tilde{u}_b + \tilde{p} \, \tilde{h}_{ab} \; , \qquad (2.54)
$$

where $\tilde{u}^a$ is the fluid 4-velocity, $\tilde{\mu}$ the total energy density, $\tilde{p}$ the pressure and $\tilde{h}^a{}_b$ the orthogonal projection tensor associated with $\tilde{u}^a$. It thus has vanishing energy current density ($\tilde{q}^a = 0$) and anisotropic pressure ($\tilde{\pi}_{ab} = 0$). If we choose to take the comoving approach ($\mathbf{e}_0 = \tilde{\mathbf{u}}$), the twice contracted Bianchi identities (2.48) and (2.49) reduce to

$$
\begin{aligned}
\mathbf{e}_0(\tilde{\mu}) &= - (\tilde{\mu} + \tilde{p}) \, \Theta \; , && (2.55) \\
0 &= \mathbf{e}_\alpha(\tilde{p}) + (\tilde{\mu} + \tilde{p}) \, \dot{u}_\alpha \; . && (2.56)
\end{aligned}
$$

However, one does not always want to choose $\mathbf{e}_0$ to coincide with the fluid 4-velocity. In the case of several perfect fluids with *different* $\mathbf{u}$ one can only adapt to one of them. One might also have some other structure one wants to adapt to (e.g., in the spatially homogeneous models below we will choose $\mathbf{e}_0$ to be the unit normal of the spatially homogeneous 3-surfaces). If the fluid is rotating one might want to use a 3 + 1 initial value formulation and choose $\mathbf{e}_0$ to be the unit normal associated with a family of spacelike 3-slices. We say that the fluid is *tilted* if $\tilde{\mathbf{u}} \neq \mathbf{e}_0$. In this case we obtain a stress-energy-momentum tensor $T_{ab} = \mu \, u_a \, u_b + 2 \, q_{(a} \, u_{b)} + p \, h_{ab} + \pi_{ab}$ with

$$
\begin{aligned}
\mu &= \Gamma^2 \, (\tilde{\mu} + \tilde{p}) - \tilde{p} \; , & p &= \tfrac{1}{3} \, (\tilde{\mu} + \tilde{p}) \, \Gamma^2 \, v^2 + \tilde{p} \; , \\
q^a &= \Gamma^2 \, (\tilde{\mu} + \tilde{p}) \, v^a \; , & \pi_{ab} &= \Gamma^2 \, (\tilde{\mu} + \tilde{p}) \, ( v_a \, v_b - \tfrac{1}{3} \, v^2 \, h_{ab} ) \; ,
\end{aligned} \qquad (2.57)
$$

where $\tilde{u}^a = \Gamma \, (u^a + v^a)$, $u_a \, v^a = 0$, $\Gamma := 1/\sqrt{1 - v^2}$, $v^2 := v_a \, v^a$. For a linear barotropic equation of state of the form $\tilde{p}(\tilde{\mu}) = (\gamma - 1) \, \tilde{\mu}$ these equations reduce to

$$
\begin{aligned}
\mu &= \Gamma^2 \, G \, \tilde{\mu} \; , & p &= \tfrac{1}{3} \, G^{-1} \, [ \, (3 - 2 \, \gamma) \, v^2 + 3 \, (\gamma - 1) \, ] \, \mu \; , \\
q^a &= \gamma \, G^{-1} \, v^a \, \mu \; , & \pi_{ab} &= \gamma \, G^{-1} \, ( v_a \, v_b - \tfrac{1}{3} \, v^2 \, h_{ab} ) \, \mu \; ,
\end{aligned} \qquad (2.58)
$$

where $G := 1 + (\gamma - 1) \, v^2$.

The source equations (2.48) and (2.49) yield evolution equations for $\mu$ and $v^a$ respectively. The one for $\mu$ is obtained directly by inserting the above expressions into the $\mu$-equation for a general stress-energy-momentum tensor. The one for $v^a$ is quite complicated and we will only show how one can obtain the $\tilde{p}(\tilde{\mu}) = (\gamma - 1)\,\tilde{\mu}$ case by taking the following linear combination of the $\mu$- and $q^a$-equations:

$$\mathbf{e}_0(v^\alpha) = \frac{G}{\gamma\,\mu\,[\,1-(\gamma-1)\,v^2\,]}\,[\,-\gamma\,v^\alpha\,\mathbf{e}_0(\mu) + [\,1-(\gamma-1)\,v^2\,]\,\mathbf{e}_0(q^\alpha) + 2\,(\gamma-1)\,v^\alpha\,v_\beta\,\mathbf{e}_0(q^\beta)\,\,] \ . \tag{2.59}$$

This equation will be evaluated explicitly in dimensionless form for spatially homogeneous perfect fluid models below (it is quite cumbersome in the general spatially inhomogeneous case). One can obtain an evolution equation for $\tilde{\mu}$ once one has one for $v^\alpha$. However, it is often preferable to keep $\mu$ and use the constraint equation (2.28) to solve for $\mu$ in terms of other variables.

### 2.4.2 Maxwell vacuum fields

An electromagnetic field is characterized by the antisymmetric field strength tensor [7]

$$F_{ab} = u_a\,E_b - u_b\,E_a + \eta_{abcd}\,H^c\,u^d \ , \tag{2.60}$$

where $E_a$ denotes the electric field and $H_a$ the magnetic field respectively and have the property $0 = E_a\,u^a = H_a\,u^a$. Treating the Maxwell vacuum as a fluid with 4-velocity $\mathbf{u}$ the contributions to its stress-energy-momentum tensor in an orthonormal frame can be expressed by

$$\mu = \tfrac{1}{2}\,(E_\alpha\,E^\alpha + H_\alpha\,H^\alpha) = 3\,p \ , \tag{2.61}$$

$$q^\alpha = \epsilon^{\alpha\beta\gamma}\,E_\beta\,H_\gamma \ , \tag{2.62}$$

$$\pi_{\alpha\beta} = -\,E_\alpha\,E_\beta - H_\alpha\,H_\beta + \tfrac{1}{3}\,\delta_{\alpha\beta}\,(E_\gamma\,E^\gamma + H_\gamma\,H^\gamma) \ . \tag{2.63}$$

The *sourcefree Maxwell equations* with respect to this orthonormal frame are given by

$$\mathbf{e}_0(E^\alpha) = -\tfrac{2}{3}\,\Theta\,E^\alpha + \sigma^\alpha{}_\beta\,E^\beta - n^\alpha{}_\beta\,H^\beta + \epsilon^{\alpha\beta\gamma}\,[\,(\mathbf{e}_\beta + \dot{u}_\beta - a_\beta)\,(H_\gamma) - (\omega_\beta - \Omega_\beta)\,E_\gamma\,] \ , \tag{2.64}$$

$$\mathbf{e}_0(H^\alpha) = -\tfrac{2}{3}\,\Theta\,H^\alpha + \sigma^\alpha{}_\beta\,H^\beta + n^\alpha{}_\beta\,E^\beta - \epsilon^{\alpha\beta\gamma}\,[\,(\mathbf{e}_\beta + \dot{u}_\beta - a_\beta)\,(E_\gamma) + (\omega_\beta - \Omega_\beta)\,H_\gamma\,] \ , \tag{2.65}$$

$$0 = (\mathbf{e}_\alpha - 2\,a_\alpha)\,(E^\alpha) + 2\,\omega_\alpha\,H^\alpha \ , \tag{2.66}$$

$$0 = (\mathbf{e}_\alpha - 2\,a_\alpha)\,(H^\alpha) - 2\,\omega_\alpha\,E^\alpha \ . \tag{2.67}$$

The covariant form of these equations was discussed in Ref. [7].

## 2.5 Introducing local coordinates

For many purposes one does not need to introduce local coordinates; it is often not even advisable to do so. However, many things *do* rely on the introduction of a local coordinate system, and if one wants to introduce coordinates one might as well do so in a geometric way reflecting the problem at hand. In the present context we have a preferred timelike congruence $\mathbf{u}$. The natural way of introducing coordinates adapted to this particular structure is the $1+3$ *threading approach*, recently discussed by Jantzen et al [16] and Boersma and Dray [30]. In this approach one expresses the orthonormal frame as follows:

$$\mathbf{e}_0 = M^{-1}\,\partial_t \ , \quad \mathbf{e}_\alpha = e_\alpha{}^i\,(M_i\,\partial_t + \partial_i) \ . \tag{2.68}$$

This leads to the following expression for the 4-geometry

$$\begin{aligned} {}^{(4)}\mathrm{g}^{-1} &= -\,\mathbf{e}_0 \otimes \mathbf{e}_0 + \delta^{\alpha\beta}\,\mathbf{e}_\alpha \otimes \mathbf{e}_\beta \\ &= -\,M^{-2}\,\partial_t \otimes \partial_t + \delta^{\alpha\beta}\,e_\alpha{}^i\,e_\beta{}^j (M_i\,\partial_t + \partial_i) \otimes (M_j\,\partial_t + \partial_j) \ , \end{aligned} \tag{2.69}$$

where $M = M(t, x^i)$ is the *threading lapse function* and $M_i\, dx^i = M_i(t, x^i)\, dx^i$ is the *threading shift 1-form* [16, 30]. The triad vector components ${e_\alpha}^i = {e_\alpha}^i(t, x^i)$ describe the *threading metric* $\gamma^{ij} = \delta^{\alpha\beta}\, {e_\alpha}^i\, {e_\beta}^j$ ( this is just the spatial projection tensor expressed in local coordinates; for a discussion on its physical significance, see Refs. [16], [30] and [31] ). Thus, in the present 1 + 3 threading approach, $t$ is a coordinate along the timelike congruence, while the $x^i$ parametrize the different flow lines of the timelike congruence. This is the "dual" formulation of the more familiar 3 + 1 decomposition. In that formulation one imposes a causal condition on 3-surfaces, which are assumed to be spacelike, while *no* condition is imposed on the flow lines threading the 3-surfaces. In the threading approach one imposes a causal condition on the congruence, which is assumed to be timelike, while *no* causal condition is imposed on the 3-surfaces.

Note that $M_i \neq 0$ if one has a rotating congruence, $\omega^\alpha \neq 0$, which is the case when the local rest 3-spaces are not hypersurface forming. Of course, one still has a considerable freedom in choosing the threading shift $M_i$; a freedom which just corresponds to the freedom of choosing a foliation such that the timelike congruence is nowhere tangent to it.

Expressing the commutators (2.9) and (2.10) in terms of the variables $M$, $M_i$ and ${e_\alpha}^i$ yields the relations

$$
\begin{aligned}
{e_\alpha}^i\, [\; (M_i)_{,t} + M^{-1}\, (M_{,i} + M_i\, M_{,t})\; ] &= \dot u_\alpha\ , && (2.70)\\
({e_\alpha}^i)_{,t} &= -\, M\, {e_\beta}^i\, [\; \tfrac{1}{3}\,\Theta\, {\delta^\beta}_\alpha + {\sigma^\beta}_\alpha + {\epsilon^\beta}_{\alpha\gamma}\, (\omega^\gamma - \Omega^\gamma)\; ]\ , && (2.71)\\
M\, {e_{[\alpha}}^i\, {e_{\beta]}}^j\, (M_{i,j} + M_{i,t}\, M_j) &= \epsilon_{\alpha\beta\gamma}\, \omega^\gamma\ , && (2.72)\\
2\, {e_{[\alpha}}^i\, [\; ({e_{\beta]}}^j)_{,i} + ({e_{\beta]}}^j)_{,t}\, M_i\; ]\, {e^\gamma}_j &= 2\, a_{[\alpha}\, {\delta^\gamma}_{\beta]} + \epsilon_{\alpha\beta\delta}\, n^{\delta\gamma}\ , && (2.73)
\end{aligned}
$$

where $f_{,t} := \partial_t f$, $f_{,i} := \partial_i f$, and where ${e^\alpha}_i$ is defined through the relation

$$
{e^\alpha}_i\, {e_\alpha}^j = {\delta^j}_i\ . \tag{2.74}
$$

Note that one can insert the expression for $({e_\alpha}^i)_{,t}$ given in Eq. (2.71) in Eq. (2.73), thus obtaining an expression involving only spatial derivatives.

As an example of the threading approach we consider Taub's comoving coordinates for isentropic perfect fluids with equation of state $\tilde p = \tilde p(\tilde\mu)$ [32]. Taub's comoving coordinates correspond to the following threading lapse function and shift 1-form:

$$
M = r^{-1}\ , \quad r := r_c\, \exp \int_{\tilde p_0}^{\tilde p} \frac{d\tilde p}{(\tilde\mu + \tilde p)}\ , \quad M_i = M_i(x^1, x^2, x^3)\ , \quad u^a = r\, {\delta^a}_0\ , \tag{2.75}
$$

where $r$ is related to the fluid's enthalpy density $(\tilde\mu + \tilde p)$. With this choice of threading lapse function and shift 1-form the momentum conservation equation (2.56) will be automatically satisfied, as can also be seen from Eq. (2.70).

If the congruence is nonrotating ($\omega^\alpha = 0$), it naturally gives rise to a foliation of the spacetime manifold by spacelike 3-slices. In this case one can choose to focus on the slices rather than the congruence. This situation suggests to take the 3 + 1 *slicing point of view* in which one introduces local spatial coordinates on the slices, while one introduces timelines that do not, in general, follow the original timelike congruence ( see e.g. Refs. [16] and [30], as well as the standard texts in Misner, Thorne and Wheeler [3] and by York [33] ). Here we will choose to express the spatial triad $\{\, \mathbf{e}_\alpha\, \}$ in a noncoordinate basis, since this is useful in the context of, for example, Bianchi cosmology (a similar formulation could have been introduced in the threading approach; but we will not pursue this any further here). In the slicing approach the orthonormal frame is expressed as follows (Note that, deviating from our conventions, here $i$, $j$, $k$ are also used as spatial indices for the noncoordinate basis.):

$$
\mathbf{e}_0 = N^{-1}\, (\partial_t - \overline{N}^i\, \overline{\mathbf{e}}_i)\ , \quad \mathbf{e}_\alpha = {\overline{e}_\alpha}^i\, \overline{\mathbf{e}}_i\ , \tag{2.76}
$$

which leads to the following 4-geometry

$$
\begin{aligned}
{}^{(4)}\mathrm{g}^{-1} &= -\, \mathbf{e}_0 \otimes \mathbf{e}_0 + \delta^{\alpha\beta}\, \mathbf{e}_\alpha \otimes \mathbf{e}_\beta \\
&= -\, N^{-2}\, (\partial_t - \overline{N}^i\, \overline{\mathbf{e}}_i) \otimes (\partial_t - \overline{N}^j\, \overline{\mathbf{e}}_j) + \delta^{\alpha\beta}\, {\overline{e}_\alpha}^i\, {\overline{e}_\beta}^j\, \overline{\mathbf{e}}_i \otimes \overline{\mathbf{e}}_j\ , && (2.77)
\end{aligned}
$$

where $N = N(t, x^i)$ is the *lapse function* and $\overline{N}^i = \overline{N}^i(t, x^i)$ is the *shift vector.* The inverse $\overline{e}^\alpha{}_i(t, x^k)$ to $\overline{e}_\alpha{}^j(t, x^k)$ ($\overline{e}^\alpha{}_i\,\overline{e}_\alpha{}^j = \delta^j{}_i$) yields the spatial metric tensor $g_{ij} = \delta_{\alpha\beta}\,\overline{e}^\alpha{}_i\,\overline{e}^\beta{}_j$ on the spacelike 3-slices expressed in the basis $\overline{\mathbf{e}}_i = \overline{e}_i{}^j(x^k)\,\partial_j$.

The above basis yields the commutation relations

$$[\,\partial_t, \overline{\mathbf{e}}_i\,] = 0\ , \quad [\,\overline{\mathbf{e}}_i, \overline{\mathbf{e}}_j\,] = \overline{\gamma}^k{}_{ij}(x^l)\,\overline{\mathbf{e}}_k\ , \tag{2.78}$$

which, when compared to the commutators (2.9) and (2.10) and setting $\omega^\alpha = 0$, leads to

$$N^{-1}\ \overline{e}_\alpha{}^i\,\overline{\mathbf{e}}_i(N) \ = \ \dot{u}_\alpha\ , \tag{2.79}$$

$$\begin{aligned}(\overline{e}_\alpha{}^i)_{,t} \ &= \ \overline{N}^j\ \overline{\mathbf{e}}_j(\overline{e}_\alpha{}^i) - \overline{e}_\alpha{}^j\,\overline{\mathbf{e}}_j(\overline{N}^i) + \overline{N}^j\,\overline{e}_\alpha{}^k\,\overline{\gamma}^i{}_{jk} \\ &\quad - N\,\overline{e}_\beta{}^i\ [\ \tfrac{1}{3}\,\Theta\,\delta^\beta{}_\alpha + \sigma^\beta{}_\alpha - \epsilon^\beta{}_{\alpha\gamma}\,\Omega^\gamma\ ]\ ,\end{aligned} \tag{2.80}$$

$$\overline{e}^\gamma{}_k\ [\ 2\,\overline{e}_{[\alpha}{}^i\,\overline{\mathbf{e}}_i(\overline{e}_{\beta]}{}^k) + \overline{e}_{[\alpha}{}^i\,\overline{e}_{\beta]}{}^j\,\overline{\gamma}^k{}_{ij}\ ] \ = \ 2\,a_{[\alpha}\,\delta^\gamma{}_{\beta]} + \epsilon_{\alpha\beta\delta}\,n^{\delta\gamma}\ . \tag{2.81}$$

Geodesic and nonrotating congruences ( $0 = \dot{u}^\alpha = \omega^\alpha$ ) imply that one can introduce a synchronous local coordinate system ( $N = 1$, $\overline{N}^i = 0$; $M = 1$, $M_i = 0$ ). These congruences have several interesting consequences. Equations (1.9) and (2.26) imply

$$q = -\,\frac{\ell\,\mathbf{e}_0[\,\mathbf{e}_0(\ell)\,]}{[\,\mathbf{e}_0(\ell)\,]^2} = 3\,\mathbf{e}_0\left[\,\frac{1}{\Theta}\,\right] - 1 = \frac{3}{\Theta^2}\,[\ 2\sigma^2 + \tfrac{1}{2}\,(\mu + 3\,p) - \Lambda\ ]\ . \tag{2.82}$$

Thus $(\mu + 3\,p) \geq 0$ and $\Lambda \leq 0$ imply that $q \geq 0$, and that a singularity forms in the past (future) with $\ell = 0$ if $\Theta > 0$ ($\Theta < 0$). The singularity may be a curvature singularity or just a crushing singularity ( see e.g. de Felice and Clarke [13] and MacCallum [10] ). Equation (2.28) and the above expression for $q$ lead to

$$q = 2 - \frac{3}{\Theta^2}\,[\ \tfrac{3}{2}\,(\mu - p) + 3\Lambda - {}^*\!R\ ]\ . \tag{2.83}$$

Thus if $\Lambda \geq 0$, $\mu \geq p$, ${}^*\!R\ \leq 0$, this leads to the inequality $q \leq 2$. Hence, if $\Lambda = 0$, $(\mu + 3\,p) \geq 0$, $(\mu - p) \geq 0$, ${}^*\!R \leq 0$, it follows that $0 \leq q \leq 2$.

Of course, there are other ways local coordinates can be introduced. For example, one might want to focus on events in the neighborhood of a single line of the timelike congruence one is interested in. In this case it is natural to introduce *Fermi normal coordinates* ( for a discussion see e.g. Ref. [3] ).

## 2.6 Evolution equations and constraints

The usual initial value problem in General Relativity is associated with a $3+1$ slicing formulation. The $1+3$ approach, with its associated threading formulation, is uncharted territory ( for a discussion see [23] ). We will thus instead discuss the $3+1$ approach in the present tetrad formulation. In the $3+1$ initial value formulation one usually uses the spatial metric tensor of the 3-slices and its extrinsic curvature tensor as variables. The lapse function represents the freedom in choosing a spatial foliation (choice of time), while the shift vector represents the freedom in choosing a threading (choice of spatial gauge). In the present formulation the triad components, $\overline{e}_\alpha{}^i$, replace the spatial metric ( $g_{ij} = \delta_{\alpha\beta}\,\overline{e}^\alpha{}_i\,\overline{e}^\beta{}_j \Rightarrow \delta_{\alpha\beta}\,\overline{e}^\alpha{}_{[i}\,\overline{e}^\beta{}_{j]} = 0$ ), while the expansion scalar and the shear tensor correspond to the trace and tracefree part of the extrinsic curvature tensor respectively; the frame derivative $\mathbf{e}_0$ is just the normal derivative to the spacelike 3-slices. The role of derivatives in the tetrad formulation becomes trivial, since all tetrad components are spacetime scalars. Note that it is only $\mathbf{e}_0$ that contains a temporal partial derivative ( $\mathbf{e}_0 = N^{-1}\,(\partial_t - \overline{N}^i\,\overline{\mathbf{e}}_i)$ , $\mathbf{e}_\alpha = \overline{e}_\alpha{}^i\,\overline{\mathbf{e}}_i$ ). Thus one obtains an evolution equation in standard form when one has a $\mathbf{e}_0$-derivative on the lefthand side simply by moving the shift part to the righthand side. Thus equations involving $\mathbf{e}_0$ will be referred to as *evolution equations*, while equations not involving $\mathbf{e}_0$ are said to be *constraint equations* in analogy to the usage of Ellis in the covariant formulation [6, 7] (even though the interpretation in the rotating $1+\ 3$ case is not clear).

It is seen that by extending the number of variables, from the triad, expansion and shear variables to all commutation functions and curvature variables, one obtains additional evolution equations as well as constraint equations. However, there are no evolution equations directly obtained for the acceleration, $\dot{u}^\alpha$, and the Fermi-rotation, $\Omega^\alpha$.

This is not surprising, since one can choose the timelike congruence so that one can obtain any convenient value for the acceleration $\dot{u}^\alpha$. This freedom is also seen when choosing local coordinates, in particular, one can choose any lapse function (recall Eq. 2.79). However, if one has a source which one wants to adapt to, this might provide an evolution equation for the acceleration. For example, in the comoving perfect fluid case the pressure gradient determines the acceleration. If one has a barotropic equation of state, $p = p(\mu)$, one can obtain an evolution equation by using the commutators in conjunction with the source equations (2.48) and (2.49). This leads to

$$\mathbf{e}_0(\dot{u}_\alpha) = \mathbf{e}_\alpha(\tfrac{dp}{d\mu}\Theta) + (\tfrac{dp}{d\mu} - \tfrac{1}{3})\,\Theta\,\dot{u}_\alpha - [\,\sigma^\beta{}_\alpha + \epsilon^\beta{}_{\alpha\gamma}\,(\omega^\gamma - \Omega^\gamma)\,]\,\dot{u}_\beta\ , \tag{2.84}$$

which for a linear barotropic equation of state of the form $p(\mu) = (\gamma - 1)\mu$ reduces to

$$\mathbf{e}_0(\dot{u}_\alpha) = (\gamma - 1)\,\mathbf{e}_\alpha(\Theta) + \tfrac{1}{3}\,(3\,\gamma - 4)\,\Theta\,\dot{u}_\alpha - [\,\sigma^\beta{}_\alpha + \epsilon^\beta{}_{\alpha\gamma}\,(\omega^\gamma - \Omega^\gamma)\,]\,\dot{u}_\beta\ . \tag{2.85}$$

Neither is it surprising that one does not have an evolution equation for the Fermi-rotation, $\Omega^\alpha$, since one can choose $\Omega^\alpha$ to be anything one wants (e.g., one can choose a Fermi-transported spatial frame, $\Omega^\alpha = 0$). Thus $\Omega^\alpha$ represents the freedom of choosing the spatial frame $\{\,\mathbf{e}_\alpha\,\}$ and hence plays a role analogous to the shift vector, which represent the freedom in choosing spatial local coordinates. The Fermi-frame is not the only interesting choice for $\Omega^\alpha$. Other possible choices are those which diagonalize $\sigma_{\alpha\beta}$ (shear eigenframes) or $E_{\alpha\beta}$. Of considerable interest is also the "co-rotating" frame, defined by $\Omega^\alpha = \omega^\alpha$.

We also note that we only have an evolution equation for the combination $(E^{\alpha\beta} + \frac{1}{2}\pi^{\alpha\beta})$, and no one for the pressure $p$. To obtain individual evolution equations for these quantities requires further assumptions about the source. In the case of perfect fluids with, for example, $\tilde{p} = \tilde{p}(\tilde{\mu})$ as an equation of state, the pressure $p$ and $\pi_{\alpha\beta}$ are determined by the energy density $\tilde{\mu}$ and the velocity $v^\alpha$ ( see Eq. (2.57) ) which have evolutions equations. For Maxwell fields one obtains $\mu$, $p$, $\pi_{\alpha\beta}$ from $E_\alpha$ and $H_\alpha$ ( see Eqs. (2.61) - (2.63) ) which also have evolution equations. Suggestions for a fully relativistic thermodynamic treatment of general viscous fluid matter sources with nonzero $q^\alpha$ and $\pi_{\alpha\beta}$ were given, for example, by Israel and Stewart [34, 35].

If one expresses the frame $\{\,\mathbf{e}_a\,\}$ explicitly in terms of the lapse function, shift vector and the triad variables and calculates the Einstein field equations, one obtains a second-order system of partial differential equations. The commutation functions are defined through the commutators, and the Jacobi identities are identically satisfied. However, one can instead view the commutation functions as a set of new independent variables. In this case one obtains a first-order system of equations in *all* derivatives described by the commutators, the Einstein field equations and the Jacobi identities, which are now "elevated" to nontrivial equations. The Ricci identities provide a definition for the Riemann curvature tensor. Using this definition one finds that the Bianchi identities are identically satisfied.

However, the curvature is a directly measurable object. This makes it desirable to consider the curvature components as variables as well. Thus we can continue and view the curvature components as independent variables. In this case the Bianchi identities are "elevated" to nontrivial field equations. Note that the system of equations for tetrad variables, commutation functions and curvature variables is highly redundant (we could have stopped at, e.g., the commutation function level with the Jacobi identities and the Einstein field equations). From a physical point of view, in the present formulation it would be desirable to start at the bottom of the derivative ladder with the curvature and the Bianchi identities. Can we go "upwards" towards lower derivatives and only use, say curvature and kinematical quantities? Do these form a complete and closed system? In general the answer is, unfortunately, no. In general we always need some tetrad variables, commutation functions and commutator equations. The redundancy, in the general case, is described by the Papapetrou identities [36, 37] ( for a pedagogical review see Ref. [14] ). Thus one can pick out interesting subsets of variables and equations in the general case. For practical reasons we are usually interested in special cases, for example, in special Petrov types. Unfortunately these special cases need further consideration and one might have to consider higher derivatives for checking consistency. The "silent" models discussed below are examples of this.

However, it should be pointed out that one can take the opposite route and go "downwards" towards higher derivatives, since it is possible to start at the curvature (or commutation function) level and take a *finite* number of covariant derivatives and obtain a complete description of the geometry. This is the equivalence problem approach along the lines of work by Cartan [38] and Karlhede [39]. In this approach one obtains, step by step, the information contained in the tetrad and the commutators, and after a finite number of covariant derivatives one has obtained all the information needed to construct the full tetrad, that is, the metric [40].

## 2.7 Irreducible tracefree decomposition

In the case of the covariantly defined spacelike symmetric tracefree tensor fields $\sigma_{\alpha\beta}$, $E_{\alpha\beta}$, $H_{\alpha\beta}$ and $\pi_{\alpha\beta}$, it is sometimes convenient to introduce a new set of variables adapted to the tracefree condition and the invariant quadratic forms $\sigma^2$, $E^2$, $H^2$ and $\pi^2$. For example, for $\sigma_{\alpha\beta}$ we define the irreducible frame components

$$\begin{aligned} &\sigma_+ := -\tfrac{3}{2}\,\sigma_{11} \quad = \tfrac{3}{2}\,(\sigma_{22}+\sigma_{33})\ , \quad \sigma_- := \tfrac{\sqrt{3}}{2}\,(\sigma_{22}-\sigma_{33}) \\ &\sigma_1 := \sqrt{3}\,\sigma_{23}\ , \quad \sigma_2 := \sqrt{3}\,\sigma_{31}\ , \qquad \sigma_3 := \sqrt{3}\,\sigma_{12}\ , \end{aligned} \tag{2.86}$$

leading to

$$\sigma^2 = \tfrac{1}{3}\,[\,(\sigma_+)^2 + (\sigma_-)^2 + (\sigma_1)^2 + (\sigma_2)^2 + (\sigma_3)^2\,]\ , \tag{2.87}$$

where the definition is motivated by the desire to simplify the kinematic part of the $\mu$-equation (2.28) as far as possible (this equation plays a key role in determining the dynamics of, for example, spatially homogeneous models). If one decomposes the shear tensor one should also similarly decompose the other spacelike traceless symmetric tensors as well. In principle one could extend this procedure to also include the spatial commutation functions $n_{\alpha\beta}$, that is, splitting them into a trace and a tracefree part, where the latter is further subdivided according to the scheme above. However, we will not take this step within this work as $n_{\alpha\beta}$ usually does *not* have a covariant meaning in the $1+3$ picture, but very much depends on the choice of the spatial frame $\{\,\mathbf{e}_\alpha\,\}$.

# 3 Dimensionless Formulation

Dimensionless variables have at least a three decade long history in General Relativity. Bondi was perhaps the first to use scale-invariant variables in order to obtain a dimensional reduction of a problem. He did this for static spherically symmetric perfect fluid models with $\tilde{p}(\tilde{\mu}) = (\gamma-1)\,\tilde{\mu}$ in 1964 [41] (see also Ref. [42]). Independently Collins used the scale invariance of the Einstein field equations to produce reduced sets of equations in Bianchi cosmology [43]. Wainwright and collaborators have introduced the concept of "expansion-normalized" dimensionless variables in Bianchi cosmology and inhomogeneous models with two commuting spacelike Killing vector fields [44, 2] (closely related variables have also been used by Rosquist and Jantzen [45]). Expansion-normalized dimensionless variables lead to relatively simple equations and are natural in a cosmological or gravitational collapse scenario (for a recent review see Ref. [2]). On the other hand, in other situations, like for example for static models or for maximal slicings, one is forced to consider other possibilities. For special cases there might also be mathematical reasons for considering other choices (see e.g. Ref. [46]). However, for problems in cosmology and gravitational collapse, expansion-normalized dimensionless variables seem to be a natural first candidate, and this is how we will proceed. Here we will generalize the work by Hewitt and Wainwright [44] on comoving perfect fluids in a geometry with two spacelike commuting Killing vector fields to general timelike congruences and geometries.

We first introduce expansion-normalized dimensionless differential operators

$$\boldsymbol{\partial}_a := \frac{3\,\mathbf{e}_a}{\Theta}\ , \tag{3.1}$$

then expansion-normalized dimensionless commutation functions

$$\left\{\ \dot{U}_\alpha, \Sigma_{\alpha\beta}, W_\alpha, R_\alpha, A_\alpha, N_{\alpha\beta}\ \right\} := \{\ \dot{u}_\alpha, \sigma_{\alpha\beta}, \omega_\alpha, \Omega_\alpha, a_\alpha, n_{\alpha\beta}\ \}\,/\,\Theta\ , \tag{3.2}$$

and finally expansion-normalized dimensionless source and curvature variables

$$\{ \; \Omega, P, Q_\alpha, \Pi_{\alpha\beta}, \Omega_\Lambda, \mathcal{E}_{\alpha\beta}, \mathcal{H}_{\alpha\beta} \; \} := 3 \; \{ \; \mu, p, q_\alpha, \pi_{\alpha\beta}, \Lambda, E_{\alpha\beta}, H_{\alpha\beta} \; \} \, / \, \Theta^2 \; . \tag{3.3}$$

For the traceless decomposition variables of the shear tensor, Eq. (2.86), one can use the following definition, which is aimed at simplifying the $\Omega$-equation arising from Eq. (2.28):

$$\{ \; \Sigma_\pm, \Sigma_i \; \} := \{ \; \sigma_\pm, \sigma_i \; \} \, / \, \Theta \; , \tag{3.4}$$

where $i = 1, 2, 3$, and from which it follows that

$$\Sigma^2 := \tfrac{3}{2} \, \Sigma_{\alpha\beta} \, \Sigma^{\alpha\beta} = (\Sigma_+)^2 + (\Sigma_-)^2 + (\Sigma_1)^2 + (\Sigma_2)^2 + (\Sigma_3)^2 \; . \tag{3.5}$$

Then one can make a similar transformation for the other spacelike symmetric tracefree tensors:

$$\{ \; \Pi_\pm, \Pi_i, \mathcal{E}_\pm, \mathcal{E}_i, \mathcal{H}_\pm, \mathcal{H}_i \; \} := 3 \; \{ \; \pi_\pm, \pi_i, E_\pm, E_i, H_\pm, H_i \; \} \, / \, \Theta^2 \; . \tag{3.6}$$

Since the only variable that carries dimension is $\Theta$, with dimension $[\,\text{length}\,]^{-1}$, the equations associated with these variables will decouple from the remaining ones. The decoupled equations are given by

**Decoupled equations**

$$\boldsymbol{\partial}_0 \Theta \;\; := \;\; - \, (1 + q) \, \Theta \tag{3.7}$$

$$\boldsymbol{\partial}_\alpha \Theta \;\; := \;\; - \, r_\alpha \, \Theta \; , \tag{3.8}$$

where $q$ is the deceleration parameter defined in equation (1.9). These expansion-normalized dimensionless variables lead to the following expressions for the commutators, curvature equations, and Bianchi identities.

## 3.1 The commutators

$$[ \boldsymbol{\partial}_0, \boldsymbol{\partial}_\alpha ] \;\; = \;\; - \, [ \; r_\alpha - 3 \, \dot{U}_\alpha \; ] \, \boldsymbol{\partial}_0 + 3 \; [ \; \tfrac{1}{3} \, q \, \delta^\beta{}_\alpha - \Sigma^\beta{}_\alpha - \epsilon^\beta{}_{\alpha\gamma} \, ( W^\gamma - R^\gamma ) \; ] \; \boldsymbol{\partial}_\beta \; , \tag{3.9}$$

$$[ \boldsymbol{\partial}_\alpha, \boldsymbol{\partial}_\beta ] \;\; = \;\; - \, 6 \, \epsilon_{\alpha\beta\gamma} \, W^\gamma \; \boldsymbol{\partial}_0 + [ \; 2 \, ( r_{[\alpha} + 3 \, A_{[\alpha} ) \, \delta^\gamma{}_{\beta]} + 3 \, \epsilon_{\alpha\beta\delta} \, N^{\delta\gamma} \; ] \; \boldsymbol{\partial}_\gamma \; . \tag{3.10}$$

## 3.2 The curvature

**The field equations**

$$q \;\; = \;\; - \, ( \boldsymbol{\partial}_\alpha - r_\alpha + 3 \, \dot{U}_\alpha - 6 \, A_\alpha ) \, \dot{U}^\alpha + 2 \, \Sigma^2 - 6 \, W^2 + \tfrac{1}{2} \, ( \Omega + 3 P ) - \Omega_\Lambda \; , \tag{3.11}$$

$$\begin{aligned} \boldsymbol{\partial}_0 \Sigma^{\alpha\beta} \;\; = \;\; & ( q - 2 ) \, \Sigma^{\alpha\beta} + ( \delta^{\gamma(\alpha} \, \boldsymbol{\partial}_\gamma - r^{(\alpha} + 3 \, \dot{U}^{(\alpha} + 3 \, A^{(\alpha} ) \, \dot{U}^{\beta)} + 6 \, W^{(\alpha} \, R^{\beta)} + \Pi^{\alpha\beta} - \mathcal{S}^{\alpha\beta} \\ & - \, \tfrac{1}{3} \, \delta^{\alpha\beta} \, \Big[ \; ( \boldsymbol{\partial}_\gamma - r_\gamma + 3 \, \dot{U}_\gamma + 3 \, A_\gamma ) \, \dot{U}^\gamma + 6 \, W_\gamma \, R^\gamma \; \Big] \\ & + \; 3 \, \epsilon^{\gamma\delta(\alpha} \, \Big[ \; 2 \, R_\gamma \, \Sigma^{\beta)}{}_\delta - N^{\beta)}{}_\gamma \, \dot{U}_\delta \; \Big] \; , \end{aligned} \tag{3.12}$$

$$\Omega \;\; = \;\; 1 - \Sigma^2 + 3 \, W^2 + 6 \, W_\alpha \, R^\alpha - \mathcal{K} - \Omega_\Lambda \; , \tag{3.13}$$

$$\begin{aligned} 0 \;\; = \;\; & Q^\alpha + \tfrac{2}{3} \, r^\alpha + ( \boldsymbol{\partial}_\beta - r_\beta - 9 \, A_\beta ) \, \Sigma^{\alpha\beta} + 3 \, N^\alpha{}_\beta \, W^\beta \\ & - \, \epsilon^{\alpha\beta\gamma} \, \Big[ \; ( \boldsymbol{\partial}_\beta - r_\beta + 6 \, \dot{U}_\beta - 3 \, A_\beta ) \, W_\gamma + 3 \, N_{\beta\delta} \, \Sigma^\delta{}_\gamma \; \Big] \; , \end{aligned} \tag{3.14}$$

where

$$\begin{aligned} \mathcal{S}_{\alpha\beta} &:= 3\,{}^*S_{\alpha\beta} / \Theta^2 \\ &= (\boldsymbol{\partial}_{(\alpha} - r_{(\alpha})\, A_{\beta)} + 3\, B_{\alpha\beta} - \tfrac{1}{3}\,\delta_{\alpha\beta}\, [\, (\boldsymbol{\partial}_\gamma - r_\gamma)\, A^\gamma + 3\, B^\gamma{}_\gamma \,] \\ &\quad -\, \epsilon^{\gamma\delta}{}_{(\alpha}\, (\boldsymbol{\partial}_{|\gamma|} - r_{|\gamma|} - 6\, A_{|\gamma|})\, N_{\beta)\delta} \ , \end{aligned} \tag{3.15}$$

$$\mathcal{K} \ := \ -(3\,{}^*R)/(2\,\Theta^2) \ = \ -\,(2\,\boldsymbol{\partial}_\alpha - 2\, r_\alpha - 9\, A_\alpha)\, A^\alpha + \tfrac{3}{4}\, B^\alpha{}_\alpha \ , \tag{3.16}$$

$$B_{\alpha\beta} \ = \ 2\, N_{\alpha\gamma}\, N^\gamma{}_\beta - N^\gamma{}_\gamma\, N_{\alpha\beta} \ , \qquad\qquad W^2 := W_\alpha\, W^\alpha \ . \tag{3.17}$$

**The Jacobi identities**

$$\begin{aligned} \boldsymbol{\partial}_0 A^\alpha &= q\, A^\alpha + \tfrac{1}{3}\, r^\alpha - \dot{U}^\alpha + \tfrac{1}{2}\, (\boldsymbol{\partial}_\beta - r_\beta + 3\, \dot{U}_\beta - 6\, A_\beta)\, \Sigma^{\alpha\beta} \\ &\quad -\, \tfrac{1}{2}\, \epsilon^{\alpha\beta\gamma}\, (\boldsymbol{\partial}_\beta - r_\beta + 3\, \dot{U}_\beta - 6\, A_\beta)\, (W_\gamma - R_\gamma) \ , \end{aligned} \tag{3.18}$$

$$\begin{aligned} \boldsymbol{\partial}_0 N^{\alpha\beta} &= q\, N^{\alpha\beta} - (\delta^{\gamma(\alpha}\, \boldsymbol{\partial}_\gamma - r^{(\alpha} + 3\, \dot{U}^{(\alpha})\, (W^{\beta)} - R^{\beta)}) + 6\, \Sigma^{(\alpha}{}_\gamma\, N^{\beta)\gamma} \\ &\quad +\, \delta^{\alpha\beta}\, (\boldsymbol{\partial}_\gamma - r_\gamma + 3\, \dot{U}_\gamma)\, (W^\gamma - R^\gamma) \\ &\quad -\, \epsilon^{\gamma\delta(\alpha} \Big[\, (\boldsymbol{\partial}_\gamma - r_\gamma + 3\, \dot{U}_\gamma)\, \Sigma^{\beta)}{}_\delta - 6\, N^{\beta)}{}_\gamma\, (W_\delta - R_\delta) \,\Big] \ , \end{aligned} \tag{3.19}$$

$$\boldsymbol{\partial}_0 W^\alpha \ = \ (q-1)\, W^\alpha + 3\, \Sigma^\alpha{}_\beta\, W^\beta + \tfrac{3}{2}\, N^\alpha{}_\beta\, \dot{U}^\beta - \epsilon^{\alpha\beta\gamma} \Big[\, \tfrac{1}{2}\, (\boldsymbol{\partial}_\beta - r_\beta - 3\, A_\beta)\, \dot{U}_\gamma + 3\, W_\beta\, R_\gamma \,\Big] \ , \tag{3.20}$$

$$0 \ = \ (\boldsymbol{\partial}_\beta - r_\beta - 6\, A_\beta)\, N^{\alpha\beta} - 2\, W^\alpha - 6\, \Sigma^\alpha{}_\beta\, W^\beta + \epsilon^{\alpha\beta\gamma}\, [\, (\boldsymbol{\partial}_\beta - r_\beta)\, A_\gamma + 6\, W_\beta\, R_\gamma \,] \ , \tag{3.21}$$

$$0 \ = \ (\boldsymbol{\partial}_\alpha - r_\alpha - 3\, \dot{U}_\alpha - 6\, A_\alpha)\, W^\alpha \ . \tag{3.22}$$

**The “electric” and “magnetic parts” of the Weyl tensor**

$$\begin{aligned} \mathcal{E}_{\alpha\beta} &= -\,(\boldsymbol{\partial}_0 - q + 1)\, \Sigma_{\alpha\beta} + (\boldsymbol{\partial}_{(\alpha} - r_{(\alpha} + 3\, \dot{U}_{(\alpha} + 3\, A_{(\alpha})\, \dot{U}_{\beta)} - 3\, \Sigma_{\alpha\gamma}\, \Sigma^\gamma{}_\beta - 3\, W_\alpha\, W_\beta + \tfrac{1}{2}\, \Pi_{\alpha\beta} \\ &\quad -\, \tfrac{1}{3}\, \delta_{\alpha\beta} \Big[\, (\boldsymbol{\partial}_\gamma - r_\gamma + 3\, \dot{U}_\gamma + 3\, A_\gamma)\, \dot{U}^\gamma - 6\, \Sigma^2 - 3\, W^2 \,\Big] \\ &\quad +\, 3\, \epsilon^{\gamma\delta}{}_{(\alpha} \Big[\, 2\, R_{|\gamma|}\, \Sigma_{\beta)\delta} - N_{\beta)\gamma}\, \dot{U}_\delta \,\Big] \ . \end{aligned} \tag{3.23}$$

The field equation (3.12) combined with the above expression (3.23) for $\mathcal{E}_{\alpha\beta}$ leads to

$$\begin{aligned} \mathcal{E}_{\alpha\beta} + \tfrac{1}{2}\, \Pi_{\alpha\beta} &= \Sigma_{\alpha\beta} - 3\, \Sigma_{\alpha\gamma}\, \Sigma^\gamma{}_\beta - 3\, W_\alpha\, W_\beta - 6\, W_{(\alpha}\, R_{\beta)} \\ &\quad +\, \tfrac{1}{3}\, \delta_{\alpha\beta}\, [\, 2\, \Sigma^2 + 3\, W^2 + 6\, W_\gamma\, R^\gamma \,] + \mathcal{S}_{\alpha\beta} \ . \end{aligned} \tag{3.24}$$

The dimensionless version of Eq. (2.41) is

$$\begin{aligned} \mathcal{H}_{\alpha\beta} &= (\boldsymbol{\partial}_{(\alpha} - r_{(\alpha} + 6\, \dot{U}_{(\alpha} + 3\, A_{(\alpha})\, W_{\beta)} + \tfrac{3}{2}\, N^\gamma{}_\gamma\, \Sigma_{\alpha\beta} - 9\, N^\gamma{}_{(\alpha}\, \Sigma_{\beta)\gamma} \\ &\quad -\, \tfrac{1}{3}\, \delta_{\alpha\beta} \Big[\, (\boldsymbol{\partial}_\gamma - r_\gamma + 6\, \dot{U}_\gamma + 3\, A_\gamma)\, W^\gamma - 9\, N_{\gamma\delta}\, \Sigma^{\gamma\delta} \,\Big] \\ &\quad +\, \epsilon^{\gamma\delta}{}_{(\alpha}\, [\, (\boldsymbol{\partial}_{|\gamma|} - r_{|\gamma|} - 3\, A_{|\gamma|})\, \Sigma_{\beta)\delta} - 3\, N_{\beta)\gamma}\, W_\delta \,] \ . \end{aligned} \tag{3.25}$$

### 3.3 The Bianchi identities

**Bianchi identities for the Weyl tensor**

$$\begin{aligned}
\boldsymbol{\partial}_0(\mathcal{E}^{\alpha\beta}+\tfrac{1}{2}\,\Pi^{\alpha\beta}) &= (2\,q-1)\,\mathcal{E}^{\alpha\beta}+\left(q+\tfrac{1}{2}\right)\Pi^{\alpha\beta}-\tfrac{3}{2}\,(\Omega+P)\,\Sigma^{\alpha\beta} \\
&\quad -\tfrac{1}{2}\,(\delta^{\gamma(\alpha}\,\boldsymbol{\partial}_\gamma-2\,r^{(\alpha}+6\,\dot{U}^{(\alpha}+3\,A^{(\alpha)})\,Q^{\beta)} \\
&\quad +9\,\Sigma^{(\alpha}{}_\gamma\,(\mathcal{E}^{\beta)\gamma}-\tfrac{1}{6}\,\Pi^{\beta)\gamma})+\tfrac{3}{2}\,N^\gamma{}_\gamma\,\mathcal{H}^{\alpha\beta}-9\,N^{(\alpha}{}_\gamma\,\mathcal{H}^{\beta)\gamma} \\
&\quad +\tfrac{1}{3}\,\delta^{\alpha\beta}\left[\tfrac{1}{2}\,(\boldsymbol{\partial}_\gamma-2\,r_\gamma+6\,\dot{U}_\gamma+3\,A_\gamma)\,Q^\gamma-9\,\Sigma_{\gamma\delta}\,(\mathcal{E}^{\gamma\delta}-\tfrac{1}{6}\,\Pi^{\gamma\delta})+9\,N_{\gamma\delta}\,\mathcal{H}^{\gamma\delta}\right] \\
&\quad +\epsilon^{\gamma\delta(\alpha}\left[(\boldsymbol{\partial}_\gamma-2\,r_\gamma+6\,\dot{U}_\gamma-3\,A_\gamma)\,\mathcal{H}^{\beta)}{}_\delta\right. \\
&\qquad\qquad \left.-3\,(W_\gamma-2\,R_\gamma)\,(\mathcal{E}^{\beta)}{}_\delta+\tfrac{1}{2}\,\Pi^{\beta)}{}_\delta)+\tfrac{3}{2}\,N^{\beta)}{}_\gamma\,Q_\delta\right], \qquad (3.26)
\end{aligned}$$

$$\begin{aligned}
\boldsymbol{\partial}_0 H^{\alpha\beta} &= (2\,q-1)\,\mathcal{H}^{\alpha\beta}+9\,\Sigma^{(\alpha}{}_\gamma\,\mathcal{H}^{\beta)\gamma}-\tfrac{9}{2}\,W^{(\alpha}\,Q^{\beta)}-\tfrac{3}{2}\,N^\gamma{}_\gamma\,(\mathcal{E}^{\alpha\beta}-\tfrac{1}{2}\,\Pi^{\alpha\beta}) \\
&\quad +9\,N^{(\alpha}{}_\gamma\,(\mathcal{E}^{\beta)\gamma}-\tfrac{1}{2}\,\Pi^{\beta)\gamma}) \\
&\quad -3\,\delta^{\alpha\beta}\left[\Sigma_{\gamma\delta}\,\mathcal{H}^{\gamma\delta}-\tfrac{1}{2}\,W_\gamma\,Q^\gamma+N_{\gamma\delta}\,(\mathcal{E}^{\gamma\delta}-\tfrac{1}{2}\,\Pi^{\gamma\delta})\right] \\
&\quad -\epsilon^{\gamma\delta(\alpha}\left[(\boldsymbol{\partial}_\gamma-2\,r_\gamma-3\,A_\gamma)\,(\mathcal{E}^{\beta)}{}_\delta-\tfrac{1}{2}\,\Pi^{\beta)}{}_\delta)+6\,\dot{U}_\gamma\,\mathcal{E}^{\beta)}{}_\delta\right. \\
&\qquad\qquad \left.-\tfrac{3}{2}\,\Sigma^{\beta)}{}_\gamma\,Q_\delta+3\,(W_\gamma-2\,R_\gamma)\,\mathcal{H}^{\beta)}{}_\delta\right], \qquad (3.27)
\end{aligned}$$

$$\begin{aligned}
0 &= (\boldsymbol{\partial}_\beta-2\,r_\beta-9\,A_\beta)\,(\mathcal{E}^{\alpha\beta}+\tfrac{1}{2}\,\Pi^{\alpha\beta})-\tfrac{1}{3}\,\delta^{\alpha\beta}\,(\boldsymbol{\partial}_\beta-2\,r_\beta)\,\Omega+Q^\alpha-\tfrac{3}{2}\,\Sigma^\alpha{}_\beta\,Q^\beta \\
&\quad +9\,W_\beta\,\mathcal{H}^{\alpha\beta}-3\,\epsilon^{\alpha\beta\gamma}\left[\Sigma_{\beta\delta}\,\mathcal{H}^\delta{}_\gamma+\tfrac{3}{2}\,W_\beta\,Q_\gamma+N_{\beta\delta}\,(\mathcal{E}^\delta{}_\gamma+\tfrac{1}{2}\,\Pi^\delta{}_\gamma)\right], \qquad (3.28)
\end{aligned}$$

$$\begin{aligned}
0 &= (\boldsymbol{\partial}_\beta-2\,r_\beta-9\,A_\beta)\,\mathcal{H}^{\alpha\beta}-3\,(\Omega+P)\,W^\alpha-9\,W_\beta\,(\mathcal{E}^{\alpha\beta}-\tfrac{1}{6}\,\Pi^{\alpha\beta})-\tfrac{3}{2}\,N^\alpha{}_\beta\,Q^\beta \\
&\quad +3\,\epsilon^{\alpha\beta\gamma}\left[\tfrac{1}{6}\,(\boldsymbol{\partial}_\beta-2\,r_\beta-3\,A_\beta)\,Q_\gamma+\Sigma_{\beta\delta}\,(\mathcal{E}^\delta{}_\gamma+\tfrac{1}{2}\,\Pi^\delta{}_\gamma)-N_{\beta\delta}\,\mathcal{H}^\delta{}_\gamma\right]. \qquad (3.29)
\end{aligned}$$

**Bianchi identities for the source terms**

$$\boldsymbol{\partial}_0\Omega = (2\,q-1)\,\Omega-3\,P-(\boldsymbol{\partial}_\alpha-2\,r_\alpha+6\,\dot{U}_\alpha-6\,A_\alpha)\,Q^\alpha-3\,\Sigma_{\alpha\beta}\,\Pi^{\alpha\beta}, \qquad (3.30)$$

$$\begin{aligned}
\boldsymbol{\partial}_0 Q^\alpha &= 2\,(q-1)\,Q^\alpha-\delta^{\alpha\beta}\,(\boldsymbol{\partial}_\beta-2\,r_\beta)\,P-3\,(\Omega+P)\,\dot{U}^\alpha-(\boldsymbol{\partial}_\beta-2\,r_\beta+3\,\dot{U}_\beta-9\,A_\beta)\,\Pi^{\alpha\beta} \\
&\quad -3\,\Sigma^\alpha{}_\beta\,Q^\beta+3\,\epsilon^{\alpha\beta\gamma}\left[(W_\beta+R_\beta)\,Q_\gamma+N_{\beta\delta}\,\Pi^\delta{}_\gamma\right]. \qquad (3.31)
\end{aligned}$$

Eqs. (2.53) and (2.52) convert into

$$\begin{aligned}
\mathcal{C}_{\alpha\beta} &:= 9\,{}^*C_{\alpha\beta}\,/\,\Theta^3 \\
&= \epsilon^{\gamma\delta}{}_{(\alpha}\,(\boldsymbol{\partial}_{|\gamma|}-2\,r_{|\gamma|}-3\,A_{|\gamma|})\,\mathcal{S}_{\beta)\delta}-9\,N^\gamma{}_{(\alpha}\,\mathcal{S}_{\beta)\gamma}+\tfrac{3}{2}\,N^\gamma{}_\gamma\,\mathcal{S}_{\alpha\beta}+3\,\delta_{\alpha\beta}\,N_{\gamma\delta}\,\mathcal{S}^{\gamma\delta} \qquad (3.32) \\
&= -(\boldsymbol{\partial}_0-2\,q+2)\,\mathcal{H}_{\alpha\beta}+9\,\Sigma^\gamma{}_{(\alpha}\,\mathcal{H}_{\beta)\gamma}-27\,N^\gamma{}_{(\alpha}\,\Sigma^\delta{}_{\beta)}\,\Sigma_{\gamma\delta}+\tfrac{9}{2}\,N^\gamma{}_\gamma\,\Sigma_{\alpha\delta}\,\Sigma^\delta{}_\beta+6\,N_{\alpha\beta}\,\Sigma^2 \\
&\quad -9\,N^\gamma{}_{(\alpha}\,\Pi_{\beta)\gamma}+\tfrac{3}{2}\,N^\gamma{}_\gamma\,\Pi_{\alpha\beta} \\
&\quad -3\,\delta_{\alpha\beta}\left[\Sigma_{\gamma\delta}\,\mathcal{H}^{\gamma\delta}-3\,N^\gamma{}_\delta\,\Sigma^\delta{}_\epsilon\,\Sigma^\epsilon{}_\gamma+N^\gamma{}_\gamma\,\Sigma^2-3\,N_{\gamma\delta}\,\Pi^{\gamma\delta}\right] \\
&\quad +3\,\epsilon^{\gamma\delta}{}_{(\alpha}\left[(\boldsymbol{\partial}_{|\gamma|}-2\,r_{|\gamma|}-3\,A_{|\gamma|})\,(\Sigma^\epsilon{}_{\beta)}\,\Sigma_{\delta\epsilon}+\tfrac{1}{3}\,\Pi_{\beta)\delta})+r_{|\gamma|}\,\Sigma_{\beta)\delta}-2\,\dot{U}_{|\gamma|}\,\mathcal{E}_{\beta)\delta}\right. \\
&\qquad\qquad \left.+\tfrac{1}{2}\,\Sigma_{\beta)\gamma}\,Q_\delta+2\,R_{|\gamma|}\,\mathcal{H}_{\beta)\delta}\right]. \qquad (3.33)
\end{aligned}$$

## 3.4 The source

### 3.4.1 Perfect fluids

For the comoving ($\mathbf{e}_0 = \mathbf{u}$) perfect fluid case we have:

$$
\begin{aligned}
\boldsymbol{\partial}_0 \tilde{\Omega} &= (2\,q - 1)\,\tilde{\Omega} - 3\,\tilde{P}\ , && (3.34)\\
0 &= (\boldsymbol{\partial}_\alpha - 2\,r_\alpha)\,\tilde{P} + 3\,(\tilde{\Omega} + \tilde{P})\,\dot{U}_\alpha\ . && (3.35)
\end{aligned}
$$

The evolution equation for the acceleration , Eq. (2.85), now takes the form

$$
\boldsymbol{\partial}_0 \dot{U}_\alpha = -\,(\gamma - 1)\,r_\alpha + [\ 3\,(\gamma - 1) + q\ ]\,\dot{U}_\alpha - 3\,[\ \Sigma^\beta{}_\alpha + \epsilon^\beta{}_{\alpha\gamma}\,(W^\gamma - R^\gamma)\ ]\,\dot{U}_\beta\ . \qquad (3.36)
$$

For a tilted perfect fluid we have

$$
\begin{array}{rclrcl}
\Omega &=& \Gamma^2\,(\tilde{\Omega} + \tilde{P}) - \tilde{P}\ , & P &=& \frac{1}{3}\,(\tilde{\Omega} + \tilde{P})\,\Gamma^2\,v^2 + \tilde{P}\ , \\
Q^a &=& \Gamma^2\,(\tilde{\Omega} + \tilde{P})\,v^a\ , & \Pi_{ab} &=& \Gamma^2\,(\tilde{\Omega} + \tilde{P})\,(\,v_a\,v_b - \frac{1}{3}\,v^2\,h_{ab}\,)\ .
\end{array} \qquad (3.37)
$$

For a linear barotropic equation of state of the form $\tilde{p}(\tilde{\mu}) = (\gamma - 1)\,\tilde{\mu}$ these equations reduce to

$$
\begin{array}{rclrcl}
\Omega &=& \Gamma^2\,G\,\tilde{\Omega}\ , & P &=& \frac{1}{3}\,G^{-1}\,[\ (3 - 2\gamma)\,v^2 + 3\,(\gamma - 1)\ ]\,\Omega\ , \\
Q^a &=& \gamma\,G^{-1}\,v^a\,\Omega\ , & \Pi_{ab} &=& \gamma\,G^{-1}\,(\,v_a\,v_b - \frac{1}{3}\,v^2\,h_{ab}\,)\,\Omega\ ,
\end{array} \qquad (3.38)
$$

where $G := 1 + (\gamma - 1)v^2$.

The source equations yield evolution equations for $\Omega$ and $v^a$. The one for $\Omega$ is obtained directly by inserting the above expressions into the $\Omega$-equation, Eq. (3.30). The one for $v^a$ can be obtained for the $\tilde{p}(\tilde{\mu}) = (\gamma - 1)\,\tilde{\mu}$ case by taking the following linear combination of the $\Omega$- and $Q^a$-equations, Eqs. (3.30) and (3.31):

$$
\begin{aligned}
\boldsymbol{\partial}_0 v^\alpha = \frac{G}{\gamma\,\Omega\,[\ 1 - (\gamma - 1)\,v^2\ ]}\,\Big[ &-\gamma\,v^\alpha\,\boldsymbol{\partial}_0\Omega + [\ 1 - (\gamma - 1)\,v^2\ ]\,\boldsymbol{\partial}_0 Q^\alpha + 2\,(\gamma - 1)\,v^\alpha\,v_\beta\,\boldsymbol{\partial}_0 Q^\beta \\
&+\ 2\,(1 + q)\,[\ \gamma\,\Omega\,v^\alpha - (\,1 - (\gamma - 1)\,v^2\,)\,Q^\alpha - 2\,(\gamma - 1)\,v^\alpha\,v_\beta\,Q^\beta\ ]\ \Big]\ . \qquad (3.39)
\end{aligned}
$$

This equation will be evaluated explicitly for the spatially homogeneous perfect fluid models below.

### 3.4.2 Maxwell vacuum fields

We now introduce the following expansion-normalized dimensionless electric and magnetic field variables:

$$
\{\ \mathcal{E}_\alpha, \mathcal{H}_\alpha\ \} := \sqrt{3}\ \{\ E_\alpha, H_\alpha\ \}\,/\,\Theta\ . \qquad (3.40)
$$

This leads to the contributions to the Maxwell stress-energy-momentum tensor

$$
\begin{aligned}
\Omega &= \tfrac{1}{2}\,(\mathcal{E}_\alpha\,\mathcal{E}^\alpha + \mathcal{H}_\alpha\,\mathcal{H}^\alpha) \quad = \quad 3\,P\ , && (3.41)\\
Q^\alpha &= \epsilon^{\alpha\beta\gamma}\,\mathcal{E}_\beta\,\mathcal{H}_\gamma\ , && (3.42)\\
\Pi_{\alpha\beta} &= -\,\mathcal{E}_\alpha\,\mathcal{E}_\beta - \mathcal{H}_\alpha\,\mathcal{H}_\beta + \tfrac{1}{3}\,\delta_{\alpha\beta}\,(\mathcal{E}_\gamma\,\mathcal{E}^\gamma + \mathcal{H}_\gamma\,\mathcal{H}^\gamma)\ . && (3.43)
\end{aligned}
$$

The sourcefree Maxwell equations are given by

$$
\begin{aligned}
\boldsymbol{\partial}_0 \mathcal{E}^\alpha &= (q - 1)\,\mathcal{E}^\alpha + 3\,\Sigma^\alpha{}_\beta\,\mathcal{E}^\beta - 3\,N^\alpha{}_\beta\,\mathcal{H}^\beta \\
&\quad +\ \epsilon^{\alpha\beta\gamma}\,\Big[\ (\boldsymbol{\partial}_\beta - r_\beta + 3\,\dot{U}_\beta - 3\,A_\beta)\,\mathcal{H}_\gamma - 3\,(W_\beta - R_\beta)\,\mathcal{E}_\gamma\ \Big]\ , && (3.44)\\
\boldsymbol{\partial}_0 \mathcal{H}^\alpha &= (q - 1)\,\mathcal{H}^\alpha + 3\,\Sigma^\alpha{}_\beta\,\mathcal{H}^\beta + 3\,N^\alpha{}_\beta\,\mathcal{E}^\beta \\
&\quad -\ \epsilon^{\alpha\beta\gamma}\,\Big[\ (\boldsymbol{\partial}_\beta - r_\beta + 3\,\dot{U}_\beta - 3\,A_\beta)\,\mathcal{E}_\gamma + 3\,(W_\beta - R_\beta)\,\mathcal{H}_\gamma\ \Big]\ , && (3.45)\\
0 &= (\boldsymbol{\partial}_\alpha - r_\alpha - 6\,A_\alpha)\,\mathcal{E}^\alpha + 6\,W_\alpha\,\mathcal{H}^\alpha\ , && (3.46)\\
0 &= (\boldsymbol{\partial}_\alpha - r_\alpha - 6\,A_\alpha)\,\mathcal{H}^\alpha - 6\,W_\alpha\,\mathcal{E}^\alpha\ . && (3.47)
\end{aligned}
$$

### 3.5 Introducing local coordinates

In the $1+3$ threading approach the quantities $M$ and $e_\alpha{}^i$ carry dimensions [length] and [length]$^{-1}$ respectively, while $M_i$ is dimensionless. The relation $\boldsymbol{\partial}_0 = 3\,\mathbf{e}_0/\Theta$ leads to the introduction of the *relative threading lapse function*

$$\mathcal{M} := M\,\Theta \ , \tag{3.48}$$

while the relation $\boldsymbol{\partial}_\alpha = 3\,\mathbf{e}_\alpha/\Theta$ leads to the introduction of the scaled triad

$$E_\alpha{}^i := \frac{e_\alpha{}^i}{\Theta} \ , \quad E^\alpha{}_i := \Theta\, e^\alpha{}_i \ . \tag{3.49}$$

Thus we have

$$\mathbf{e}_0 = \Theta\,\mathcal{M}^{-1}\,\partial_t \quad \Rightarrow \quad \boldsymbol{\partial}_0 = 3\,\mathcal{M}^{-1}\,\partial_t \ , \tag{3.50}$$

$$\mathbf{e}_\alpha = \Theta\,E_\alpha{}^i\,(M_i\,\partial_t + \partial_i) \quad \Rightarrow \quad \boldsymbol{\partial}_\alpha = 3\,E_\alpha{}^i\,(M_i\,\partial_t + \partial_i) \ , \tag{3.51}$$

which leads to the possibility of expressing the 4-geometry as follows

$${}^{(4)}\mathrm{g}^{-1} = \Theta^2\,[\, -\mathcal{M}^{-2}\,\partial_t \otimes \partial_t + \delta^{\alpha\beta}\,E_\alpha{}^i\,E_\beta{}^j\,(M_i\,\partial_t + \partial_i) \otimes (M_j\,\partial_t + \partial_j)\,] \ . \tag{3.52}$$

Expressing the "dimensionless commutators" in terms of $\mathcal{M}$, $M_i$ and $E_\alpha{}^i$ leads to the dimensionless form of Eqs. (2.70) - (2.73):

$$3\,E_\alpha{}^i\,[\,(M_i)_{,t} + \mathcal{M}^{-1}\,(\mathcal{M}_{,i} + M_i\,\mathcal{M}_{,t})\,] = -[\,r_\alpha - 3\,\dot{U}_\alpha\,] \ , \tag{3.53}$$

$$(E_\alpha{}^i)_{,t} = \mathcal{M}\,E_\beta{}^i\,[\,\tfrac{1}{3}\,q\,\delta^\beta{}_\alpha - \Sigma^\beta{}_\alpha - \epsilon^\beta{}_{\alpha\gamma}\,(W^\gamma - R^\gamma)\,] \ , \tag{3.54}$$

$$\mathcal{M}\,E_{[\alpha}{}^i\,E_{\beta]}{}^j\,(M_{i,j} + M_{i,t}\,M_j) = \epsilon_{\alpha\beta\gamma}\,W^\gamma \ , \tag{3.55}$$

$$2\,E_{[\alpha}{}^i\,[\,(E_{\beta]}{}^j)_{,i} + (E_{\beta]}{}^j)_{,t}\,M_i\,]\,E^\gamma{}_j = 2\,(A_{[\alpha} + r_{[\alpha})\delta^\gamma{}_{\beta]} + \epsilon_{\alpha\beta\delta}\,N^{\delta\gamma} \ , \tag{3.56}$$

where one can insert the expression for $(E_\alpha{}^i)_{,t}$ given in Eq. (3.54) into Eq. (3.56), thus obtaining an expression only involving spatial derivatives.

In the slicing approach to the nonrotating ($\omega^\alpha = 0$) case we assign dimensions [length] and [length]$^{-1}$ to the quantities $N$ and $\bar{e}_\alpha{}^i$, while the shift vector $\overline{N}^i$ and the spatial frame vectors $\bar{\mathbf{e}}_i$ are taken to be dimensionless (the components of $\bar{\mathbf{e}}_i$ are assumed to be given functions of the spatial coordinates $x^k$, which themselves are taken to be dimensionless; for a discussion of coordinates and dimensions of spacetime tensors see Ref. [47]). The relation $\boldsymbol{\partial}_0 = 3\,\mathbf{e}_0/\Theta$ leads to the introduction of the *relative lapse function*

$$\mathcal{N} := N\,\Theta \ , \tag{3.57}$$

while the relation $\boldsymbol{\partial}_\alpha = 3\,\mathbf{e}_\alpha/\Theta$ gives

$$\overline{E}_\alpha{}^i := \frac{\bar{e}_\alpha{}^i}{\Theta} \ , \quad \overline{E}^\alpha{}_i := \Theta\,\bar{e}^\alpha{}_i \ . \tag{3.58}$$

Thus we have

$$\mathbf{e}_0 = \Theta\,\mathcal{N}^{-1}\,(\partial_t - \overline{N}^i\,\bar{\mathbf{e}}_i) \quad \Rightarrow \quad \boldsymbol{\partial}_0 = 3\,\mathcal{N}^{-1}\,(\partial_t - \overline{N}^i\,\bar{\mathbf{e}}_i) \tag{3.59}$$

$$\mathbf{e}_\alpha = \Theta\,\overline{E}_\alpha{}^i\,\bar{\mathbf{e}}_i \quad \Rightarrow \quad \boldsymbol{\partial}_\alpha = 3\,\overline{E}_\alpha{}^i\,\bar{\mathbf{e}}_i \ , \tag{3.60}$$

which leads to the 4-geometry

$${}^{(4)}\mathrm{g}^{-1} = \Theta^2\,\Big[\, -\mathcal{N}^{-2}\,(\partial_t - \overline{N}^i\,\bar{\mathbf{e}}_i) \otimes (\partial_t - \overline{N}^j\,\bar{\mathbf{e}}_j) + \delta^{\alpha\beta}\,\overline{E}_\alpha{}^i\,\overline{E}_\beta{}^j\,\bar{\mathbf{e}}_i \otimes \bar{\mathbf{e}}_j\,\Big] \ . \tag{3.61}$$

The relations (2.79) - (2.81) take the form:

$$3\,\overline{E}_\alpha{}^i\,\overline{\mathbf{e}}_i(\ln\mathcal{N}) \;=\; -\,[\,r_\alpha - 3\,\dot{U}_\alpha\,]\;, \qquad (3.62)$$

$$\begin{aligned}(\overline{E}_\alpha{}^i)_{,t} \;=\;& \overline{\mathcal{N}}^j\,\overline{\mathbf{e}}_j(\overline{E}_\alpha{}^i) - \overline{E}_\alpha{}^j\,\overline{\mathbf{e}}_j(\overline{\mathcal{N}}^i) + \overline{\mathcal{N}}^j\,\overline{E}_\alpha{}^k\,\overline{\gamma}^i{}_{jk}\\ &+\,\mathcal{N}\,\overline{E}_\beta{}^i\,[\,\tfrac{1}{3}\,q\,\delta^\beta{}_\alpha - \Sigma^\beta{}_\alpha + \epsilon^\beta{}_{\alpha\gamma}\,R^\gamma\,]\;, \qquad (3.63)\end{aligned}$$

$$\overline{E}^\gamma{}_k\,[\,2\,\overline{E}_{[\alpha}{}^i\,\overline{\mathbf{e}}_i\,(\overline{E}_{\beta]}{}^k) + \overline{E}_{[\alpha}{}^i\,\overline{E}_{\beta]}{}^j\,\overline{\gamma}^k{}_{ij}\,] \;=\; 2\,(A_{[\alpha} + r_{[\alpha})\delta^\gamma{}_{\beta]} + \epsilon_{\alpha\beta\delta}\,N^{\delta\gamma}\;. \qquad (3.64)$$

As mentioned earlier, one can introduce synchronous local coordinate systems ($N = 1$, $\overline{N}^i = 0$) for nonaccelerating and nonrotating congruences. With this choice of lapse the time variable $t_p$ measures proper time (also sometimes denoted as "clock time"). However, if one is so inclined, one can choose a new time variable characterized by $N = N(t)$. Note that individual curves of the congruence have constant spatial coordinates (since $\overline{N}^i = 0$). Thus one can choose $N = \pm\,3/\Theta$, $\mathcal{N} = \pm\,3$, on an individual curve (where $+$ is chosen when $\Theta > 0$ and $-$ is chosen when $\Theta < 0$). With this choice Eq. (3.59) leads to the relation $\boldsymbol{\partial}_0 = \pm\,\partial_\tau$, where $\tau$ is the time variable associated with the above lapse. From the definition of $\Theta$ in terms of $\ell$ it follows that $\ell^{-1}\,\boldsymbol{\partial}_\tau\ell = \pm\,1$ which in turn yields

$$\ell = \ell_*\,e^{\pm(\tau-\tau_*)}\;. \qquad (3.65)$$

Thus it follows that $\tau \to \mp\infty$ at the singularity where $\ell \to 0$.[2] The equation $\boldsymbol{\partial}_0\Theta = -\,(1+q)\,\Theta$ yields

$$\Theta = \Theta_*\,e^{\mp\int_{\tau_*}^{\tau}(1+q)\,d\tau'}\;, \qquad (3.66)$$

for an individual curve of the congruence. From $N = \pm\,3/\Theta$ it follows that

$$t_p(*) - t_p(s) = \Delta t_p = \pm\int_{\mp\infty}^{\tau_*}\frac{3}{\Theta}\,d\tau = \pm\,\frac{3}{\Theta_*}\int_{\mp\infty}^{\tau_*} e^{\pm\int_{\tau_*}^{\tau}(1+q)\,d\tau'}\,d\tau\;, \qquad (3.67)$$

where $t_p(s)$ is the time at the singularity. This leads to

$$\tfrac{1}{3}\,\Theta_*\,\Delta t_p = H_*\,\Delta t_p = \pm\int_{\mp\infty}^{\tau_*} e^{\pm\int_{\tau_*}^{\tau}(1+q)\,d\tau'}\,d\tau\;. \qquad (3.68)$$

Hence the inequality $0 \le q \le 2$ leads to (dropping the $*$)

$$\tfrac{1}{3} \le H\Delta t_p \le 1\;, \qquad (3.69)$$

where $H\,\Delta t_p \le 1$ requires $\Lambda \le 0$, $(\mu + 3p) \ge 0$, while $1/3 \le H\,\Delta t$ requires $\Lambda \ge 0$, $(\mu - p) \ge 0$, ${}^*R \le 0$, and thus both inequalities are satisfied if $\Lambda = 0$, $(\mu + 3p) \ge 0$, $(\mu - p) \ge 0$, ${}^*R \le 0$ along a given curve. The present discussion generalizes the results of Wainwright in [2] from orthogonal spatially homogeneous perfect fluid models to general nonaccelerating nonrotating congruences. As an example one can mention nonrotating inhomogeneous dust models. Current observations of the age of the Universe, $\Delta t_p$, and the present value of the Hubble parameter, $H$, suggest that $H\Delta t_p$ may satisfy the inequality $H\Delta t_p \ge 1$ (see e.g. Ref. [2]). Thus, observations may cast doubt on these cosmologies in the near future. Apparently one may thus be forced to consider other sources, a positive cosmological constant, or rotation.

Of considerable interest is the situation when $\mathbf{e}_0(q) = 0$. In this case one obtains

$$H\,[\,t_p - t_p(s)\,] = H\,\Delta t_p = (1+q)^{-1}\;, \qquad (3.70)$$

which generalizes the result of Wainwright [2] for orthogonal spatially homogeneous models exhibiting an additional self-similar symmetry. In that case one can choose $t_p(s) = 0$ for all fluid lines since one has spatial homogeneity, thus reducing the above expression to $H\,t_p = (1+q)^{-1}$.

[2] Often the notation $\tau$ is used for proper time which here is denoted by $t_p$. We have followed the conventions of Wainwright and Ellis [2] when it comes to the definition of $\tau$. Note that this definition differs by a factor of 3 compared to that used by Bruni et al [18].

### 3.6 Discussion

Above we said that the $\Theta$-equations were decoupled, since $\Theta$ was the only variable that was not dimensionless. However, this is not true if one introduces constants carrying dimension through the matter source or if one introduces a cosmological constant (which also carries dimension). Such constants effectively reintroduces $\Theta$. For example, for a cosmological constant we have that $\Omega_\Lambda = 3\,\Lambda/\Theta^2$. This definition leads to the following equations:

$$\begin{aligned} \boldsymbol{\partial}_0 \Omega_\Lambda &= 2\,(q+1)\,\Omega_\Lambda \ , && (3.71) \\ \boldsymbol{\partial}_\alpha \Omega_\Lambda &= 2\,r_\alpha\,\Omega_\Lambda \ . && (3.72) \end{aligned}$$

For a perfect fluid with $\tilde{p} = \tilde{p}(\tilde{\mu})$ as an equation of state it is only $\tilde{p}(\tilde{\mu}) = (\gamma - 1)\,\tilde{\mu}$ that does *not* lead to a reintroduction of $\Theta$. For example, for a polytropic equation of state, $\tilde{p}(\tilde{\mu}) = K\,\tilde{\mu}^\eta$, one obtains $\tilde{P}(\tilde{\Omega}) = 3^{1-\eta}\,K\,\Theta^{2\,(\eta-1)}\,\tilde{\Omega}^\eta$.

Note also that it is necessary to choose a lapse function $\mathcal{M}$ (or $\mathcal{N}$) and shift $M_i$ (or $\overline{N}^i$) depending only on dimensionless variables if one wants to decouple the $\Theta$-equations. This might not be possible if one wants to use, for example, comoving local coordinates for the perfect fluid case. Thus dust, for example, leads to $\mathcal{M} \propto \Theta$. However, in special cases like spatially homogeneous models and "silent" models of the Universe (to be discussed below) it turns out to be possible to obtain a decoupling of the evolution equation for $\Theta$, leading to a reduced dynamical problem. Even in the cases where one does not obtain a reduction it might still be useful to make a $\Theta$-normalization, since this makes it possible to probe the regime where $\Theta \to \pm\infty$.

The quantity $q$ appearing in the $\boldsymbol{\partial}_0\Theta$-equation can be obtained from Eq. (3.11) (which is the dimensionless version of the Raychaudhuri equation). The quantity $r_\alpha$ in the $\boldsymbol{\partial}_\alpha\Theta$-equation is more problematic. It can be obtained from Eq. (3.14) if the matrix $(\,2/3\,\delta^{\alpha\beta} - \Sigma^{\alpha\beta} + \epsilon^{\alpha\beta\gamma}\,W_\gamma\,)$ is invertible. If one considers the problem with two commuting spacelike Killing vector fields, this matrix reduces to a scalar quantity (two components of $r_\alpha$ are identically zero if one adapts the frame to the Killing vectors). This is the case considered by Hewitt and Wainwright [44]. They claim that this quantity typically is nonzero. Even if the above matrix would be degenerate at some point one might still obtain $r_\alpha$ from some of the constraints in the Jacobi identities (these constraints are identically satisfied in the case of two spacelike Killing vector fields [44]). Thus it should probably be possible, generically, to obtain the components of $r_\alpha$ (i.e., those that are nonzero even in the presence of Killing vector spacetime symmetries; such symmetries lead to a linear dependence between the components $r_\alpha$ if the symmetry has no timelike component).

## 4 Special Examples

Perhaps the main application of the above outlined formalism is that it allows one to obtain many interesting special cases in a useful unified and compact form just by deleting terms. A number of such cases will be discussed in this section.

### 4.1 Silent models of the Universe including magnetic Maxwell fields

In analogy with what is often done in the Newman–Penrose formalism [15, 48] we can, for example, kinematically specialize the preferred timelike congruence $\mathbf{u}$. Here we want to consider nonrotating congruences without acceleration, $0 = \omega^\alpha = \dot{u}^\alpha$. This means that we can impose a synchronous local coordinate system with a timelike coordinate affinely parametrizing the congruence.

Then one can ask the question of what restrictions one needs to impose on the curvature variables in order to obtain a decoupled system of evolution equations. This is again in close analogy with the Newman–Penrose approach where the restriction to an algebraically special Weyl curvature tensor leads to a decoupling of a set of equations which can be solved (see e.g. Refs. [49] and [50] for further references). It turns out that in the present case one can obtain a decoupling of a set of evolution equations if one assumes that the "magnetic part" of the Weyl curvature tensor vanishes, $H_{\alpha\beta} = 0$, and that the matter source defining the preferred timelike direction be a nonrotating perfect fluid with dust equation of state, $0 = p = q^\alpha = \pi_{\alpha\beta}$. These conditions lead to the generically spatially inhomogeneous "silent" cosmological models of Matarrese, Bruni and collaborators [17] - [19].

Here, different from the original models, we consider the possibility of also including a sourcefree magnetic Maxwell field, noninteracting with the dust fluid, which is aligned along one of the eigendirections of the "electric part" of the Weyl curvature tensor. Thus we will assume that $0 = E_\alpha = H_2 = H_3$, which, by Eq. (2.62), implies that the Poynting vector will be zero. A Maxwell field restricted in this way (together with the above mentioned kinematical specialization of the timelike congruence) will have covariantly vanishing spatial divergence and spatial rotation [7]. The total stress-energy-momentum tensor of such a matter configuration will have the following nonzero contributions:

$$\mu = \mu_{dust} + \tfrac{1}{2}(H_1)^2 \ , \tag{4.1}$$

$$p = \tfrac{1}{6}(H_1)^2 \ , \tag{4.2}$$

$$\pi_{11} = -\tfrac{2}{3}(H_1)^2 = -2\,\pi_{22} = -2\,\pi_{33} = -\tfrac{2}{3}\,\pi_+ \ . \tag{4.3}$$

Barnes and Rowlingson [51] showed that for irrotational perfect fluid matter with vanishing "magnetic part" of the Weyl curvature tensor, a canonical orthonormal frame comoving with $\mathbf{u}$ can be chosen such that the shear tensor of the fluid congruence, $\sigma_{\alpha\beta}$, and the "electric part" of the Weyl curvature tensor, $E_{\alpha\beta}$, are simultaneously diagonalized. This property follows from Eq. (2.47), which reduces to the algebraic constraint

$$0 = \epsilon^{\alpha\beta\gamma}\,\sigma_{\beta\delta}\,(E^{\delta}{}_{\gamma} + \tfrac{1}{2}\,\pi^{\delta}{}_{\gamma}) \ , \tag{4.4}$$

where we have already taken account of the proposed generalization. The diagonal components of Eqs. (2.41) and (2.45) yield that $0 = n_{11} = n_{22} = n_{33}$, which in turn implies $n^{\alpha}{}_{\alpha} = 0$. The off-diagonal components of Eqs. (2.27) and (2.44) yield that $\Omega^\alpha = 0$. Hence, it follows from Eq. (2.11) and $\omega^\alpha = 0$ (since $\sigma_{\alpha\beta}$ and $E_{\alpha\beta}$ can be chosen to be diagonal) that $\mathbf{e}_0$, $\mathbf{e}_1$, $\mathbf{e}_2$ and $\mathbf{e}_3$ are HSO ( a property which enables one to diagonalize the metric tensor ), and that the shear eigenframe is Fermi-propagated along $\mathbf{u}$. Thus we have generalized the results of Barnes and Rowlingson [51] to the proposed "silent" configuration with nonzero magnetic Maxwell field. We also remark that the momentum conservation equation (2.49) is identically satisfied by virtue of the Maxwell field equations. Note also that, since $0 = \dot{u}^\alpha = \omega^\alpha$, the earlier discussion related to inequalities for the values of the deceleration parameter $q$ and the dimensionless product $H\,\Delta t_p$ applies.

As a result of these specializations, and applying an irreducible decompositions of $\sigma_{\alpha\beta}$, $E_{\alpha\beta}$, $\pi_{\alpha\beta}$ and $^*S_{\alpha\beta}$ according to Eq. (2.86), the relations presented in Section 2 reduce to the following set:

The commutators:

$$[\,\mathbf{e}_0, \mathbf{e}_1\,] = -\tfrac{1}{3}(\Theta - 2\,\sigma_+)\,\mathbf{e}_1 \tag{4.5}$$

$$[\,\mathbf{e}_0, \mathbf{e}_2\,] = -\tfrac{1}{3}(\Theta + \sigma_+ + \sqrt{3}\,\sigma_-)\,\mathbf{e}_2 \tag{4.6}$$

$$[\,\mathbf{e}_0, \mathbf{e}_3\,] = -\tfrac{1}{3}(\Theta + \sigma_+ - \sqrt{3}\,\sigma_-)\,\mathbf{e}_3 \tag{4.7}$$

$$[\,\mathbf{e}_1, \mathbf{e}_2\,] = -(a_2 - n_{31})\,\mathbf{e}_1 + (a_1 + n_{23})\,\mathbf{e}_2 \tag{4.8}$$

$$[\,\mathbf{e}_2, \mathbf{e}_3\,] = -(a_3 - n_{12})\,\mathbf{e}_2 + (a_2 + n_{31})\,\mathbf{e}_3 \tag{4.9}$$

$$[\,\mathbf{e}_3, \mathbf{e}_1\,] = -(a_1 - n_{23})\,\mathbf{e}_3 + (a_3 + n_{12})\,\mathbf{e}_1 \ . \tag{4.10}$$

Since we have an irrotational geodesic congruence and we demonstrated how a Fermi-transported HSO spatial frame $\{\,\mathbf{e}_\alpha\,\}$ can be chosen, we can now write the orthonormal frame vectors as

$$\begin{array}{ll} \mathbf{e}_0 = \partial_{t_p} \ , & \mathbf{e}_2 = e^{-\beta^0 - \beta^+ - \sqrt{3}\,\beta^-}\,\partial_y \ , \\ \mathbf{e}_1 = e^{-\beta^0 + 2\,\beta^+}\,\partial_x \ , & \mathbf{e}_3 = e^{-\beta^0 - \beta^+ + \sqrt{3}\,\beta^-}\,\partial_z \ , \end{array} \tag{4.11}$$

where we have used a "Misner parametrization" of the diagonal 3-metric [52]. Of particular interest are the commutators involving $\mathbf{e}_0$ which give the following relations:

$$\mathbf{e}_0(\beta^0) = \tfrac{1}{3}\,\Theta \ , \qquad \mathbf{e}_0(\beta^\pm) = \tfrac{1}{3}\,\sigma_\pm \ . \tag{4.12}$$

Decoupled subsystem of ordinary differential (evolution) equations:

$$
\begin{aligned}
\mathbf{e}_0(\Theta) &= -\tfrac{1}{3}\,\Theta^2 - 2\,\sigma^2 - \tfrac{1}{2}\,\mu_{dust} - \tfrac{1}{2}\,(H_1)^2 + \Lambda && (4.13)\\
\mathbf{e}_0(\sigma_+) &= -\tfrac{1}{3}\,(2\,\Theta\, - \sigma_+)\,\sigma_+ - \tfrac{1}{3}\,(\sigma_-)^2 - E_+ + \tfrac{1}{2}\,(H_1)^2 && (4.14)\\
\mathbf{e}_0(\sigma_-) &= -\tfrac{2}{3}\,(\Theta\, + \sigma_+)\,\sigma_- - E_- && (4.15)\\
\mathbf{e}_0(E_+) &= -\tfrac{1}{2}\,\mu_{dust}\,\sigma_+ - (\Theta + \sigma_+)\,E_+ + \sigma_-\,E_- + \tfrac{1}{2}\,(\Theta + \sigma_+)\,(H_1)^2 && (4.16)\\
\mathbf{e}_0(E_-) &= -\tfrac{1}{2}\,\mu_{dust}\,\sigma_- - (\Theta - \sigma_+)\,E_- + \sigma_-\,E_+ - \tfrac{1}{2}\,\sigma_-\,(H_1)^2 && (4.17)\\
\mathbf{e}_0(\mu_{dust}) &= -\,\mu_{dust}\,\Theta && (4.18)\\
\mathbf{e}_0(H_1) &= -\tfrac{2}{3}\,(\Theta + \sigma_+)\,H_1\ . && (4.19)
\end{aligned}
$$

Remaining system of evolution equations:

$$
\begin{aligned}
\mathbf{e}_0(a_1) &= -\tfrac{1}{3}\,\mathbf{e}_1(\Theta + \sigma_+) - \tfrac{1}{3}\,(\Theta - 2\,\sigma_+)\,a_1 && (4.20)\\
\mathbf{e}_0(a_2) &= -\tfrac{1}{3}\,\mathbf{e}_2(\Theta - \tfrac{1}{2}\,\sigma_+ - \tfrac{\sqrt{3}}{2}\,\sigma_-) - \tfrac{1}{3}\,(\Theta + \sigma_+ + \sqrt{3}\,\sigma_-)\,a_2 && (4.21)\\
\mathbf{e}_0(a_3) &= -\tfrac{1}{3}\,\mathbf{e}_3(\Theta - \tfrac{1}{2}\,\sigma_+ + \tfrac{\sqrt{3}}{2}\,\sigma_-) - \tfrac{1}{3}\,(\Theta + \sigma_+ - \sqrt{3}\,\sigma_-)\,a_3 && (4.22)\\
\mathbf{e}_0(n_{23}) &= -\tfrac{1}{3}\,(\Theta - 2\,\sigma_+)\,n_{23} - \tfrac{1}{\sqrt{3}}\,\mathbf{e}_1(\sigma_-) && (4.23)\\
\mathbf{e}_0(n_{31}) &= -\tfrac{1}{3}\,(\Theta + \sigma_+ + \sqrt{3}\,\sigma_-)\,n_{31} - \tfrac{1}{2}\,\mathbf{e}_2(\sigma_+ - \tfrac{1}{\sqrt{3}}\,\sigma_-) && (4.24)\\
\mathbf{e}_0(n_{12}) &= -\tfrac{1}{3}\,(\Theta + \sigma_+ - \sqrt{3}\,\sigma_-)\,n_{12} + \tfrac{1}{2}\,\mathbf{e}_3(\sigma_+ + \tfrac{1}{\sqrt{3}}\,\sigma_-)\ , && (4.25)
\end{aligned}
$$

where

$$
\sigma^2 = \tfrac{1}{3}\,[\ (\sigma_+)^2 + (\sigma_-)^2\ ]\ . \qquad (4.26)
$$

Tracefree part and trace of 3-Ricci curvature of spacelike 3-surfaces orthogonal to $\mathbf{u}$ (Gauß equation):

$$
\begin{aligned}
{}^*S_+ &= -\tfrac{1}{2}\,[\ 2\,\mathbf{e}_1(a_1) - \mathbf{e}_2(a_2) - \mathbf{e}_3(a_3) - 4\,(n_{23})^2 + 2\,(n_{31})^2 + 2\,(n_{12})^2 \\
&\qquad - 3\,(\mathbf{e}_2 - 2\,a_2)\,(n_{31}) + 3\,(\mathbf{e}_3 - 2\,a_3)\,(n_{12})\ ] \\
&= E_+ - \tfrac{1}{3}\,(\Theta + \sigma_+)\,\sigma_+ + \tfrac{1}{3}\,(\sigma_-)^2 + \tfrac{1}{2}\,(H_1)^2 && (4.27)\\
{}^*S_- &= \tfrac{\sqrt{3}}{2}\,[\ \mathbf{e}_2(a_2) - \mathbf{e}_3(a_3) - 2\,(n_{31})^2 + 2\,(n_{12})^2 + 2\,(\mathbf{e}_1 - 2\,a_1)\,(n_{23}) \\
&\qquad - (\mathbf{e}_2 - 2\,a_2)\,(n_{31}) - (\mathbf{e}_3 - 2\,a_3)\,(n_{12})\ ] \\
&= E_- - \tfrac{1}{3}\,(\Theta - 2\,\sigma_+)\,\sigma_- && (4.28)\\
{}^*R &= 2\,[\ (2\,\mathbf{e}_1 - 3\,a_1)\,(a_1) + (2\,\mathbf{e}_2 - 3\,a_2)\,(a_2) + (2\,\mathbf{e}_3 - 3\,a_3)\,(a_3) \\
&\qquad - (n_{23})^2 - (n_{31})^2 - (n_{12})^2\ ] \\
&= -\tfrac{2}{3}\,\Theta^2 + 2\,\mu_{dust} + 2\,\sigma^2 + (H_1)^2 + 2\,\Lambda\ . && (4.29)
\end{aligned}
$$

The constraint equations:

$$
\begin{aligned}
0 &= \mathbf{e}_1(\Theta) + (\mathbf{e}_1 - 3\,a_1)\,(\sigma_+) - \sqrt{3}\,n_{23}\,\sigma_- && (4.30)\\
0 &= \mathbf{e}_2(\Theta) - \tfrac{1}{2}\,(\mathbf{e}_2 - 3\,a_2)\,(\sigma_+ + \sqrt{3}\,\sigma_-) - \tfrac{3}{2}\,n_{31}\,(\sigma_+ - \tfrac{1}{\sqrt{3}}\,\sigma_-) && (4.31)\\
0 &= \mathbf{e}_3(\Theta) - \tfrac{1}{2}\,(\mathbf{e}_3 - 3\,a_3)\,(\sigma_+ - \sqrt{3}\,\sigma_-) + \tfrac{3}{2}\,n_{12}\,(\sigma_+ + \tfrac{1}{\sqrt{3}}\,\sigma_-) && (4.32)\\
0 &= (\mathbf{e}_1 - 2\,a_1)\,(n_{31}) + (\mathbf{e}_2 - 2\,a_2)\,(n_{23}) + \mathbf{e}_1(a_2) - \mathbf{e}_2(a_1) && (4.33)\\
0 &= (\mathbf{e}_2 - 2\,a_2)\,(n_{12}) + (\mathbf{e}_3 - 2\,a_3)\,(n_{31}) + \mathbf{e}_2(a_3) - \mathbf{e}_3(a_2) && (4.34)\\
0 &= (\mathbf{e}_3 - 2\,a_3)\,(n_{23}) + (\mathbf{e}_1 - 2\,a_1)\,(n_{12}) + \mathbf{e}_3(a_1) - \mathbf{e}_1(a_3) && (4.35)\\
0 &= (\mathbf{e}_1 - a_1)\,(\sigma_-) - \sqrt{3}\,n_{23}\,\sigma_+ && (4.36)\\
0 &= (\mathbf{e}_2 - a_2)\,(\sigma_+ - \tfrac{1}{\sqrt{3}}\,\sigma_-) + n_{31}\,(\sigma_+ + \sqrt{3}\,\sigma_-) && (4.37)\\
0 &= (\mathbf{e}_3 - a_3)\,(\sigma_+ + \tfrac{1}{\sqrt{3}}\,\sigma_-) - n_{12}\,(\sigma_+ - \sqrt{3}\,\sigma_-) && (4.38)
\end{aligned}
$$

$$
\begin{aligned}
0 &= (\mathbf{e}_1 - 3\,a_1)\,(E_+) + \tfrac{1}{2}\,\mathbf{e}_1(\mu_{dust}) - \sqrt{3}\,n_{23}\,E_- + \tfrac{1}{2}\,a_1\,(H_1)^2 && (4.39)\\
0 &= (\mathbf{e}_2 - 3\,a_2)\,(E_+ + \sqrt{3}\,E_-) - \mathbf{e}_2(\mu_{dust}) + 3\,n_{31}\,(E_+ - \tfrac{1}{\sqrt{3}}\,E_-) - \tfrac{1}{2}\,(a_2 - n_{31})\,(H_1)^2 && (4.40)\\
0 &= (\mathbf{e}_3 - 3\,a_3)\,(E_+ - \sqrt{3}\,E_-) - \mathbf{e}_3(\mu_{dust}) - 3\,n_{12}\,(E_+ + \tfrac{1}{\sqrt{3}}\,E_-) - \tfrac{1}{2}\,(a_3 + n_{12})\,(H_1)^2 && (4.41)\\
0 &= (\mathbf{e}_1 - a_1)\,(E_-) - \sqrt{3}\,n_{23}\,E_+ + \tfrac{\sqrt{3}}{2}\,n_{23}\,(H_1)^2 && (4.42)\\
0 &= (\mathbf{e}_2 - a_2)\,(E_+ - \tfrac{1}{\sqrt{3}}\,E_-) + n_{31}\,(E_+ + \sqrt{3}\,E_-) - \tfrac{1}{2}\,(a_2 - n_{31})\,(H_1)^2 && (4.43)\\
0 &= (\mathbf{e}_3 - a_3)\,(E_+ + \tfrac{1}{\sqrt{3}}\,E_-) - n_{12}\,(E_+ - \sqrt{3}\,E_-) - \tfrac{1}{2}\,(a_3 + n_{12})\,(H_1)^2 && (4.44)\\
0 &= (\mathbf{e}_1 - 2\,a_1)\,(H_1) && (4.45)\\
0 &= (\mathbf{e}_2 - a_2 + n_{31})\,(H_1) && (4.46)\\
0 &= (\mathbf{e}_3 - a_3 - n_{12})\,(H_1)\;. && (4.47)
\end{aligned}
$$

In order to derive Eqs. (4.39) - (4.44) we made use of the Maxwell equations (4.45) - (4.47). Lesame et al [53] recently reported that *no* new constraints arise from propagating the conditions (4.30) - (4.44) along the fluid flow lines in the zero magnetic Maxwell field case ($H_1 = 0$). Hence, any $H_1 = 0$ initial data, which satisfy the constraints (4.30) - (4.44), evolve according to the evolution equations (4.13) - (4.18) and provide a consistent solution to the Einstein field equations with dust matter and $0 = \omega^\alpha = \dot{u}^\alpha$ and $H_{\alpha\beta} = 0$. Without proof we here assume that this consistency is not being violated by the introduction of the nonzero magnetic Maxwell field, $H_1 \neq 0$, and leave a rigorous consistency check for a future project.

It has been conjectured by Berger et al [54] that for irrotational matter flows there might be a link between the presence of gravitational radiation in a spacetime and the nonvanishing of the conformal curvature of the spacelike 3-surfaces orthogonal to $\mathbf{u}$. Their hypothesis arose from an investigation of the Petrov type D spatially inhomogeneous dust spacetimes of Szekeres [55], in which they demonstrated that in those models the spacelike 3-surfaces have vanishing conformal curvature. Here we want to reverse the argument. As a *nonzero* "magnetic part" of the Weyl curvature tensor as measured with respect to fundamental observers comoving with $\mathbf{u}$ is assumed to be a necessary condition for the presence of gravitational radiation in a particular spacetime geometry, we are interested in whether the 3-Cotton–York tensor is zero for (generalized) "silent" models and thus further support is given to the Berger et al conjecture. Since the canonical orthonormal frame for "silent" models employed here diagonalizes the shear tensor as well as $\pi_{\alpha\beta}$, and only off-diagonal terms of the commutation functions $n_{\alpha\beta}$ are nonzero, it follows directly from Eq. (2.53) that the on-diagonal terms of ${}^*C_{\alpha\beta}$ vanish. However, by use of the $(0\alpha)$-equations (4.30) - (4.32), the $H$-constraint equations (4.36) - (4.38), and the Maxwell equations (4.45) - (4.47), we obtain from Eq. (2.53) for the off-diagonal terms of ${}^*C_{\alpha\beta}$

$$
\begin{aligned}
{}^*C_1 &= \tfrac{1}{\sqrt{3}}\,\sigma_-\,(\mathbf{e}_1 - a_1)\,(\sigma_+) + n_{23}\,[\,(\sigma_+)^2 - \tfrac{2}{3}\,(\sigma_-)^2 - (H_1)^2\,] && (4.48)\\
{}^*C_2 &= -\tfrac{1}{\sqrt{3}}\,(\sigma_+ - \tfrac{1}{\sqrt{3}}\,\sigma_-)\,(\mathbf{e}_2 - a_2)\,(\sigma_-) + \sqrt{3}\,n_{31}\,(\sigma_+ + \tfrac{1}{3\sqrt{3}}\,\sigma_-)\,\sigma_- + \tfrac{1}{2}\,(a_2 - n_{31})\,(H_1)^2 && (4.49)\\
{}^*C_3 &= -\tfrac{1}{\sqrt{3}}\,(\sigma_+ + \tfrac{1}{\sqrt{3}}\,\sigma_-)\,(\mathbf{e}_3 - a_3)\,(\sigma_-) - \sqrt{3}\,n_{12}\,(\sigma_+ - \tfrac{1}{3\sqrt{3}}\,\sigma_-)\,\sigma_- - \tfrac{1}{2}\,(a_3 + n_{12})\,(H_1)^2\;. && (4.50)
\end{aligned}
$$

We do *not* see any a priori reason why the righthand sides of these expressions should be zero. Hence, against our expectations, for a general representative within the (generalized) class of "silent" models the zero "magnetic part" of the Weyl tensor condition does *not* imply the vanishing of the 3-Cotton–York tensor. Nevertheless, the result of Berger et al as regards the dust spacetimes of Szekeres is contained in Eqs. (4.48) - (4.50): for $0 = \sigma_- = n_{23}$ (see Ref. [56]) and $H_1 = 0$ it follows that $0 = {}^*C_1 = {}^*C_2 = {}^*C_3$.

Dimensionless formulation:

In terms of the expansion-normalized dimensionless variables introduced in Section 3 the sets of equations (4.1) - (4.3), (4.5) - (4.10), (4.14) - (4.19), (4.20) - (4.25), (4.27) - (4.29) and (4.30) - (4.47) become respectively:

$$
\begin{aligned}
\Omega &= \Omega_{dust} + \tfrac{1}{2}\,(\mathcal{H}_1)^2 && (4.51)\\
P &= \tfrac{1}{6}\,(\mathcal{H}_1)^2 && (4.52)
\end{aligned}
$$

$$\Pi_+ = (\mathcal{H}_1)^2 \quad (4.53)$$
$$0 = Q_\alpha = \Pi_- \, . \quad (4.54)$$

The commutators:

$$[\boldsymbol{\partial}_0, \boldsymbol{\partial}_1] = -r_1\, \boldsymbol{\partial}_0 + (q + 2\,\Sigma_+)\, \boldsymbol{\partial}_1 \quad (4.55)$$
$$[\boldsymbol{\partial}_0, \boldsymbol{\partial}_2] = -r_2\, \boldsymbol{\partial}_0 + (q - \Sigma_+ - \sqrt{3}\,\Sigma_-)\, \boldsymbol{\partial}_2 \quad (4.56)$$
$$[\boldsymbol{\partial}_0, \boldsymbol{\partial}_3] = -r_3\, \boldsymbol{\partial}_0 + (q - \Sigma_+ + \sqrt{3}\,\Sigma_-)\, \boldsymbol{\partial}_3 \quad (4.57)$$
$$[\boldsymbol{\partial}_1, \boldsymbol{\partial}_2] = -(r_2 + 3\,A_2 - 3\,N_{31})\, \boldsymbol{\partial}_1 + (r_1 + 3\,A_1 + 3\,N_{23})\, \boldsymbol{\partial}_2 \quad (4.58)$$
$$[\boldsymbol{\partial}_2, \boldsymbol{\partial}_3] = -(r_3 + 3\,A_3 - 3\,N_{12})\, \boldsymbol{\partial}_2 + (r_2 + 3\,A_2 + 3\,N_{31})\, \boldsymbol{\partial}_3 \quad (4.59)$$
$$[\boldsymbol{\partial}_3, \boldsymbol{\partial}_1] = -(r_1 + 3\,A_1 - 3\,N_{23})\, \boldsymbol{\partial}_3 + (r_3 + 3\,A_3 + 3\,N_{12})\, \boldsymbol{\partial}_1 \, . \quad (4.60)$$

On an individual fluid flow line one can choose $N = \pm 3\,\Theta$, $N^\alpha = 0$. With this choice Eq. (4.12) (with which one can replace the above $\boldsymbol{\partial}_0$-commutators, if one is so inclined) takes the form

$$\boldsymbol{\partial}_0 \beta^0 = 1 \, , \qquad \boldsymbol{\partial}_0 \beta^+ = \Sigma_\pm \, . \quad (4.61)$$

Decoupled subsystem of ordinary differential (evolution) equations:

$$\boldsymbol{\partial}_0 \Sigma_+ = (q - 1 + \Sigma_+)\,\Sigma_+ - (\Sigma_-)^2 - \mathcal{E}_+ + \tfrac{1}{2}\,(\mathcal{H}_1)^2 \quad (4.62)$$
$$\boldsymbol{\partial}_0 \Sigma_- = (q - 1 - 2\,\Sigma_+)\,\Sigma_- - \mathcal{E}_- \quad (4.63)$$
$$\boldsymbol{\partial}_0 \mathcal{E}_+ = (2\,q - 1 - 3\,\Sigma_+)\,\mathcal{E}_+ - \tfrac{3}{2}\,\Omega_{dust}\,\Sigma_+ + 3\,\Sigma_-\,\mathcal{E}_- + \tfrac{3}{2}\,(1 + \Sigma_+)\,(\mathcal{H}_1)^2 \quad (4.64)$$
$$\boldsymbol{\partial}_0 \mathcal{E}_- = (2\,q - 1 + 3\,\Sigma_+)\,\mathcal{E}_- - \tfrac{3}{2}\,\Omega_{dust}\,\Sigma_- + 3\,\Sigma_-\,\mathcal{E}_+ - \tfrac{3}{2}\,\Sigma_-\,(\mathcal{H}_1)^2 \quad (4.65)$$
$$\boldsymbol{\partial}_0 \Omega_{dust} = (2\,q - 1)\,\Omega_{dust} \quad (4.66)$$
$$\boldsymbol{\partial}_0 \mathcal{H}_1 = (q - 1 - 2\,\Sigma_+)\,\mathcal{H}_1 \quad (4.67)$$
$$\boldsymbol{\partial}_0 \Omega_\Lambda = 2\,(q + 1)\,\Omega_\Lambda \, . \quad (4.68)$$

As before ( cf. Eq. (3.7) ) we have an evolution equation for $\Theta$, which in itself is decoupled from the set (4.62) - (4.68):

$$\boldsymbol{\partial}_0 \Theta = -(1 + q)\,\Theta \, , \quad (4.69)$$

and the parameter $q$ is given by

$$q = 2\,[\,(\Sigma_+)^2 + (\Sigma_-)^2\,] + \tfrac{1}{2}\,\Omega_{dust} + \tfrac{1}{2}\,(\mathcal{H}_1)^2 - \Omega_\Lambda \, . \quad (4.70)$$

Remaining system of evolution equations:

$$\boldsymbol{\partial}_0 A_1 = q\,A_1 + \tfrac{1}{3}\,r_1 - \tfrac{1}{3}\,(\boldsymbol{\partial}_1 - r_1 - 6\,A_1)\,\Sigma_+ \quad (4.71)$$
$$\boldsymbol{\partial}_0 A_2 = q\,A_2 + \tfrac{1}{3}\,r_2 + \tfrac{1}{6}\,(\boldsymbol{\partial}_2 - r_2 - 6\,A_2)\,(\Sigma_+ + \sqrt{3}\,\Sigma_-) \quad (4.72)$$
$$\boldsymbol{\partial}_0 A_3 = q\,A_3 + \tfrac{1}{3}\,r_3 + \tfrac{1}{6}\,(\boldsymbol{\partial}_3 - r_3 - 6\,A_3)\,(\Sigma_+ - \sqrt{3}\,\Sigma_-) \quad (4.73)$$
$$\boldsymbol{\partial}_0 N_{23} = (q + 2\,\Sigma_+)\,N_{23} - \tfrac{1}{\sqrt{3}}\,(\boldsymbol{\partial}_1 - r_1)\,\Sigma_- \quad (4.74)$$
$$\boldsymbol{\partial}_0 N_{31} = (q - \Sigma_+ - \sqrt{3}\,\Sigma_-)\,N_{31} - \tfrac{1}{2}\,(\boldsymbol{\partial}_2 - r_2)\,(\Sigma_+ - \tfrac{1}{\sqrt{3}}\,\Sigma_-) \quad (4.75)$$
$$\boldsymbol{\partial}_0 N_{12} = (q - \Sigma_+ + \sqrt{3}\,\Sigma_-)\,N_{12} + \tfrac{1}{2}\,(\boldsymbol{\partial}_3 - r_3)\,(\Sigma_+ + \tfrac{1}{\sqrt{3}}\,\Sigma_-) \, . \quad (4.76)$$

Tracefree part and trace of 3-Ricci curvature of spacelike 3-surfaces orthogonal to $\mathbf{u}$ (Gauß equation):

$$\begin{aligned} \mathcal{S}_+ &= -\tfrac{1}{2}\,[\,2\,(\boldsymbol{\partial}_1 - r_1)\,A_1 - (\boldsymbol{\partial}_2 - r_2)\,A_2 - (\boldsymbol{\partial}_3 - r_3)A_3 - 12\,(N_{23})^2 + 6\,(N_{31})^2 + 6\,(N_{12})^2 \\ &\quad - 3\,(\boldsymbol{\partial}_2 - r_2 - 6\,A_2)\,N_{31} + 3\,(\boldsymbol{\partial}_3 - r_3 - 6\,A_3)\,N_{12}\,] \\ &= \mathcal{E}_+ - (1 + \Sigma_+)\,\Sigma_+ + (\Sigma_-)^2 + \tfrac{1}{2}\,(\mathcal{H}_1)^2 \end{aligned} \quad (4.77)$$
$$\mathcal{S}_- = \tfrac{\sqrt{3}}{2}\,[\,(\boldsymbol{\partial}_2 - r_2)\,A_2 - (\boldsymbol{\partial}_3 - r_3)\,A_3 - 6\,(N_{31})^2 + 6\,(N_{12})^2 + 2\,(\boldsymbol{\partial}_1 - r_1 - 6\,A_1)\,N_{23}$$

$$
\begin{aligned}
& \quad - (\boldsymbol{\partial}_2 - r_2 - 6\,A_2)\,N_{31} - (\boldsymbol{\partial}_3 - r_3 - 6\,A_3)\,N_{12}\ ] \\
&= \mathcal{E}_- - (1 - 2\,\Sigma_+)\,\Sigma_- && (4.78)\\
\mathcal{K} &= -(2\,\boldsymbol{\partial}_1 - 2\,r_1 - 9\,A_1)\,A_1 - (2\,\boldsymbol{\partial}_2 - 2\,r_2 - 9\,A_2)\,A_2 - (2\,\boldsymbol{\partial}_3 - 2\,r_3 - 9\,A_3)\,A_3 \\
& \quad + 3\,(N_{23})^2 + 3\,(N_{31})^2 + 3\,(N_{12})^2 \\
&= 1 - \Omega_{dust} - [\ (\Sigma_+)^2 + (\Sigma_-)^2\ ] - \tfrac{1}{2}\,(\mathcal{H}_1)^2 - \Omega_\Lambda\ . && (4.79)
\end{aligned}
$$

The constraint equations:

$$
\begin{aligned}
0 &= (\boldsymbol{\partial}_1 + r_1)\,\Theta && (4.80)\\
0 &= (\boldsymbol{\partial}_2 + r_2)\,\Theta && (4.81)\\
0 &= (\boldsymbol{\partial}_3 + r_3)\,\Theta && (4.82)\\
0 &= r_1 - (\boldsymbol{\partial}_1 - r_1 - 9\,A_1)\,\Sigma_+ + 3\sqrt{3}\,N_{23}\,\Sigma_- && (4.83)\\
0 &= r_2 + \tfrac{1}{2}\,(\boldsymbol{\partial}_2 - r_2 - 9\,A_2)\,(\Sigma_+ + \sqrt{3}\Sigma_-) + \tfrac{9}{2}\,N_{31}\,(\Sigma_+ - \tfrac{1}{\sqrt{3}}\,\Sigma_-) && (4.84)\\
0 &= r_3 + \tfrac{1}{2}\,(\boldsymbol{\partial}_3 - r_3 - 9\,A_3)\,(\Sigma_+ - \sqrt{3}\Sigma_-) - \tfrac{9}{2}\,N_{12}\,(\Sigma_+ + \tfrac{1}{\sqrt{3}}\,\Sigma_-) && (4.85)\\
0 &= (\boldsymbol{\partial}_1 - r_1 - 6\,A_1)\,N_{31} + (\boldsymbol{\partial}_2 - r_2 - 6\,A_2)\,N_{23} + (\boldsymbol{\partial}_1 - r_1)\,A_2 - (\boldsymbol{\partial}_2 - r_2)\,A_1 && (4.86)\\
0 &= (\boldsymbol{\partial}_2 - r_2 - 6\,A_2)\,N_{12} + (\boldsymbol{\partial}_3 - r_3 - 6\,A_3)\,N_{31} + (\boldsymbol{\partial}_2 - r_2)\,A_3 - (\boldsymbol{\partial}_3 - r_3)\,A_2 && (4.87)\\
0 &= (\boldsymbol{\partial}_3 - r_3 - 6\,A_3)\,N_{23} + (\boldsymbol{\partial}_1 - r_1 - 6\,A_1)\,N_{12} + (\boldsymbol{\partial}_3 - r_3)\,A_1 - (\boldsymbol{\partial}_1 - r_1)\,A_3 && (4.88)\\
0 &= (\boldsymbol{\partial}_1 - r_1 - 3\,A_1)\,\Sigma_- - 3\sqrt{3}\,N_{23}\,\Sigma_+ && (4.89)\\
0 &= (\boldsymbol{\partial}_2 - r_2 - 3\,A_2)\,(\Sigma_+ - \tfrac{1}{\sqrt{3}}\,\Sigma_-) + 3\,N_{31}\,(\Sigma_+ + \sqrt{3}\,\Sigma_-) && (4.90)\\
0 &= (\boldsymbol{\partial}_3 - r_3 - 3\,A_3)\,(\Sigma_+ + \tfrac{1}{\sqrt{3}}\,\Sigma_-) - 3\,N_{12}\,(\Sigma_+ - \sqrt{3}\,\Sigma_-) && (4.91)\\
0 &= (\boldsymbol{\partial}_1 - 2\,r_1 - 9\,A_1)\,\mathcal{E}_+ + \tfrac{1}{2}\,(\boldsymbol{\partial}_1 - 2\,r_1)\,\Omega - 3\sqrt{3}\,N_{23}\,\mathcal{E}_- + \tfrac{3}{2}\,A_1\,(\mathcal{H}_1)^2 && (4.92)\\
0 &= (\boldsymbol{\partial}_2 - 2\,r_2 - 9\,A_2)\,(\mathcal{E}_+ + \sqrt{3}\,\mathcal{E}_-) - (\boldsymbol{\partial}_2 - 2\,r_2)\,\Omega + 9\,N_{31}\,(\mathcal{E}_+ - \tfrac{1}{\sqrt{3}}\,\mathcal{E}_-) \\
& \quad - \tfrac{3}{2}\,(A_2 - N_{31})\,(\mathcal{H}_1)^2 && (4.93)\\
0 &= (\boldsymbol{\partial}_3 - 2\,r_3 - 9\,A_3)\,(\mathcal{E}_+ - \sqrt{3}\,\mathcal{E}_-) - (\boldsymbol{\partial}_3 - 2\,r_3)\,\Omega - 9\,N_{12}\,(\mathcal{E}_+ + \tfrac{1}{\sqrt{3}}\,\mathcal{E}_-) \\
& \quad - \tfrac{3}{2}\,(A_3 + N_{12})\,(\mathcal{H}_1)^2 && (4.94)\\
0 &= (\boldsymbol{\partial}_1 - 2\,r_1 - 3\,A_1)\,\mathcal{E}_- - 3\sqrt{3}\,N_{23}\,\mathcal{E}_+ + \tfrac{3\sqrt{3}}{2}\,N_{23}\,(\mathcal{H}_1)^2 && (4.95)\\
0 &= (\boldsymbol{\partial}_2 - 2\,r_2 - 3\,A_2)\,(\mathcal{E}_+ - \tfrac{1}{\sqrt{3}}\,\mathcal{E}_-) + 3\,N_{31}\,(\mathcal{E}_+ + \sqrt{3}\,\mathcal{E}_-) - \tfrac{3}{2}\,(A_2 - N_{31})\,(\mathcal{H}_1)^2 && (4.96)\\
0 &= (\boldsymbol{\partial}_3 - 2\,r_3 - 3\,A_3)\,(\mathcal{E}_+ + \tfrac{1}{\sqrt{3}}\,\mathcal{E}_-) - 3\,N_{12}\,(\mathcal{E}_+ - \sqrt{3}\,\mathcal{E}_-) - \tfrac{3}{2}\,(A_3 + N_{12})\,(\mathcal{H}_1)^2 && (4.97)\\
0 &= (\boldsymbol{\partial}_1 - r_1 - 6\,A_1)\,\mathcal{H}_1 && (4.98)\\
0 &= (\boldsymbol{\partial}_2 - r_2 - 3\,A_2 + 3\,N_{31})\,\mathcal{H}_1 && (4.99)\\
0 &= (\boldsymbol{\partial}_3 - r_3 - 3\,A_3 - 3\,N_{12})\,\mathcal{H}_1 && (4.100)\\
0 &= (\boldsymbol{\partial}_1 - 2\,r_1)\,\Omega_\Lambda && (4.101)\\
0 &= (\boldsymbol{\partial}_2 - 2\,r_2)\,\Omega_\Lambda && (4.102)\\
0 &= (\boldsymbol{\partial}_3 - 2\,r_3)\,\Omega_\Lambda\ . && (4.103)
\end{aligned}
$$

The evolution equation (4.68) arises from Eq. (3.71), while the extra constraints (4.80) - (4.82) and (4.101) - (4.103) derive respectively from Eqs. (3.8) and (3.72).

There are a number of ways one can generalize the "silent" models. For example, one can relax the conditions on the kinematic quantities and allow rotation. Another possibility is to allow parallel electric and magnetic Maxwell fields that are *not* aligned with the shear or "electric part" eigendirections. The key to generalizations seems to be to respect the "silent" conditions: $0 = H_{\alpha\beta} = q^\alpha$, and to use a geodesic congruence, $\dot{u}^\alpha = 0$. Perhaps it is not so surprising that, in such cases, one obtains ordinary differential equations, since the "silent" conditions are assumed to correspond to the fact that no information is exchanged between different fluid flow lines, motivating the notion "silent". The above mentioned rotational and electromagnetic generalizations will be discussed further elsewhere [57].

## 4.2 Locally rotationally symmetric models

LRS symmetry of a fluid spacetime configuration is given when there exists a (normalized) spacelike congruence $\mathbf{e}$ orthogonal to $\mathbf{u}$, covariantly defined by for example a nonzero vorticity vector field, an eigendirection of a nonzero degenerate rate of shear tensor field, or a nonvanishing acceleration vector field. As all physical measurements have to be invariant under spatial rotations about $\mathbf{e}$ and hence identical in all spatial directions orthogonal to $\mathbf{e}$, this congruence must be geodesic and shearfree in the local rest 3-spaces orthogonal to $\mathbf{u}$ [9, 20]. Given a choice of orthonormal frame with $\mathbf{e}_0 = \mathbf{u}$, $\mathbf{e}_1 = \mathbf{e}$, *all* covariantly defined spacelike *vector fields* orthogonal to $\mathbf{u}$ are consequently colinear with $\mathbf{e}_1$, and *all* covariantly defined spacelike *symmetric tracefree tensor fields* orthogonal to $\mathbf{u}$ have coinciding eigenframes with two equal eigenvalues ($T_- = T_1 = T_2 = T_3 = 0$) [9, 20, 58]. Furthermore, one can make a specialization of the orthonormal frame such that only the following commutation functions are nonzero [9, 20]:

$$\Theta,\ \dot{u}_1 := \dot{u},\ \sigma_{11} = -2\,\sigma_{22} = -2\,\sigma_{33} := -\tfrac{2}{3}\,\sigma_+,\ \omega_1 = \Omega_1 := \omega,\ a_1 := a,\ a_2 = n_{31},\ n_{11}\ . \tag{4.104}$$

The spatial frame derivatives of those variables $f$, which are proportional to covariantly defined scalar quantities, have to satisfy the condition $0 = \mathbf{e}_2(f) = \mathbf{e}_3(f)$.

The commutators:

$$\begin{aligned}
[\,\mathbf{e}_0, \mathbf{e}_1\,] &= \dot{u}\,\mathbf{e}_0 - \tfrac{1}{3}\,(\Theta - 2\,\sigma_+)\,\mathbf{e}_1 && (4.105)\\
[\,\mathbf{e}_0, \mathbf{e}_2\,] &= -\,\tfrac{1}{3}\,(\Theta + \sigma_+)\,\mathbf{e}_2 && (4.106)\\
[\,\mathbf{e}_0, \mathbf{e}_3\,] &= -\,\tfrac{1}{3}\,(\Theta + \sigma_+)\,\mathbf{e}_3 && (4.107)\\
[\,\mathbf{e}_1, \mathbf{e}_2\,] &= a\,\mathbf{e}_2 && (4.108)\\
[\,\mathbf{e}_2, \mathbf{e}_3\,] &= -\,2\,\omega\,\mathbf{e}_0 + n_{11}\,\mathbf{e}_1 + 2\,n_{31}\,\mathbf{e}_3 && (4.109)\\
[\,\mathbf{e}_3, \mathbf{e}_1\,] &= -\,a\,\mathbf{e}_3\ , && (4.110)
\end{aligned}$$

where $\mathbf{e}_3(n_{31}) = 0$, and $n_{11}$ is proportional to the magnitude of the spatial rotation of the covariantly defined spacelike vector field $\mathbf{e}$. From the LRS spacetime symmetry and a particular use of the freedom of fixing the orthonormal frame it follows that $\mathbf{e}_2$ and $\mathbf{e}_3$ are HSO [9, 20]. In the zero-vorticity LRS case, $\omega = 0$, it follows that the spatial frame is actually Fermi-propagated along $\mathbf{u}$, and if *both* $0 = \omega = n_{11}$, all four frame fields are HSO (cf. Eq.(2.11)).

Table 1: Notational differences. Note that $K$ ($^2K$) is actually a covariantly defined variable *only* if $\omega = 0 = n_{11}$.

| Stewart and Ellis (1968) [20] | van Elst and Ellis (1995) [58] | van Elst and Uggla |
|---|---|---|
| $\alpha$ | $[\,(\Theta/c) + 2\sqrt{3}\,(\sigma/c)\,]\,/\,3$ | $(\Theta - 2\,\sigma_+)\,/\,3$ |
| $\beta$ | $[\,(\Theta/c) - \sqrt{3}\,(\sigma/c)\,]\,/\,3$ | $(\Theta + \sigma_+)\,/\,3$ |
| $a$ | $-\,a/2$ | $a$ |
| $k$ | $-\,k$ | $-\,n_{11}$ |
| $s$ | — | $2\,n_{31}$ |
| $B^2\,K$ | $K$ | $2\,(\mathbf{e}_2 - 2\,n_{31})\,(n_{31}),\ ^2K$ |

The evolution equations:

$$\begin{aligned}
\mathbf{e}_0(\Theta) &= -\tfrac{1}{3}\,\Theta^2 + (\mathbf{e}_1 + \dot{u} - 2\,a)\,(\dot{u}) - \tfrac{2}{3}\,(\sigma_+)^2 + 2\,\omega^2 - \tfrac{1}{2}\,(\mu + 3p) + \Lambda && (4.111)\\
\mathbf{e}_0(\sigma_+) &= -\tfrac{1}{3}\,(2\,\Theta - \sigma_+)\,\sigma_+ - (\mathbf{e}_1 + \dot{u} + a)\,(\dot{u}) + \omega^2 - (E_+ - \tfrac{1}{2}\,\pi_+) && (4.112)\\
\mathbf{e}_0(\omega) &= -\tfrac{2}{3}\,(\Theta + \sigma_+)\,\omega + \tfrac{1}{2}\,n_{11}\,\dot{u} && (4.113)\\
\mathbf{e}_0(a) &= -\tfrac{1}{3}\,(\Theta + \sigma_+)\,(a + \dot{u}) - \tfrac{1}{2}\,n_{11}\,\omega - \tfrac{1}{2}\,q_1 && (4.114)\\
\mathbf{e}_0(n_{11}) &= -\tfrac{1}{3}\,(\Theta + 4\,\sigma_+)\,n_{11} && (4.115)\\
\mathbf{e}_0(n_{31}) &= -\tfrac{1}{3}\,(\Theta + \sigma_+)\,n_{31} && (4.116)\\
\mathbf{e}_0(E_+ + \tfrac{1}{2}\,\pi_+) &= -\tfrac{1}{2}\,(\mu + p)\,\sigma_+ - \Theta\,(E_+ + \tfrac{1}{6}\,\pi_+) + \tfrac{1}{2}\,(\mathbf{e}_1 + 2\,\dot{u} + a)\,(q_1)\\
&\quad - \sigma_+\,(E_+ - \tfrac{1}{6}\,\pi_+) - \tfrac{3}{2}\,n_{11}\,H_+ && (4.117)\\
\mathbf{e}_0(H_+) &= -(\Theta + \sigma_+)\,H_+ + \tfrac{3}{2}\,\omega\,q_1 + \tfrac{3}{2}\,n_{11}\,(E_+ - \tfrac{1}{2}\,\pi_+) && (4.118)\\
\mathbf{e}_0(\mu) &= -(\mu + p)\,\Theta - (\mathbf{e}_1 + 2\,\dot{u} - 2\,a)\,(q_1) - \tfrac{2}{3}\,\sigma_+\,\pi_+ && (4.119)\\
\mathbf{e}_0(q_1) &= -\tfrac{4}{3}\,\Theta\,q_1 - \mathbf{e}_1(p) - (\mu + p)\,\dot{u} + \tfrac{2}{3}\,(\mathbf{e}_1 + \dot{u} - 3\,a)\,(\pi_+) + \tfrac{2}{3}\,\sigma_+\,q_1\ . && (4.120)
\end{aligned}$$

The constraint equations:

$$\begin{aligned}
0 &= (\mathbf{e}_1 - \dot{u} - 2\,a)\,(\omega) && (4.121)\\
0 &= q_1 - \tfrac{2}{3}\,\mathbf{e}_1(\Theta) - \tfrac{2}{3}\,(\mathbf{e}_1 - 3\,a)\,(\sigma_+) + n_{11}\,\omega && (4.122)\\
0 &= (\mathbf{e}_1 - a)\,(a) + \tfrac{1}{9}\,\Theta^2 - \tfrac{1}{9}\,(\Theta + 2\,\sigma_+)\sigma_+ + \tfrac{1}{4}\,(n_{11})^2 + \tfrac{1}{3}\,(E_+ + \tfrac{1}{2}\,\pi_+) - \tfrac{1}{3}\,\mu - \tfrac{1}{3}\,\Lambda && (4.123)\\
0 &= (\mathbf{e}_1 - 2\,a)\,(n_{11}) - \tfrac{2}{3}\,(\Theta - 2\,\sigma_+)\,\omega && (4.124)\\
0 &= (\mathbf{e}_1 - a)\,(n_{31}) && (4.125)\\
0 &= H_+ + 3\,(\dot{u} + a)\,\omega + \tfrac{3}{2}\,n_{11}\,\sigma_+ && (4.126)\\
0 &= (\mathbf{e}_1 - 3\,a)\,(E_+ + \tfrac{1}{2}\,\pi_+) + \tfrac{1}{2}\,\mathbf{e}_1(\mu) - \tfrac{1}{2}\,(\Theta + \sigma_+)\,q_1 + 3\,\omega\,H_+ && (4.127)\\
0 &= (\mathbf{e}_1 - 3\,a)\,(H_+) + \tfrac{3}{2}\,(\mu + p)\,\omega - 3\,\omega\,(E_+ - \tfrac{1}{6}\,\pi_+) + \tfrac{3}{4}\,n_{11}\,q_1 && (4.128)
\end{aligned}$$

Tracefree part and trace of ${}^*S_{\alpha\beta}$ (3-Ricci curvature of spacelike 3-surfaces orthogonal to $\mathbf{u}$ when $\omega = 0$):

$$\begin{aligned}
{}^*S_+ &= -\,\mathbf{e}_1(a) - (n_{11})^2 + 2\,(\mathbf{e}_2 - 2\,n_{31})\,(n_{31})\\
&= (E_+ + \tfrac{1}{2}\,\pi_+) - \tfrac{1}{3}\,(\Theta + \sigma_+)\,\sigma_+ - 3\,\omega^2 && (4.129)\\
{}^*R &= 2\,(2\,\mathbf{e}_1 - 3\,a)\,(a) - \tfrac{1}{2}\,(n_{11})^2 + 4\,(\mathbf{e}_2 - 2\,n_{31})\,(n_{31})\\
&= -\tfrac{2}{3}\,\Theta^2 + 2\,\mu + \tfrac{2}{3}\,(\sigma_+)^2 - 6\,\omega^2 + 2\,\Lambda\ . && (4.130)
\end{aligned}$$

The variables $\Theta$, $\sigma_+$, $\omega$, $a$, $n_{11}$, $E_+$ and $H_+$ correspond directly (up to a constant factor) to covariantly defined scalar quantities (cf. Ref. [58]), and therefore, due to the imposed LRS spacetime symmetry, have vanishing frame derivatives in the $\mathbf{e}_2$- and $\mathbf{e}_3$-directions. Thus, by application of the commutator relation (4.109) to this set of variables, and subsequent substitution from the evolution and constraint equations (4.111) - (4.120) and (4.121) - (4.128), one can successively derive consistency conditions which need to be satisfied by any solution to the LRS dynamical equations [20]. In this way one obtains from $\omega$ the condition

$$0 = \tfrac{2}{3}\,(\Theta + \sigma_+)\,\omega + a\,n_{11}\ , \qquad (4.131)$$

which is reproduced when (4.109) acts upon $n_{11}$; from $(\Theta + \sigma_+)$, using (4.131), the condition

$$\begin{aligned}
0 &= \omega\,[\,(E_+ - \tfrac{1}{2}\,\pi_+) + \tfrac{1}{3}\,(\Theta - \sigma_+)\,\Theta - \tfrac{2}{3}\,(\sigma_+)^2 - 3\,\omega^2 + 3\,a\,\dot{u} + \tfrac{3}{4}\,(n_{11})^2\\
&\quad + \tfrac{1}{2}\,(\mu + 3\,p) - \Lambda\,] + \tfrac{3}{4}\,n_{11}\,q_1\ ; && (4.132)
\end{aligned}$$

and from $a$, using (4.131), the condition

$$\begin{aligned}
0 &= n_{11}\,[\,(E_+ + \tfrac{1}{2}\,\pi_+) + \tfrac{1}{3}\,(\Theta - \sigma_+)\,\Theta - \tfrac{2}{3}\,(\sigma_+)^2 - 3\,\omega^2 + 3\,a\,\dot{u} + \tfrac{3}{4}\,(n_{11})^2\\
&\quad - \mu - \Lambda\,] - 3\,\omega\,q_1\ . && (4.133)
\end{aligned}$$

Combining (4.132) and (4.133) then yields the condition

$$0 \;=\; [\,2\,\omega^2 + \tfrac{1}{2}\,(n_{11})^2\,]\,q_1 + [\,(\mu+p) - \tfrac{2}{3}\,\pi_+\,]\,\omega\,n_{11}\ . \tag{4.134}$$

This shows that if hypersurface orthogonality properties of the two existing congruences is chosen as a classification criterion for consistent LRS spacetime geometries, either $\omega = 0$ or $n_{11} = 0$ impose strong restrictions on the matter field source, since in this case the energy current density is demanded to vanish, $q_1 = 0$. This is, for example, possible for magnetic Maxwell fields as discussed in the previous subsection. No such restriction arises for $0 = \omega = n_{11}$. Finally, applying (4.109) to $E_+$, we get

$$\begin{aligned} 0 \;=\;& \omega\,[\,(\mu+p)\,\sigma_+ - \tfrac{2}{3}\,(\Theta + 2\,\sigma_+)\,\pi_+ - (\mathbf{e}_1 + 2\,\dot{u} + a)\,(q_1)\,] \\ &+ n_{11}\,[\,\mathbf{e}_1(\mu) - (\Theta + \sigma_+)\,q_1\,]\ , \end{aligned} \tag{4.135}$$

while applied to $H_+$ it reproduces Eq. (4.134).

Under the assumption that the matter source be a comoving *perfect fluid* ($0 = q_1 = \pi_+$) with $(\mu + p) > 0$, the consistency condition (4.134) reduces to the simple relation

$$0 = \omega\,n_{11}\ , \tag{4.136}$$

which allows a neat geometric classification of the related LRS spacetime geometries into three distinct classes [9, 20]:

- **LRS class I:** $\omega \neq 0 \quad \Longrightarrow \quad 0 = n_{11} = \Theta = \sigma_+, \quad \mathbf{e}_0(f) = 0$ .
  $\mathbf{e}_0 = \mathbf{u}$ *not* HSO.
- **LRS class II:** $0 = \omega = n_{11}$ .
  *All* frame vectors HSO.
- **LRS class III:** $n_{11} \neq 0 \quad \Longrightarrow \quad 0 = \omega = \dot{u} = a, \quad \mathbf{e}_1(f) = 0$ .
  $\mathbf{e}_1$ *not* HSO.

Solutions within LRS class I are invariant under the transformations of a multiply-transitive $G_4$ with 3-D timelike orbits, those within the spatially homogeneous LRS class III under transformations of a multiply-transitive $G_4$ with 3-D spacelike orbits. The underlying spacetime symmetry group of LRS class II is a multiply-transitive $G_3$ with 2-D spacelike orbits of either constant positive, zero, or negative Gaußian curvature

$$^2K := 2\,(\mathbf{e}_2 - 2\,n_{31})\,(n_{31})\ . \tag{4.137}$$

Contained within the three LRS perfect fluid classes are a number of prominent cosmological models such as the Gödel model (class I: $\mathbf{e}_1(f) = 0$, $0 = \Theta = \sigma_+ = \dot{u} = a = n_{11}$, $0 = p = H_+$, $\Lambda < 0$), spatially homogeneous models (classes II and III: $\mathbf{e}_1(f) = 0$, $0 = \omega = \dot{u} = a$) with Kantowski–Sachs (class II: $n_{11} = 0$, $H_+ = 0$, $^2K > 0$) and Friedmann–Lemaître–Robertson–Walker (classes II and III: $\sigma_+ = 0$, $0 = E_+ = H_+$) models as special cases, and spherically symmetric Lemaître–Tolman–Bondi dust models (class II: $0 = \omega = a = n_{11}$, $0 = p = H_+$, $^2K > 0$) (see [9, 20, 58] and references therein).

Note that Eq. (4.126) implies that LRS class II models have zero “magnetic part” of the Weyl curvature tensor. Thus the corresponding dust models are examples of special “silent” models contained in the previous subsection. Apart from the Kantowski–Sachs models, further spatially homogeneous specializations contained in LRS class II are of LRS Bianchi Type–III ($^2K < 0$) and LRS Bianchi Type–I ($^2K = 0$). The conformal properties of spacelike 3-surfaces orthogonal to $\mathbf{u}$ in LRS models of classes II ($^*C_{\alpha\beta} = 0$) and III ($^*C_{\alpha\beta} \neq 0$) were investigated by Wainwright [59]. Various exact solutions with LRS spacetime symmetries have been discussed in Refs. [9], [20] and [48].

Dimensionless formulation:

In terms of the expansion-normalized dimensionless variables introduced in Section 3 the sest of equations (4.105) - (4.110), (4.112) - (4.120), (4.121) - (4.128), (4.129) - (4.130) and (4.131) - (4.135) become respectively:

The commutators:

$$
\begin{aligned}
[\boldsymbol{\partial}_0, \boldsymbol{\partial}_1] &= -(r - 3\dot{U})\,\boldsymbol{\partial}_0 + (q + 2\Sigma_+)\,\boldsymbol{\partial}_1 && (4.138)\\
[\boldsymbol{\partial}_0, \boldsymbol{\partial}_2] &= (q - \Sigma_+)\,\boldsymbol{\partial}_2 && (4.139)\\
[\boldsymbol{\partial}_0, \boldsymbol{\partial}_3] &= (q - \Sigma_+)\,\boldsymbol{\partial}_3 && (4.140)\\
[\boldsymbol{\partial}_1, \boldsymbol{\partial}_2] &= (r + 3A)\,\boldsymbol{\partial}_2 && (4.141)\\
[\boldsymbol{\partial}_2, \boldsymbol{\partial}_3] &= -6W\,\boldsymbol{\partial}_0 + 3N_{11}\,\boldsymbol{\partial}_1 + 6N_{31}\,\boldsymbol{\partial}_3 && (4.142)\\
[\boldsymbol{\partial}_3, \boldsymbol{\partial}_1] &= -(r + 3A)\,\boldsymbol{\partial}_3\ . && (4.143)
\end{aligned}
$$

The evolution equations:

$$
\begin{aligned}
\boldsymbol{\partial}_0\Sigma_+ &= (q - 1 + \Sigma_+)\,\Sigma_+ - (\boldsymbol{\partial}_1 - r + 3\dot{U} + 3A)\,\dot{U} + 3W^2 - (\mathcal{E}_+ - \tfrac{1}{2}\Pi_+) && (4.144)\\
\boldsymbol{\partial}_0 W &= (q - 1 - 2\Sigma_+)\,W + \tfrac{3}{2}N_{11}\,\dot{U} && (4.145)\\
\boldsymbol{\partial}_0 A &= q\,A - \dot{U} - (\dot{U} + A)\,\Sigma_+ - \tfrac{3}{2}N_{11}\,W - \tfrac{1}{2}Q_1 && (4.146)\\
\boldsymbol{\partial}_0 N_{11} &= (q - 4\Sigma_+)\,N_{11} && (4.147)\\
\boldsymbol{\partial}_0 N_{31} &= (q - \Sigma_+)\,N_{31} && (4.148)\\
\boldsymbol{\partial}_0(\mathcal{E}_+ + \tfrac{1}{2}\Pi_+) &= (2q - 1 - 3\Sigma_+)\,\mathcal{E}_+ + \left(q + \tfrac{1}{2} + \tfrac{1}{2}\Sigma_+\right)\Pi_+ - \tfrac{3}{2}(\Omega + P)\,\Sigma_+ \\
&\quad + \tfrac{1}{2}(\boldsymbol{\partial}_1 - 2r + 6\dot{U} + 3A)\,Q_1 - \tfrac{9}{2}N_{11}\,\mathcal{H}_+ && (4.149)\\
\boldsymbol{\partial}_0 H_+ &= (2q - 1 - 3\Sigma_+)\,\mathcal{H}_+ + \tfrac{9}{2}W\,Q_1 + \tfrac{9}{2}N_{11}\,(\mathcal{E}_+ - \tfrac{1}{2}\Pi_+) && (4.150)\\
\boldsymbol{\partial}_0\Omega &= (2q - 1)\,\Omega - 3P - (\boldsymbol{\partial}_1 - 2r + 6\dot{U} - 6A)\,Q_1 - 2\Sigma_+\Pi_+ && (4.151)\\
\boldsymbol{\partial}_0 Q_1 &= 2(q - 1)\,Q_1 - (\boldsymbol{\partial}_1 - 2r)\,P - 3(\Omega + P)\,\dot{U} + \tfrac{2}{3}(\boldsymbol{\partial}_1 - 2r + 3\dot{U} - 9A)\,\Pi_+ \\
&\quad + 2\Sigma_+\,Q_1 && (4.152)\\
\boldsymbol{\partial}_0\Omega_\Lambda &= 2(q + 1)\,\Omega_\Lambda\ . && (4.153)
\end{aligned}
$$

As before (cf. Eq. (3.7)) we have an evolution equation for $\Theta$, which in itself is decoupled from the set (4.144) - (4.153):

$$\boldsymbol{\partial}_0\Theta = -(1 + q)\,\Theta\ , \qquad (4.154)$$

and

$$q = -(\boldsymbol{\partial}_1 - r + 3\dot{U} - 6A)\,\dot{U} + 2(\Sigma_+)^2 - 6W^2 + \tfrac{1}{2}(\Omega + 3P) - \Omega_\Lambda\ . \qquad (4.155)$$

The constraint equations:

$$
\begin{aligned}
0 &= (\boldsymbol{\partial}_1 + r)\,\Theta && (4.156)\\
0 &= (\boldsymbol{\partial}_1 - r - 3\dot{U} - 6A)\,W && (4.157)\\
0 &= Q_1 + \tfrac{2}{3}r - \tfrac{2}{3}(\boldsymbol{\partial}_1 - r - 9A)\,\Sigma_+ + 3N_{11}\,W && (4.158)\\
0 &= (\boldsymbol{\partial}_1 - r - 3A)\,A + \tfrac{1}{3} - \tfrac{1}{3}(1 + 2\Sigma_+)\,\Sigma_+ + \tfrac{3}{4}(N_{11})^2 + \tfrac{1}{3}(\mathcal{E}_+ + \tfrac{1}{2}\Pi_+) - \tfrac{1}{3}\Omega - \tfrac{1}{3}\Omega_\Lambda && (4.159)\\
0 &= (\boldsymbol{\partial}_1 - r - 6A)\,N_{11} - 2(1 - 2\Sigma_+)\,W && (4.160)\\
0 &= (\boldsymbol{\partial}_1 - r - 3A)\,N_{31} && (4.161)\\
0 &= \mathcal{H}_+ + 9(\dot{U} + A)\,W + \tfrac{9}{2}N_{11}\,\Sigma_+ && (4.162)\\
0 &= (\boldsymbol{\partial}_1 - 2r - 9A)\,(\mathcal{E}_+ + \tfrac{1}{2}\Pi_+) + \tfrac{1}{2}(\boldsymbol{\partial}_1 - 2r)\,\Omega - \tfrac{3}{2}(1 + \Sigma_+)\,Q_1 + 9W\mathcal{H}_+ && (4.163)\\
0 &= (\boldsymbol{\partial}_1 - 2r - 9A)\,\mathcal{H}_+ + \tfrac{9}{2}(\Omega + P)\,W - 9W\,(\mathcal{E}_+ - \tfrac{1}{6}\Pi_+) + \tfrac{9}{4}N_{11}\,Q_1 && (4.164)\\
0 &= (\boldsymbol{\partial}_1 - 2r)\,\Omega_\Lambda\ . && (4.165)
\end{aligned}
$$

The evolution equation (4.153) arises from Eq. (3.71), while the extra constraints (4.156) and (4.165) derive respectively from Eqs. (3.8) and (3.72).

Tracefree part and trace of $\mathcal{S}_{\alpha\beta}$ (3-Ricci curvature of spacelike 3-surfaces orthogonal to $\mathbf{u}$ when $W = 0$):

$$\begin{aligned} \mathcal{S}_{+} &= -(\boldsymbol{\partial}_1 - r)\,A - 3\,(N_{11})^2 + 2\,(\boldsymbol{\partial}_2 - 6\,N_{31})\,N_{31} && (4.166)\\ &= (\mathcal{E}_{+} + \tfrac{1}{2}\,\Pi_{+}) - (1+\Sigma_{+})\,\Sigma_{+} - 9\,W^2 && (4.167)\\ \mathcal{K} &= -(2\,\boldsymbol{\partial}_1 - 2\,r - 9\,A)\,A + \tfrac{3}{4}\,(N_{11})^2 - 2\,(\boldsymbol{\partial}_2 - 6\,N_{31})\,N_{31} && (4.168)\\ &= 1 - \Omega - (\Sigma_{+})^2 + 9\,W^2 - \Omega_{\Lambda}\ . && (4.169) \end{aligned}$$

Consistency conditions:

$$\begin{aligned} 0 &= \tfrac{2}{3}\,(1+\Sigma_{+})\,W + A\,N_{11} && (4.170)\\ 0 &= W\,[\,(\mathcal{E}_{+} - \tfrac{1}{2}\,\Pi_{+}) + (1-\Sigma_{+}) - 2\,(\Sigma_{+})^2 - 9\,W^2 + 9\,A\,\dot{U} + \tfrac{9}{4}\,(N_{11})^2 \\ &\qquad + \tfrac{1}{2}\,(\Omega + 3\,P) - \Omega_{\Lambda}\,] + \tfrac{3}{4}\,N_{11}\,Q_1 && (4.171)\\ 0 &= N_{11}\,[\,(\mathcal{E}_{+} + \tfrac{1}{2}\,\Pi_{+}) + (1-\Sigma_{+}) - 2\,(\Sigma_{+})^2 - 9\,W^2 + 9\,A\,\dot{U} + \tfrac{9}{4}\,(N_{11})^2 \\ &\qquad - \Omega - \Omega_{\Lambda}\,] - 3\,W\,Q_1 && (4.172)\\ 0 &= [\,2\,W^2 + \tfrac{1}{2}\,(N_{11})^2\,]\,Q_1 + [\,(\Omega + P) - \tfrac{2}{3}\,\Pi_{+}\,]\,N_{11}\,W && (4.173)\\ 0 &= W\,[\,3\,(\Omega + P)\,\Sigma_{+} - 2\,(1 + 2\,\Sigma_{+})\,\Pi_{+} - (\boldsymbol{\partial}_1 - 2\,r + 6\,\dot{U} + 3\,A)\,Q_1\,] \\ &\quad + 3\,N_{11}\,[\,(\boldsymbol{\partial}_1 - 2\,r)\,\Omega - (1+\Sigma_{+})\,Q_1\,]\ . && (4.174) \end{aligned}$$

### 4.2.1 LRS class II perfect fluid models

Although they have been discussed many times in the past (and one will probably continue to do so in the future), we think that in the context of the overall framework presented here it is instructive to take a closer look at models within the generically spatially inhomogeneous LRS class II that contain a perfect fluid matter source with an equation of state $p = p(\mu)$. Using the covariantly defined Gaußian 2-curvature of the symmetry group orbits, ${}^2K$, in favor of $E_{+}$ as a dynamical variable, the sets of equations (4.105) - (4.110), (4.111) - (4.120), and (4.121) - (4.128) specialize to:

$$\begin{aligned} [\,\mathbf{e}_0, \mathbf{e}_1\,] &= \dot{u}\,\mathbf{e}_0 - \tfrac{1}{3}\,(\Theta - 2\,\sigma_{+})\,\mathbf{e}_1 && (4.175)\\ [\,\mathbf{e}_0, \mathbf{e}_2\,] &= -\tfrac{1}{3}\,(\Theta + \sigma_{+})\,\mathbf{e}_2 && (4.176)\\ [\,\mathbf{e}_0, \mathbf{e}_3\,] &= -\tfrac{1}{3}\,(\Theta + \sigma_{+})\,\mathbf{e}_3 && (4.177)\\ [\,\mathbf{e}_1, \mathbf{e}_2\,] &= a\,\mathbf{e}_2 && (4.178)\\ [\,\mathbf{e}_2, \mathbf{e}_3\,] &= 2\,n_{31}\,\mathbf{e}_3 && (4.179)\\ [\,\mathbf{e}_3, \mathbf{e}_1\,] &= -\,a\,\mathbf{e}_3\ , && (4.180) \end{aligned}$$

$$\begin{aligned} \mathbf{e}_0(\Theta) &= -\tfrac{1}{3}\,\Theta^2 + (\mathbf{e}_1 + \dot{u} - 2\,a)\,(\dot{u}) - \tfrac{2}{3}\,(\sigma_{+})^2 - \tfrac{1}{2}\,(\mu + 3p) + \Lambda && (4.181)\\ \mathbf{e}_0(\sigma_{+}) &= -\tfrac{1}{6}\,\Theta^2 - (\Theta - \tfrac{1}{6}\,\sigma_{+})\,\sigma_{+} - (\mathbf{e}_1 + \dot{u} + a)\,(\dot{u}) + \tfrac{3}{2}\,a^2 - \tfrac{3}{2}\,{}^2K + \tfrac{1}{2}\,\mu + \tfrac{1}{2}\,\Lambda && (4.182)\\ \mathbf{e}_0(\dot{u}) &= \mathbf{e}_1(\,\tfrac{dp}{d\mu}\,\Theta\,) + (\,\tfrac{dp}{d\mu} - \tfrac{1}{3}\,)\,\Theta\,\dot{u} + \tfrac{2}{3}\,\sigma_{+}\,\dot{u} && (4.183)\\ \mathbf{e}_0(a) &= -\tfrac{1}{3}\,(\Theta + \sigma_{+})\,(a + \dot{u}) && (4.184)\\ \mathbf{e}_0({}^2K) &= -\tfrac{2}{3}\,(\Theta + \sigma_{+})\,{}^2K && (4.185)\\ \mathbf{e}_0(\mu) &= -\,(\mu + p)\,\Theta\ , && (4.186) \end{aligned}$$

(Note that we have added a $\dot{u}$ evolution equation since we are considering a nontilted perfect fluid.) and

$$\begin{aligned} 0 &= \mathbf{e}_1(\Theta) + (\mathbf{e}_1 - 3\,a)\,(\sigma_{+}) && (4.187)\\ 0 &= (\mathbf{e}_1 - \tfrac{3}{2}\,a)\,(a) + \tfrac{1}{6}\,\Theta^2 - \tfrac{1}{6}\,(\sigma_{+})^2 + \tfrac{1}{2}\,{}^2K - \tfrac{2}{3}\,\mu - \tfrac{2}{3}\,\Lambda && (4.188)\\ 0 &= (\mathbf{e}_1 - 2\,a)\,({}^2K) && (4.189)\\ 0 &= \mathbf{e}_1(p) + (\mu + p)\,\dot{u}\ . && (4.190) \end{aligned}$$

$E_+$ and ${}^2\!K$ are related by

$$E_+ = \tfrac{3}{2}\,{}^2\!K + \tfrac{1}{6}\,(\Theta+\sigma_+)^2 - \tfrac{3}{2}\,a^2 - \tfrac{1}{2}\,\mu - \tfrac{1}{2}\,\Lambda\ . \tag{4.191}$$

Now taking the slicing point of view we can choose local coordinates $t, x, y, z$ and express the orthonormal frame vectors according to Eq. (2.76) as

$$\begin{aligned} \mathbf{e}_0 &= N^{-1}\,(\partial_t - N_x\,\partial_x)\ , \qquad & \mathbf{e}_2 &= Y^{-1}\,\partial_y\ , \\ \mathbf{e}_1 &= X^{-1}\,\partial_x\ , & \mathbf{e}_3 &= (YZ)^{-1}\,\partial_z\ , \end{aligned} \tag{4.192}$$

where $N = N(t,x)$, $N_x = N_x(t,x)$, $X = X(t,x)$, $Y = Y(t,x)$, and $Z = Z(y)$. Inserting these expressions into the commutation relations (4.175) - (4.180) we obtain

$$\dot{u} = N^{-1}\,\mathbf{e}_1(N) \tag{4.193}$$

$$\mathbf{e}_0(X) = \tfrac{1}{3}\,(\Theta - 2\,\sigma_+)\,X + X^2\,N^{-1}\,\mathbf{e}_1(N_x) \tag{4.194}$$

$$\mathbf{e}_0(Y) = \tfrac{1}{3}\,(\Theta+\sigma_+)\,Y \tag{4.195}$$

$$0 = (\mathbf{e}_1 + a)\,(Y) \tag{4.196}$$

$$0 = (\mathbf{e}_2 + 2\,n_{31})\,(Z)\ . \tag{4.197}$$

Then, demanding that ${}^2\!K$ be constant for $t = \text{const} = x$, we get from Eq. (4.137) the condition

$$0 = Z_{,yy} + C_1\,Z\ , \qquad C_1 = \text{const}\ , \tag{4.198}$$

from which it follows that ${}^2\!K = C_1\,/\,Y^2$. If $C_1 > 0$ the group orbits are spherically symmetric 2-surfaces, if $C_1 = 0$ they are flat 2-planes, and if $C_1 < 0$ the 2-surfaces have hyperbolic geometry.

How much information about the metric does the system of equations (4.181) - (4.190) contain? Through the commutators we find that the metric variable $Y$ occurs implicitly through ${}^2\!K$ in the spherically and hyperbolically symmetric cases. However, there is *no* information about $X$, which is implicitly appearing in these equations through $\mathbf{e}_1$. Thus to obtain a complete system one has to include the commutator equation corresponding to Eq. (4.194). It is also worth noting that the time and space frame derivatives of $Y$ are encoded through the kinematic quantities (by the combination $(\Theta+\sigma_+)$) and the variable $a$. This is in stark contrast to the metric variable $X$, where the only information we can obtain is that its $\mathbf{e}_0$-derivative is determined by the combination $(\Theta - 2\,\sigma_+)$, if we have specified a shift vector (of course the shift also occurs implicitly in $\mathbf{e}_0$ in the above equations). The fact that we do not have a variable corresponding to an $\mathbf{e}_1$-derivative of $X$ is completely analogous to the nonexistence of an $\mathbf{e}_0$-derivative of the lapse; the nonexistence of an $\mathbf{e}_1$-derivative of $X$ is associated with the freedom of making a coordinate reparametrization $x = x(\overline{x})$.

In terms of dimensionless variables the LRS class II perfect fluid models are described by the sets of equations:

$$[\boldsymbol{\partial}_0, \boldsymbol{\partial}_1] = -\,(r - 3\,\dot{U})\,\boldsymbol{\partial}_0 + (q + 2\,\Sigma_+)\,\boldsymbol{\partial}_1 \tag{4.199}$$

$$[\boldsymbol{\partial}_0, \boldsymbol{\partial}_2] = (q - \Sigma_+)\,\boldsymbol{\partial}_2 \tag{4.200}$$

$$[\boldsymbol{\partial}_0, \boldsymbol{\partial}_3] = (q - \Sigma_+)\,\boldsymbol{\partial}_3 \tag{4.201}$$

$$[\boldsymbol{\partial}_1, \boldsymbol{\partial}_2] = (r + 3\,A)\,\boldsymbol{\partial}_2 \tag{4.202}$$

$$[\boldsymbol{\partial}_2, \boldsymbol{\partial}_3] = 6\,N_{31}\,\boldsymbol{\partial}_3 \tag{4.203}$$

$$[\boldsymbol{\partial}_3, \boldsymbol{\partial}_1] = -\,(r + 3\,A)\,\boldsymbol{\partial}_3\ , \tag{4.204}$$

$$\boldsymbol{\partial}_0\Sigma_+ = (q - 2 + \tfrac{1}{2}\,\Sigma_+)\,\Sigma_+ - \tfrac{1}{2} - (\boldsymbol{\partial}_1 - r + 3\,\dot{U} + 3\,A)\,\dot{U} + \tfrac{9}{2}\,A^2 - \tfrac{3}{2}\,{}^2\mathcal{K} + \tfrac{1}{2}\,\Omega + \tfrac{1}{2}\,\Omega_\Lambda \tag{4.205}$$

$$\boldsymbol{\partial}_0\dot{U} = (\,q + 3\,\tfrac{dp}{d\mu} + 2\,\Sigma_+\,)\,\dot{U} + (\boldsymbol{\partial}_1 - r)\,(\,\tfrac{dp}{d\mu}\,) \tag{4.206}$$

$$\boldsymbol{\partial}_0 A = q\,A - \dot{U} - (\dot{U} + A)\,\Sigma_+ \tag{4.207}$$

$$\boldsymbol{\partial}_0\,{}^2\mathcal{K} = 2\,(q - \Sigma_+)\,{}^2\mathcal{K} \tag{4.208}$$

$$\boldsymbol{\partial}_0\Omega = (2\,q - 1)\,\Omega - 3\,P \tag{4.209}$$

$$\boldsymbol{\partial}_0\Omega_\Lambda = 2\,(q + 1)\,\Omega_\Lambda\ , \tag{4.210}$$

where

$$q = -\,(\boldsymbol{\partial}_1 - r + 3\,\dot{U} - 6\,A)\,\dot{U} + 2\,(\Sigma_+)^2 + \tfrac{1}{2}\,(\Omega + 3P) - \Omega_\Lambda \ , \tag{4.211}$$

and

$$\begin{aligned}
0 &= r - (\boldsymbol{\partial}_1 - r - 9\,A)\,\Sigma_+ && (4.212)\\
0 &= (\boldsymbol{\partial}_1 - r - \tfrac{9}{2}\,A)\,A + \tfrac{1}{2} - \tfrac{1}{2}\,(\Sigma_+)^2 + \tfrac{1}{2}\,{}^2\mathcal{K} - \tfrac{2}{3}\,\Omega - \tfrac{2}{3}\,\Omega_\Lambda && (4.213)\\
0 &= (\boldsymbol{\partial}_1 - 2\,r - 6\,A)\,{}^2\mathcal{K} && (4.214)\\
0 &= (\boldsymbol{\partial}_1 - 2\,r)\,P + 3\,(\Omega + P)\,\dot{U} && (4.215)\\
0 &= (\boldsymbol{\partial}_1 - 2\,r)\,\Omega_\Lambda \ . && (4.216)
\end{aligned}$$

The variables $\mathcal{E}_+$ and ${}^2\mathcal{K} := 3\,{}^2K\,/\,\Theta^2$ are related by

$$\mathcal{E}_+ = \tfrac{3}{2}\,{}^2\mathcal{K} + \tfrac{1}{2}\,(1 + \Sigma_+)^2 - \tfrac{9}{2}\,A^2 - \tfrac{1}{2}\,\Omega - \tfrac{1}{2}\,\Omega_\Lambda \ . \tag{4.217}$$

Let us now consider the special case where $p(\mu) = (\gamma - 1)\,\mu$, $P(\Omega) = (\gamma - 1)\,\Omega$, so that $\Theta$ is not "reintroduced". Equation (4.211) defines $q$ while one can use Eq. (4.213) to solve for $\Omega$. One can choose Eq. (4.212) to solve for $r$. This is indeed the equation Hewitt and Wainwright used in the case of two commuting spacelike Killing vector fields (which overlaps with the present case when ${}^2K = 0$) [44]. However, in the present case one can also choose (4.214) in order to solve for $r$. Once one has solved for $r$ the remaining equation will constitute a constraint (in contrast to the case considered by Hewitt and Wainwright which did not contain such a constraint [44]). If one includes a cosmological constant, Eq. (4.216) implies that the possibilities of solving for $r$ increase even further.

## 4.3 Spatially homogeneous models

When dealing with spatially homogeneous models it is convenient to choose $\mathbf{e}_0$ as the unit normal with respect to the spacelike 3-surfaces of homogeneity. Since one usually focuses on spatially homogeneous expanding models, in this section we will go directly to the dimensionless formulation. The condition of adapting to the spatially homogeneous 3-surfaces leads to the restrictions: $0 = \dot{U}^\alpha = W^\alpha = r_\alpha = \boldsymbol{\partial}_\alpha F$; $F$ denoting the dimensionless form of any geometrically defined variable. (Note that, since $0 = \dot{U}^\alpha = W^\alpha$, the earlier discussion related to inequalities for the values of the deceleration parameter $q$ and the dimensionless product $H\,\Delta t_p$ applies.) Spatially homogeneous models have been reviewed many times before, see, for example, Refs. [21], [22], and [2]. Since we have adapted to the spatially homogeneous 3-surfaces it is natural to take the slicing point of view. This leads to the relations

$$\boldsymbol{\partial}_0 = \mathcal{N}^{-1}\,(\partial_t - \overline{N}^i\,\overline{\mathbf{e}}_i) \ , \quad \boldsymbol{\partial}_\alpha = \overline{E}_\alpha{}^i\,\overline{\mathbf{e}}_i \ , \tag{4.218}$$

where $\overline{\mathbf{e}}_i$ are the vector fields associated with the spatial homogeneity symmetry group. Thus $\overline{\gamma}^i{}_{jk}$ are just the structure constants describing the various Bianchi groups (the Kantowski–Sachs models with multiply-transitive spacetime isometry group have already been treated in the previous LRS section). Spatial homogeneity requires $\mathcal{N} = \mathcal{N}(t)$, while the shift vector can be set to zero (however, a full use of the so-called automorphism group requires a nonzero spatially dependent shift vector satisfying certain requirements; for further discussions see Refs. [60] and [61]). In this case one can choose $\mathcal{N} = 3$, leading to a time variable $\tau = \ln(\ell/\ell_0)$ (recall Eq. (1.8)). Thus the operator $\boldsymbol{\partial}_0$ acting on a function $F(t)$ reduces to $dF/d\tau$. These restrictions lead to:

**The commutators**

$$\begin{aligned}
[\,\boldsymbol{\partial}_0, \boldsymbol{\partial}_\alpha\,] &= 3\,\left[\,\tfrac{1}{3}\,q\,\delta^\beta{}_\alpha - \Sigma^\beta{}_\alpha + \epsilon^\beta{}_{\alpha\gamma}\,R^\gamma\,\right]\,\boldsymbol{\partial}_\beta && (4.219)\\
[\,\boldsymbol{\partial}_\alpha, \boldsymbol{\partial}_\beta\,] &= 3\,\left[\,2\,A_{[\alpha}\,\delta^\gamma{}_{\beta]} + \epsilon_{\alpha\beta\delta}\,N^{\delta\gamma}\,\right]\,\boldsymbol{\partial}_\gamma && (4.220)
\end{aligned}$$

**The field equations**

*Decoupled equation*

$$\begin{aligned} \boldsymbol{\partial}_0\Theta &= -(1+q)\,\Theta\ , && (4.221)\\ q &= 2\,\Sigma^2 + \tfrac{1}{2}\,(\Omega+3P) - \Omega_\Lambda\ . && (4.222) \end{aligned}$$

*Remaining field equations*

$$\begin{aligned} \boldsymbol{\partial}_0\Sigma^{\alpha\beta} &= (q-2)\,\Sigma^{\alpha\beta} + \Pi^{\alpha\beta} - \mathcal{S}^{\alpha\beta} + 6\,\epsilon^{\gamma\delta(\alpha}\,R_\gamma\,\Sigma^{\beta)}{}_\delta && (4.223)\\ \Omega &= 1 - \Sigma^2 - \mathcal{K} - \Omega_\Lambda && (4.224)\\ 0 &= Q^\alpha - 9\,A_\beta\,\Sigma^{\alpha\beta} - 3\,\epsilon^{\alpha\beta\gamma}\,N_{\beta\delta}\,\Sigma^\delta{}_\gamma\ , && (4.225) \end{aligned}$$

where

$$\begin{aligned} \mathcal{S}_{\alpha\beta} &= 3\,B_{\alpha\beta} - \delta_{\alpha\beta}\,B^\gamma{}_\gamma + 6\,\epsilon^{\gamma\delta}{}_{(\alpha}\,A_{|\gamma|}\,N_{\beta)\delta} && (4.226)\\ \mathcal{K} &= 9\,A_\alpha\,A^\alpha + \tfrac{3}{4}\,B^\alpha{}_\alpha && (4.227)\\ B_{\alpha\beta} &= 2\,N_{\alpha\gamma}\,N^\gamma{}_\beta - N^\gamma{}_\gamma\,N_{\alpha\beta} && (4.228) \end{aligned}$$

**The Jacobi identities**

$$\begin{aligned} \boldsymbol{\partial}_0 A^\alpha &= q\,A^\alpha - 3\,A_\beta\,\Sigma^{\alpha\beta} - 3\,\epsilon^{\alpha\beta\gamma}\,A_\beta\,R_\gamma && (4.229)\\ \boldsymbol{\partial}_0 N^{\alpha\beta} &= q\,N^{\alpha\beta} + 6\,\Sigma^{(\alpha}{}_\gamma\,N^{\beta)\gamma} - 6\,\epsilon^{\gamma\delta(\alpha}\,N^{\beta)}{}_\gamma\,R_\delta && (4.230)\\ 0 &= A_\beta\,N^{\alpha\beta} && (4.231) \end{aligned}$$

**The "electric" and "magnetic parts" of the Weyl tensor**

$$\begin{aligned} \mathcal{E}_{\alpha\beta} + \tfrac{1}{2}\,\Pi_{\alpha\beta} &= \Sigma_{\alpha\beta} - 3\,\Sigma_{\alpha\gamma}\,\Sigma^\gamma{}_\beta + \tfrac{2}{3}\,\delta_{\alpha\beta}\,\Sigma^2 + \mathcal{S}_{\alpha\beta} && (4.232)\\ \mathcal{H}_{\alpha\beta} &= \tfrac{3}{2}\,N^\gamma{}_\gamma\,\Sigma_{\alpha\beta} - 9\,N^\gamma{}_{(\alpha}\,\Sigma_{\beta)\gamma} + 3\,\delta_{\alpha\beta}\,N_{\gamma\delta}\,\Sigma^{\gamma\delta} - 3\,\epsilon^{\gamma\delta}{}_{(\alpha}\,A_{|\gamma|}\,\Sigma_{\beta)\delta} && (4.233) \end{aligned}$$

**The Bianchi identities for the Weyl tensor**

$$\begin{aligned} \boldsymbol{\partial}_0(\mathcal{E}^{\alpha\beta} + \tfrac{1}{2}\,\Pi^{\alpha\beta}) &= (2\,q-1)\,\mathcal{E}^{\alpha\beta} + \big(q+\tfrac{1}{2}\big)\,\Pi^{\alpha\beta} - \tfrac{3}{2}\,(\Omega+P)\,\Sigma^{\alpha\beta} - \tfrac{3}{2}\,A^{(\alpha}\,Q^{\beta)}\\ &\quad + 9\,\Sigma^{(\alpha}{}_\gamma\,(\mathcal{E}^{\beta)\gamma} - \tfrac{1}{6}\,\Pi^{\beta)\gamma}) + \tfrac{3}{2}\,N^\gamma{}_\gamma\,\mathcal{H}^{\alpha\beta} - 9\,N^{(\alpha}{}_\gamma\,\mathcal{H}^{\beta)\gamma}\\ &\quad + \delta^{\alpha\beta}\,\big[\,\tfrac{1}{2}\,A_\gamma\,Q^\gamma - 3\,\Sigma_{\gamma\delta}\,(\mathcal{E}^{\gamma\delta} - \tfrac{1}{6}\,\Pi^{\gamma\delta}) + 3\,N_{\gamma\delta}\,\mathcal{H}^{\gamma\delta}\,\big]\\ &\quad - 3\,\epsilon^{\gamma\delta(\alpha}\,\Big[\,A_\gamma\,\mathcal{H}^{\beta)}{}_\delta - 2\,R_\gamma\,(\mathcal{E}^{\beta)}{}_\delta + \tfrac{1}{2}\,\Pi^{\beta)}{}_\delta) - \tfrac{1}{2}\,N^{\beta)}{}_\gamma\,Q_\delta\,\Big] && (4.234)\\ \boldsymbol{\partial}_0 H^{\alpha\beta} &= (2\,q-1)\,\mathcal{H}^{\alpha\beta} + 9\,\Sigma^{(\alpha}{}_\gamma\,\mathcal{H}^{\beta)\gamma} - \tfrac{3}{2}\,N^\gamma{}_\gamma\,(\mathcal{E}^{\alpha\beta} - \tfrac{1}{2}\,\Pi^{\alpha\beta}) + 9\,N^{(\alpha}{}_\gamma\,(\mathcal{E}^{\beta)\gamma} - \tfrac{1}{2}\,\Pi^{\beta)\gamma})\\ &\quad - 3\,\delta^{\alpha\beta}\,\big[\,\Sigma_{\gamma\delta}\,\mathcal{H}^{\gamma\delta} + N_{\gamma\delta}\,(\mathcal{E}^{\gamma\delta} - \tfrac{1}{2}\,\Pi^{\gamma\delta})\,\big]\\ &\quad + 3\,\epsilon^{\gamma\delta(\alpha}\,\Big[\,A_\gamma)\,(\mathcal{E}^{\beta)}{}_\delta - \tfrac{1}{2}\,\Pi^{\beta)}{}_\delta) + \tfrac{1}{2}\,\Sigma^{\beta)}{}_\gamma\,Q_\delta + 2\,R_\gamma)\,\mathcal{H}^{\beta)}{}_\delta\,\Big] && (4.235) \end{aligned}$$

Note that the constraints that arise in the Bianchi identities do not give any new information. They are just combinations of the constraints obtained in the field equations and the Jacobi identities.

**The Bianchi identities for the source terms**

$$\boldsymbol{\partial}_0 \Omega = (2\,q-1)\,\Omega - 3\,P + 6\,A_\alpha\,Q^\alpha - 3\,\Sigma_{\alpha\beta}\,\Pi^{\alpha\beta} \tag{4.236}$$

$$\boldsymbol{\partial}_0 Q^\alpha = 2\,(q-1)\,Q^\alpha + 9\,A_\beta\,\Pi^{\alpha\beta} - 3\,\Sigma^\alpha{}_\beta\,Q^\beta + 3\,\epsilon^{\alpha\beta\gamma}\,\left[\,R_\beta\,Q_\gamma + N_{\beta\delta}\,\Pi^\delta{}_\gamma\,\right]\,. \tag{4.237}$$

The dimensionless form of the 3-Cotton–York tensor for spatially homogeneous models is

$$\begin{aligned}\mathcal{C}_{\alpha\beta} = & -(\boldsymbol{\partial}_0 - 2\,q + 2)\,\mathcal{H}_{\alpha\beta} + 9\,\Sigma^\gamma{}_{(\alpha}\,\mathcal{H}_{\beta)\gamma} - 27\,N^\gamma{}_{(\alpha}\,\Sigma^\delta{}_{\beta)}\,\Sigma_{\gamma\delta} + \tfrac{9}{2}\,N^\gamma{}_\gamma\,\Sigma_{\alpha\delta}\,\Sigma^\delta{}_\beta + 6\,N_{\alpha\beta}\,\Sigma^2 \\ & - 9\,N^\gamma{}_{(\alpha}\,\Pi_{\beta)\gamma} + \tfrac{3}{2}\,N^\gamma{}_\gamma\,\Pi_{\alpha\beta} \\ & - 3\,\delta_{\alpha\beta}\,\left[\,\Sigma_{\gamma\delta}\,\mathcal{H}^{\gamma\delta} - 3\,N^\gamma{}_\delta\,\Sigma^\delta{}_\epsilon\,\Sigma^\epsilon{}_\gamma + N^\gamma{}_\gamma\,\Sigma^2 - 3\,N_{\gamma\delta}\,\Pi^{\gamma\delta}\,\right] \\ & - 3\,\epsilon^{\gamma\delta}{}_{(\alpha}\,\left[\,A_{|\gamma|}\,(3\,\Sigma^\epsilon{}_{\beta)}\,\Sigma_{\delta\epsilon} + \Pi_{\beta)\delta}) - \tfrac{1}{2}\,\Sigma_{\beta)\gamma}\,Q_\delta - 2\,R_{|\gamma|}\,\mathcal{H}_{\beta)\delta}\,\right]\,.\end{aligned} \tag{4.238}$$

The conformal properties of spacelike 3-surfaces in (orthogonal) spatially homogeneous cosmological models were investigated by Wainwright [59].

Note that the class B parameter $h$ gives rise to a constraint through its definition:

$$A_\alpha\,A_\beta = \tfrac{1}{2}\,h\,\epsilon_{\alpha\gamma\sigma}\epsilon_{\beta\delta\tau}\,N^{\gamma\delta}\,N^{\sigma\tau} \quad \Leftrightarrow \quad A^2 = \tfrac{1}{2}\,h\,\overline{N}\,, \quad \overline{N} := (N^\alpha{}_\alpha)^2 - N_{\alpha\beta}\,N^{\alpha\beta}\,. \tag{4.239}$$

Expressed in expansion-normalized variables the Bianchi–Schücking–Behr classification (see e.g. Ref. [22]) takes the form given in Tab. 2 (the variables, and therefore also the present form of the classification, breaks down when $\Theta = 0$. In that case one has to use other variables, such as for example the original dimensional ones). It is interesting to note that the Type–I models (which are characterized by $0 = N_{\alpha\beta} = A^\alpha$) are examples of "silent" models (see Eq. (4.233)).

Table 2: Classification of the Bianchi groups into classes A and B and group types I–IX. Bianchi Type–III belongs to Type–VI$_h$ with $h = -1$.

| Group Class | | Group Type | |
|---|---|---|---|
| Class A $\Leftrightarrow A_\alpha = 0$ | $\det(N^{\alpha\beta}) \neq 0$ | IX: $\mathrm{sgn}(\Theta)\det(N^{\alpha\beta}) > 0$ | VIII: $\mathrm{sgn}(\Theta)\det(N^{\alpha\beta}) < 0$ |
| | $\det(N^{\alpha\beta}) = 0$ | VII$_0$: $\overline{N} > 0$ | VI$_0$: $\overline{N} < 0$ |
| | | II: $\overline{N} = 0\,, N^\alpha{}_\alpha \neq 0$ | I: $\overline{N} = 0\,, N^\alpha{}_\alpha = 0$ |
| Class B $\Leftrightarrow A_\alpha \neq 0$ | $\det(N^{\alpha\beta}) = 0$ | VII$_h$: $\overline{N} > 0$ | VI$_h$: $\overline{N} < 0$ |
| | | IV: $\overline{N} = 0\,, N^\alpha{}_\alpha \neq 0$ | V: $\overline{N} = 0\,, N^\alpha{}_\alpha = 0$ |

Note that the above equations are ordinary differential equations. Furthermore, at first sight there seems to be no need for the commutators, since the only differential operator appearing in them ($\boldsymbol{\partial}_0$) reduces to an ordinary derivative ($d/d\tau$), which is completely known. This is a consequence of the spatial symmetry. In more general circumstances one does *not* have control of the information hidden in the differential operators. Thus in such cases one is forced to consider the commutator equations. However, even in the spatially homogeneous case one needs to consider the commutators if one wants to construct the metric. Nevertheless, in this case it turns out that one can do so *algebraically* without solving any differential equations, by choosing a shift vector adapted to the automorphism group [61]. The most common frame choice is one restricting $N_{\alpha\beta}$ and $A^\alpha$. Usually, with the exception of some special class B models, one chooses $N_{\alpha\beta}$ to be diagonal. Various exact solutions of spatially homogeneous (fluid) spacetime geometries have been discussed in, for example, Refs. [21], [22], [48] and [62].

### 4.3.1 Perfect fluids

Here we will consider a perfect fluid with equation of state $\tilde{p}(\tilde{\mu}) = (\gamma - 1)\,\tilde{\mu}$. Since we want to assume that the fluid is tilted with respect to the spatially homogeneous 3-surfaces, this leads to the dimensionless expressions given in Eq. (3.37). Then the equations which follow for $\Omega$ and $v^\alpha$ are

$$\begin{aligned} \boldsymbol{\partial}_0\Omega &= \Omega\, G^{-1}\,[\,2\,G\,q - (3\,\gamma - 2) - (2-\gamma)\,v^2 - 3\,\gamma\,\Sigma_{\alpha\beta}\,v^\alpha\,v^\beta + 6\,\gamma\,A_\alpha\,v^\alpha\,] \qquad (4.240)\\ \boldsymbol{\partial}_0 v^\alpha &= \frac{3\,v^\alpha}{\gamma\,[\,1-(\gamma-1)\,v^2\,]}\,[\,\tfrac{1}{3}\,(3\,\gamma-4)\,(1-v^2) + (2-\gamma)\,(\Sigma_{\beta\gamma}\,v^\beta\,v^\gamma) \\ &\quad + [\,(3-2\,\gamma) + (\gamma-1)\,v^2\,]\,(A_\beta\,v^\beta)\,] - 3\,\Sigma^\alpha{}_\beta\,v^\beta + 3\,\epsilon^{\alpha\beta\gamma}\,(R_\beta + N^\delta{}_\beta\,v_\delta)\,v_\gamma \\ &\quad - 3\,v^2\,A^\alpha\;. \qquad (4.241) \end{aligned}$$

It is convenient to use Eq. (4.224) to solve for $\Omega$. However, it is often of advantage not to use the constraint (4.225) to solve for $v^\alpha$ globally, since the constraint surface is quite complicated. Instead one can use Eq. (4.225) to solve for some variables (not necessarily $v^\alpha$) locally. For examples on how to deal with constraints of the type given in equation (4.225), see Refs. [63], [64] and [65]. It is usually convenient to make variable transformations to other dimensionless variables when dealing with specific problems. Examples of this can be found in Ref. [66], which addressed class B models, and Ref. [46], which dealt with LRS Type–IX models.

### 4.3.2 Maxwell vacuum fields

The dimensionless form of the contributions to the Maxwell stress-energy-momentum tensor were given in Eqs. (3.41) - (3.43), while the sourcefree Maxwell equations for spatially homogeneous fields are given by

$$\begin{aligned} \boldsymbol{\partial}_0\mathcal{E}^\alpha &= (q-1)\,\mathcal{E}^\alpha + 3\,\Sigma^\alpha{}_\beta\,\mathcal{E}^\beta - 3\,N^\alpha{}_\beta\,\mathcal{H}^\beta - 3\,\epsilon^{\alpha\beta\gamma}\,[\,A_\beta\,\mathcal{H}_\gamma - R_\beta\,\mathcal{E}_\gamma\,] \qquad (4.242)\\ \boldsymbol{\partial}_0\mathcal{H}^\alpha &= (q-1)\,\mathcal{H}^\alpha + 3\,\Sigma^\alpha{}_\beta\,\mathcal{H}^\beta + 3\,N^\alpha{}_\beta\,\mathcal{E}^\beta + 3\,\epsilon^{\alpha\beta\gamma}\,[\,A_\beta\,\mathcal{E}_\gamma + R_\beta\,\mathcal{H}_\gamma\,] \qquad (4.243)\\ 0 &= A_\alpha\,\mathcal{E}^\alpha \qquad (4.244)\\ 0 &= A_\alpha\,\mathcal{H}^\alpha\;. \qquad (4.245) \end{aligned}$$

# 5 Concluding Remarks

It has been shown how, from the general $1+3$ orthonormal frame equations presented in sections 2 and 3, one can extract equations suitable for further investigations for a large set of different problems. In a series of future papers the authors will, together with M. Bruni, present a dynamical systems analysis for rotating "silent" models and "silent" models including electromagnetic fields [57].

The dimensionless formulation presented in section 3 is adapted to the existence of homothetic Killing vector fields, since the result of acting with the associated symmetry transformation on any dimensionless quantity yields zero. It is believed that self-similar solutions act as key building blocks when it comes to the understanding of more general models not exhibiting such a symmetry. For example, this is true in spatially homogeneous cosmology and for models with two spacelike commuting Killing vector fields, where they occur as equilibrium states in the reduced dimensionless phase space (see e.g. Ref. [2]). Thus dimensionless formulations should provide a proper setting for probing the role of self-similar solutions in General Relativity.

The $1+3$ formulation which has been presented here is motivated by the existence of a preferred timelike vector field $\mathbf{u}$. Usually this field is supplied by the stress-energy-momentum tensor. When it comes to *vacuum* spacetime geometries it is perhaps more natural to use formalisms based on $2+2$ splittings rather than $1+3$ or $3+1$ splittings. In a $2+2$ splitting formalism one can choose to adapt to null congruences. This has the advantage that the formalism then is naturally adapted to the algebraic properties of the Weyl curvature tensor. Moreover, such approaches should also be useful for problems involving gravitational radiation. In this context one can mention the Newman–Penrose formalism and the Geroch–Held–Penrose formalism (see e.g. Refs. [67], [14]). Another $2+2$ formalism worth mentioning is the covariant $2+2$ formalism developed by d'Inverno and others [68], which now frequently finds application in the literature. Here the spacetime manifold is foliated by two families of spacelike 2-surfaces, which are Lie-dragged along two congruences of arbitrarily timelike, null or spacelike character. Its advantages lie in the fact that it can explicitly identify the degrees of freedom of a vacuum gravitational field as the entries of a 2-metric, and that initial data is unconstrained.

In general $2+2$ approaches seem to be preferred when it comes to vacuum spacetimes. However, it should be pointed out that there exist nontrivial vacuum and electromagnetic invariant submanifolds in the "silent" cases, particularly if one allows the congruence $\mathbf{u}$ to have rotation. Moreover, if $\sigma_-$ and $E_-$ are nonzero one has Petrov type I. Since most exact vacuum and electromagnetic solutions are algebraically special, the models arising in the $1+3$ context, which lead to a decoupled system of ordinary differential equations, may provide an alternative to the traditional $2+2$ approaches when it comes to finding out properties of models that are of Petrov type I.

In the context of "silent" models one may investigate the size of the solution space for a given proposed configuration. Some insight in this respect can be gained from a systematic study of the integrability conditions underlying the particular setting under scrutiny, that is, various consistency conditions might arise from propagating the constraint equations along the preferred timelike congruence $\mathbf{u}$. Another interesting possibility of probing the solution space is provided by an application of the equivalence problem approach [38, 39]. Further progress could arise from numerical calculations aimed at revealing the content of the related constraint equations, which, to our knowledge, to date have not been pursued. Finally, it should be pointed out that "silent" models contain a fairly large set of known exact solutions, although no new solution has been discovered within this class so far. However, it seems unlikely that there should not exist any further solution for these models, apart from the known exact ones. Even if this was so, the given geometric formulation of the class of "silent" models (and beyond) presents a remarkably attractive framework that nicely unifies otherwise scattered results.

# Acknowledgments

HvE likes to thank the Department of Physics at Stockholm University for a kind invitation and Reza Tavakol at QMW for permanent encouragement. We acknowledge Marco Bruni and Brian Edgar for useful comments and pointing out relevant references. HvE is supported by a grant from the Drapers' Society at QMW. CU is supported by the Swedish Natural Science Research Council. Throughout this work the computer algebra packages `REDUCE` and `CLASSI` have been valuable tools.

# A Appendix

When translating the equations of the $1+3$ splitting approach to General Relativity from covariant into orthonormal frame form the following relations valid for the derivative terms occurring are very useful:

## A.1 Vector derivatives

$V_\mu \, u^\mu = 0$.

$$h^\mu{}_\nu \, u^\rho \nabla_\rho V^\nu \quad \Longrightarrow \quad \mathbf{e}_0(V^\alpha) - \epsilon^{\alpha\beta\gamma} \, \Omega_\beta \, V_\gamma \ . \tag{A.1}$$

$$h^\mu{}_\nu \nabla_\mu V^\nu \quad \Longrightarrow \quad (\mathbf{e}_\alpha - 2\, a_\alpha) \, (V^\alpha) \ . \tag{A.2}$$

$$\eta^{\mu\nu\rho\sigma} (\nabla_\nu V_\rho) \, u_\sigma \quad \Longrightarrow \quad \epsilon^{\alpha\beta\gamma} \, (\mathbf{e}_\beta - a_\beta) \, (V_\gamma) - n^\alpha{}_\beta \, V^\beta \ . \tag{A.3}$$

$$h^{(\mu}{}_\rho \, h^{\nu)}{}_\sigma \nabla^\rho V^\sigma \quad \Longrightarrow \quad (\delta^{\gamma(\alpha} \, \mathbf{e}_\gamma + a^{(\alpha)}) \, (V^{\beta)}) - \delta^{\alpha\beta} \, a_\gamma \, V^\gamma - \epsilon^{\gamma\delta(\alpha} \, n^{\beta)}{}_\gamma \, V_\delta \ . \tag{A.4}$$

## A.2 Tensor derivatives

$A_{\mu\nu} = A_{(\mu\nu)}$, $A^{\mu}{}_{\mu} = 0$, $A_{\mu\nu}\, u^{\nu} = 0$.

$$h^{\mu}{}_{\rho}\, h^{\nu}{}_{\sigma}\, u^{\tau} \nabla_{\tau} A^{\rho\sigma} \quad \Longrightarrow \quad \mathbf{e}_0(A^{\alpha\beta}) - 2\,\epsilon^{\gamma\delta(\alpha}\,\Omega_{\gamma}\,A^{\beta)}{}_{\delta}\ . \tag{A.5}$$

$$h^{\mu}{}_{\rho}\, h^{\nu}{}_{\sigma} \nabla_{\nu} A^{\rho\sigma} \quad \Longrightarrow \quad (\mathbf{e}_{\beta} - 3\,a_{\beta})\,(A^{\alpha\beta}) - \epsilon^{\alpha\beta\gamma}\,n_{\beta\delta}\,A^{\delta}{}_{\gamma}\ . \tag{A.6}$$

$$h^{(\mu}{}_{\rho}\, h^{\nu)}{}_{\sigma}\, \eta^{\rho\tau\kappa\lambda}(\nabla_{\tau} A^{\sigma}{}_{\kappa})\, u_{\lambda} \quad \Longrightarrow \quad \epsilon^{\gamma\delta(\alpha}\,(\mathbf{e}_{\gamma} - a_{\gamma})\,(A^{\beta)}{}_{\delta}) - 3\,n^{(\alpha}{}_{\gamma}\,A^{\beta)\gamma} + \tfrac{1}{2}\,n^{\gamma}{}_{\gamma}\,A^{\alpha\beta} + \delta^{\alpha\beta}\,n^{\gamma}{}_{\delta}\,A^{\delta}{}_{\gamma}\ . \tag{A.7}$$

# References

[1] J. Ehlers, “Beiträge zur relativistischen Mechanik kontinuierlicher Medien”, *Akad. Wiss. Lit. Mainz, Abhandl. Math.-Nat. Kl.* **11**, (1961).
Translation: J. Ehlers, “Contributions to the Relativistic Mechanics of Continuous Media” *Gen. Rel. Grav.* **25**, 1225 (1993).

[2] J. Wainwright and G. F. R. Ellis (Eds.), *Dynamical Systems in Cosmology* (Cambridge University Press, Cambridge, 1996). To appear.

[3] C. W. Misner, K. S. Thorne and J. A. Wheeler, *Gravitation* (Freeman & Co., New York, 1973).

[4] G. F. R. Ellis, “Cosmological Models from a Covariant Viewpoint”, University of Cape Town Preprint (1996). Talk given at ICGC95. To appear in Conference Proceedings. Ed. Padmanathan et al.

[5] S. W. Hawking, “Perturbations of an Expanding Universe”, *Astrophys. J.* **145**, 544 (1966).

[6] G. F. R. Ellis, “Relativistic Cosmology”, *General Relativity and Cosmology* Proceedings of the XLVII Enrico Fermi Summer School, Ed. R. K. Sachs (Academic Press, New York, 1971).

[7] G. F. R. Ellis, “Relativistic Cosmology”, *Cargèse Lectures in Physics Vol. 6*, Ed. E. Schatzman (Gordon and Breach, New York, 1973).

[8] F. A. E. Pirani, “On the Physical Significance of the Riemann Tensor”, *Acta Phys. Polon.* **15**, 389 (1956).

[9] G. F. R. Ellis, “Dynamics of Pressure-Free Matter in General Relativity”, *J. Math. Phys.* **8**, 1171 (1967).

[10] M. A. H. MacCallum, “Cosmological Models from a Geometric Point of View”, *Cargèse Lectures in Physics Vol. 6*, Ed. E. Schatzman (Gordon and Breach, New York, 1973).

[11] M. A. H. MacCallum, private notes.

[12] R. M. Wald, *General Relativity* (University of Chicago Press, Chicago, 1984).

[13] F. de Felice and C. J. S. Clarke, *Relativity on Curved Manifolds* (Cambridge University Press, Cambridge, 1990).

[14] S. B. Edgar, “The Structure of Tetrad Formalisms in General Relativity: The General Case”, *Gen. Rel. Grav.* **12**, 347 (1980).

[15] E. Newman and R. Penrose, “An Approach to Gravitational Radiation by a Method of Spin Coefficients”, *J. Math. Phys.* **3**, 566 (1962).

[16] R. T. Jantzen, P. Carini and D. Bini, “The Many Faces of Gravitoelectromagnetism”, *Ann. Phys.* **215**, 1 (1992).

[17] S. Matarrese, O. Pantano and D. Saez, “General Relativistic Dynamics of Irrotational Dust: Cosmological Implications”, *Phys. Rev. Lett.* **72**, 320 (1994).

[18] M. Bruni, S. Matarrese and O. Pantano, “Dynamics of Silent Universes”, *Astrophys. J.* **445**, 958 (1995).

[19] M. Bruni, S. Matarrese and O. Pantano, “A Local View of the Observable Universe”, *Phys. Rev. Lett.* **74**, 1916 (1995).

[20] J. M. Stewart and G. F. R. Ellis, “Solutions of Einstein's Equations for a Fluid which Exhibits Local Rotational Symmetry”, *J. Math. Phys.* **9**, 1072 (1968).

[21] G. F. R. Ellis and M. A. H. MacCallum, “A Class of Homogeneous Cosmological Models”, *Commun. Math. Phys.* **12**, 108 (1969).

[22] A. R. King and G. F. R. Ellis, “Tilted Homogeneous Cosmological Models”, *Commun. Math. Phys.* **31**, 209 (1973).

[23] R. T. Jantzen, P. Carini and D. Bini, *Understanding Spacetime Splittings and Their Relationships*, in preparation.

[24] P. Szekeres, “The Gravitational Compass”, *J. Maths. Phys.* **6**, 1387 (1965).

[25] J. L. Synge and A. Schild, *Tensor Calculus* (University of Toronto Press, Toronto, 1949).
Reprinted: (Dover Publ., New York, 1978).

[26] W. Kundt and M. Trümper, “Beiträge zur Theorie der Gravitations-Strahlungsfelder”, *Akad. Wiss. Lit. Mainz, Abhandl. Math.-Nat. Kl.* **12**, (1962).

[27] E. Cotton, *Ann. Fac. Sci. Toulouse (II)* **1**, 385 (1899).

[28] J. W. York, Jr., “Gravitational Degrees of Freedom and the Initial-Value Problem”, *Phys. Rev. Lett.* **26**, 1656 (1971).

[29] H. Stephani, *Allgemeine Relativitätstheorie* (4. Aufl.) (Dt. Verlag d. Wissenschaften, Berlin, 1991).
Translation: H. Stephani, *General Relativity* (Cambridge University Press, Cambridge, 2nd Edn. 1990).

[30] S. Boersma and T. Dray, “Slicing, Threading and Parametric Manifolds”, *Gen. Rel. Grav.* **27**, 319 (195).

[31] L. D. Landau and E. M. Lifshitz, *The Classical Theory of Fields* (Pergamon Press, Oxford, 4th Edn. 1975).

[32] A. H. Taub, “Stability of Fluid Motions and Variational Principles”, *Proceedings of the 1967 Colloque on “Fluids et Champ Gravitationel en Relativité Générale,”* No. **170**, (Centre National de la Recherche Scientifique, Paris), 57 (1969). 1 (1992).

[33] J. W. York, Jr., “Kinematics and Dynamics of General Relativity”, *Sources of Gravitational Radiation, Procceedings of the Battelle Seattle Workshop*, Ed. L. L. Smarr (Cambridge University Press, Cambridge, 1979).

[34] W. Israel, “Nonstationary Irreversible Thermodynamics: A Causal Relativistic Theory”, *Ann. Phys. (N.Y.)* **100**, 310 (1976).

[35] W. Israel and J. M. Stewart, “Transient Relativistic Thermodynamics and Kinetic Theory”, *Ann. Phys. (N.Y.)* **118**, 341 (1979).

[36] A. Papapetrou, “Quelques Remarques sur le Formalisme de Newman–Penrose”, *C. R. Acad. Sci. (Paris)* **A 272**, 1537 (1971).

[37] A. Papapetrou, “Les Relations Identiques entre les Équations du Formalisme de Newman–Penrose”, *C. R. Acad. Sci. (Paris)* **A 272**, 1613 (1971).

[38] E. Cartan, *Leçons sur la Geomètrie des Espaces de Riemann* (Gauthier-Villars, Paris, 2nd Edn. 1946).

[39] A. Karlhede, "A Review of the Geometrical Equivalence of Metrics in General Relativity", *Gen. Rel. Grav.* **12** (1980), 693.

[40] M. Bradley and A. Karlhede, "On the Curvature Description of Gravitational Fields", *Class. Quantum Grav.* **7** (1990), 449.

[41] H. Bondi, "The Contraction of Gravitating Spheres", *Proc. Roy. Soc. London A* **281**, 39 (1964).

[42] B. K. Harrison, K. S. Thorne, M. Wakano and J. A. Wheeler, *Gravitation Theory and Gravitational Collapse* (The University of Chicago Press, Chicago and London, 1965).

[43] C. B. Collins, "More Qualitative Cosmology", *Commun. Math. Phys.* **23** (1971), 137.

[44] C. G. Hewitt and J. Wainwright, "Orthogonally Transitive $G_2$ Cosmologies", *Class. Quantum Grav.* **7**, 2295 (1990).

[45] K. Rosquist and R. T. Jantzen, "Unified Regularization of Bianchi Cosmology", *Phys. Rep.* **166**, 89 (1988).

[46] C. Uggla and H. von Zur-Mühlen, "Compactified and Reduced Dynamics for Locally Rotationally Symmetric Bianchi Type IX Perfect Fluid Models", *Class. Quantum Grav.* **7**, 1365 (1990).

[47] D. M. Eardley, "Self-Similar Spacetimes: Geometry and Dynamics", *Commun. Math. Phys.* **37**, 287 (1974).

[48] D. Kramer, H. Stephani, M. A. H. MacCallum and E. Herlt, *Exact Solutions of Einstein's Field Equations* (VEB Dt. Verlag d. Wissenschaften, Berlin, 1980).

[49] J. Stewart, *Advanced General Relativity* (Cambridge University Press, Cambridge, 1990).

[50] G. Ludwig and S. B. Edgar, "Integration in the GHP Formalism I: A Coordinate Approach with Applications to Twisting Type N Spaces". To appear in *Gen. Rel. Grav.* (1996).

[51] A. Barnes and R. R. Rowlingson, "Irrotational Perfect Fluids with a Purely Electric Weyl Tensor", *Class. Quantum Grav.* **6**, 949 (1989).

[52] C. W. Misner, "The Isotropy of the Universe", *Astrophys. J.* **151**, 431 (1968).

[53] W. M. Lesame, P. K. S. Dunsby and G. F. R. Ellis, "Integrability Conditions for Irrotational Dust with a Purely Electric Weyl Tensor: A Tetrad Analysis", *Phys. Rev.* D **52**, 3406 (1995).

[54] B. K. Berger, D. M. Eardley and D. W. Olson, "Note on the Spacetimes of Szekeres", *Phys. Rev.* D **16**, 3086 (1977).

[55] P. Szekeres, "A Class of Inhomogeneous Cosmological Models", *Commun. Math. Phys.* **41**, 56 (1975).

[56] D. A. Szafron and C. B. Collins, "A New Approach to Inhomogeneous Cosmologies: Intrinsic Symmetries. II. Conformally Flat Slices and an Invariant Classification", *J. Math. Phys.* **20**, 2354 (1979).

[57] M. Bruni, H. van Elst and C. Uggla, in preparation.

[58] H. van Elst and G. F. R. Ellis, "The Covariant Approach to LRS Perfect Fluid Spacetime Geometries", *gr-qc/9510044*, QMW/UCT Preprint, (1995). To appear in *Class. Quantum Grav.*.

[59] J. Wainwright, "A Classification Scheme for Non-Rotating Inhomogeneous Cosmologies", *J. Phys. A: Math. Gen.* **12**, 2015 (1979).

[60] R. T. Jantzen, “Spatially Homogeneous Dynamics: A Unified Picture”, in Proc. Int. Sch. Phys. “E. Fermi” Course LXXXVI on “Gamov Cosmology” (R. Ruffini, F. Melchiorri, Eds.) (North Holland, Amsterdam, 1987), and in Cosmology of the Early Universe (R. Ruffini, L.Z. Fang, Eds.) (World Scientific, Singapore, 1984).

[61] C. Uggla and R. T. Jantzen, in preparation.

[62] C. Uggla, R. T. Jantzen and K. Rosquist, “Exact Hypersurface-Homogeneous Solutions in Cosmology and Astrophysics”, *Phys. Rev.* D **51**, 5522 (1995).

[63] C. G. Hewitt and J. Wainwright, “Dynamical Systems Approach to Tilted Bianchi Cosmologies: Irrotational Models of Type V”, *Phys. Rev.* D **46**, 4242 (1992).

[64] U. Nilsson and C. Uggla, “Spatially Self-Similar Locally Rotationally Symmetric Perfect Fluid Models”, *gr-qc/9511064*. Submitted to *Class. Quantum Grav.*.

[65] U. Nilsson and C. Uggla, “Self-Similar Spherically Symmetric Perfect Fluid Models”, in preparation.

[66] C. G. Hewitt and J. Wainwright, “A Dynamical Systems Approach to Bianchi Cosmologies: Orthogonal Models of Class B”, *Class. Quantum Grav.* **10**, 99 (1993).

[67] R. Penrose and W. Rindler, *Spinors and Space-Time, Vol. I* (Cambridge University Press, Cambridge, 1984).

[68] R. A. d'Inverno and J. Smallwood, “Covariant 2+2 Formulation of the Initial-Value Problem in General Relativity”, *Phys. Rev.* D **22**, 1233 (1980).

* * *